\documentclass[12pt,preprint]{aastex}
\usepackage{float}
\usepackage{epsfig}
\usepackage{lscape}
\usepackage{rotating}
\usepackage{multirow}
\usepackage{natbib}
\newcommand{\ch}{{\it Chandra\/}}
\newcommand{\xmm}{{\it XMM\/}}
\newcommand{\xmmn}{{\it XMM-Newton}}

\def\oiii{\hbox{[O\ {\sc iii}}]}
\def\loiii{$L_{\rm [O\ III]}$}
\def\lx{$L_{\rm X}$}

\begin{document}

\title{An Archival \ch\ and \xmmn\ Survey of Type 2 Quasars}

\author{Jianjun Jia\altaffilmark{1}, Andrew Ptak\altaffilmark{2}, Timothy 
Heckman\altaffilmark{1}, Nadia L. Zakamska\altaffilmark{1}}
\affil{\altaffilmark{1}Department of Physics and Astronomy, Johns Hopkins 
University, Baltimore, MD 21218, USA\\
\altaffilmark{2}Goddard Space Flight Center, Greenbelt, MD 20771, USA}

\begin{abstract}
In order to investigate obscuration in high-luminosity type 2 AGN, we 
analyzed the \ch\ and \xmmn\ archival observations for 71 type 2
quasars detected at $0.05<z<0.73$. 
These objects were selected by cross-correlating the largest catalog of 
optically identified type 2 quasars to date selected from Sloan Digital Sky 
Survey (SDSS) with the \ch\ and \xmmn\ archives. The type 2 quasar sample was 
selected based on the high equivalent width  of $\oiii$$\lambda$5007 optical emission line which
we assume to be an 
approximate indicator of the intrinsic AGN luminosity.
The archival X-ray spectra were fitted with absorbed power-law models to 
characterize the spectral properties of each source. 
For 54 objects with good spectral fits, the observed hard X-ray luminosity
ranges from $2\times 10^{41}$ to $5.3\times 10^{44}~{\rm erg~s^{-1}}$, with
the median of $1.1\times 10^{43}~{\rm erg~s^{-1}}$.
We find that the means 
of the column density and photon index of our sample are $\log N_{\rm 
H}=22.9$ cm$^{-2}$ and $\Gamma=1.87$ respectively.
The observed ratios of hard X-ray and 
$\oiii$ line luminosities imply that the majority of our sample suffer 
significant amounts of obscuration in the hard X-ray band.
We also fit the spectra using a more physically realistic model which 
accounts for both Compton scattering and a potential partial covering of the 
central X-ray source to estimate the true absorbing column density and use 
simulations to reproduce the observed \lx/\loiii\ ratios. We find that the 
absorbing column density estimates based on simple power-law models 
significantly underestimate the actual absorption in approximately half of 
the sources. Eleven sources show a prominent Fe K$\alpha$ emission line (EW$>$100 eV in rest frame), and 
we detect this line in the other sources through a joint fit (spectral 
stacking). The correlation between the Fe K$\alpha$ and $\oiii$ fluxes and 
the inverse correlation of the equivalent width of Fe K$\alpha$ line with the 
ratio of hard X-ray and $\oiii$ fluxes is consistent with previous results 
for lower luminosity Seyfert 2 galaxies. We conclude that obscuration is the 
cause of the weak hard X-ray emission rather than intrinsically low X-ray 
luminosities. We find that about half of the population of optically-selected 
type 2 quasars are likely to be Compton-thick. We also find no evidence that 
the amount of X-ray obscuration depends on the AGN luminosity (over a range 
of more than three orders-of-magnitude in luminosity).

\end{abstract}


\section{Introduction\label{introduction}}

In the standard unification model, all active galactic nuclei (AGN) are 
powered by accretion onto supermassive black holes (SMBHs), with different 
geometries resulting in various types of AGNs \citep{antonucci93}.  That is, 
AGN are grossly classified by whether broad emission lines are (type 1) or 
are not (type 2) present in the optical and UV spectrum. In the unified 
model, the central accretion disk and surrounding retinue of high velocity 
gas is directly visible in type 1 AGN, while this region is blocked from a 
direct view by a toroidal obscuring structure in type 2 AGN. In the local 
universe, low-luminosity type 2 AGNs (type 2 Seyfert galaxies) are found to 
be as abundant as type 1 AGNs (type 1 Seyfert galaxies), and the 
applicability of the unified model is well-established
\citep[e.g.,][]{2005AJ....129.1795H}. Given the strong 
cosmic evolution of the AGN population, the most luminous AGNs are very rare 
in the local universe and this population is only well-characterized at high 
redshift. Unfortunately, the heavy obscuration by the dense gas and dust 
surrounding the SMBH makes type 2 AGNs much fainter than type 1 AGNs and they 
become difficult to discover at high redshifts. It is therefore unclear 
how well the standard unified model works for AGN of the highest luminosities 
and at high redshifts.

Indeed, X-ray surveys have shown that the ratio of type 2 to type 1 AGN 
decreases with AGN X-ray luminosity \citep{ueda03,sazonov04,barger05,
treister05, akylas06, gch07, fiore08, treister08, 2012AdAst2012E..16T}, 
but see \cite{2006MNRAS.372.1755D} for 
different results. This anti-correlation between obscuration and luminosity
is in contrast to the results from the 
infrared, radio and optical surveys (Reyes et al. 2008, see 
Lawrence \& Elvis 2010 for a review), which suggest that obscured AGNs are 
about as common as the unobscured ones at the highest probed luminosity.

In this paper we will explore the hard X-ray and optical emission-line 
properties of the largest {\it optically selected} sample available to date 
of highly luminous type 2 AGN. We will then compare these properties to those 
of typical low-luminosity AGN to test the unified model at high luminosity. 
We note that throughout the rest of our paper, we will use the term `Seyfert' 
to refer to low-luminosity AGN, and `quasar' to refer to high-luminosity AGN 
(with a dividing line at a bolometric luminosity greater than $10^{45}~{\rm 
ergs~s^{-1}}$). 

A large sample of type 2 quasars are needed in order to test how and if the 
unified model applies at high luminosities. Although the central engine is 
hidden from view in type 2 AGN, the strong UV radiation escaping along the 
polar axis of the obscuring material distribution photo-ionizes
circum-nuclear gas leading to strong  
narrow high-ionization emission-lines.  Since this narrow-line region is at 
larger radii than the bulk of the obscuring material, selection 
based on narrow optical emission lines promises to be less biased against 
type 2 AGN than hard (E $<$ 10 keV) X-ray surveys \citep[see, 
e.g.,][]{lamassa09,lamassa10}. Since the narrow line emission mechanism is 
the same for both type 1s and 2s in the standard AGN model, we can
expect that the line 
luminosity serves as an indicator of the intrinsic luminosity of the nucleus,
especially for the $\oiii\lambda 5007$ emission line, which is the strongest line in the optical spectra, 
and is not heavily contaminated by star-forming activities \citep{2004MNRAS.351.1151B,2004ApJ...613..109H}.
When compared with the observed hard X-ray luminosity, it can also serve as a diagnostic of 
X-ray obscuration \citep{bassani99,2010A&A...519A..92G}.

\citet[hereafter Z03]{zakamska03} selected 291 type 2 quasars at redshifts 
$0.3<z<0.83$ based on their optical emission line properties from the 
spectroscopic data of the Sloan Digital Sky Survey \citep[SDSS;][]{york00}. 
They found strong narrow emission lines with high-ionization line ratios but 
no broad emission lines in these objects, and therefore identified them as 
type 2 quasar candidates based on \oiii$\lambda$5007 emission-line 
luminosities greater than $10^8~L_{\odot}$. This new method has greatly 
expanded the number of type 2 quasars known, and it allows the properties of 
type 2 quasars to be studied in detail. Subsequent multi-wavelength studies 
\citep{zakamska04,zakamska05,zakamska06,ptak06,vignali06} confirmed that the 
standard models for AGNs could give good descriptions of those optically 
selected type 2 quasars. \citet[hereafter V10]{vignali10} recently studied 
the X-ray spectra of 25 type 2 quasars from \cite{zakamska03}, by comparing 
the measured hard X-ray luminosity with the intrinsic (de-absorbed) X-ray 
luminosity derived from the $\oiii$$\lambda$5007\AA~and mid-IR (5.8$\mu$m and 
12.3$\mu$m) line estimators, and concluded that about half of the SDSS type 2 
quasars with exceptionally high luminosities (\loiii$>10^{9.3}L_{\odot}$) 
might be Compton-thick (absorbing column density $N_{\rm H}>10^{24} \rm cm^{-
2}$). The bolometric luminosities of these quasars are difficult to determine accurately,
 but their high overall energetics can be gleaned from the mid-infrared data (Spitzer and WISE), 
 where obscuring material thermally re-emits much of the absorbed radiation \citep{zakamska08} 
 and monochromatic luminosities $\nu L_{\nu}$ well in excess of $10^{45}$ erg s$^{-1}$ are often seen. 
 Our estimate for bolometric luminosities based on comparison of \oiii luminosities in type 1
  and type 2 quasars is presented in \citet{2009ApJ...702.1098L}; $L_{\rm bol}$ is 
  about $10^{45}$ erg s$^{-1}$ at \loiii $=10^8L_{\odot}$ and increases approximately linearly with \loiii\ thereafter. 

By applying the same selection technique to the more recent data, a catalog 
containing 887 type 2 quasars from SDSS was released by 
\citet[hereafter R08]{reyes08}, which expanded the original sample by a 
factor of four, preferentially at higher $\oiii$ luminosities. We
selected the objects covered in X-ray archival  
observations from this pool, and investigated their X-ray properties. These 
objects provide the largest sample of X-ray type 2 quasars which have no bias 
with respect to X-ray luminosity, since they are selected on the basis of 
optical line emission. In this paper, we present our study of 71 type 2 
quasars observed by \ch\ and \xmmn. Section \ref{pipeline} describes our 
sample selection and data analysis. Section \ref{spec} gives the X-ray 
spectral analysis. We discuss our results in Section \ref{results} and come 
to conclusions in Section \ref{con}. An $h=0.7$, $\Omega_m=0.3$ 
and $\Omega_{\Lambda}=0.7$ cosmology is assumed throughout this paper 
\citep{spergel03}. 

\section{Sample Description and Data Analysis\label{pipeline}}

By correlating those 887 optically selected type 2 quasars with the public 
\ch\ (within an 8$\arcmin$ search radius) and \xmmn\ (within a 15$\arcmin$ 
search radius) archives, 71 quasars were found to be covered by \ch\ or 
\xmmn\ or both as of February 2011
\footnote{This work was performed using the High-Energy 
Astrophysics Science Archive (HEASARC), http://heasarc.gsfc.nasa.gov}. The 
list of the coordinates, Galactic column density, redshift, 
observed $\oiii\lambda$5007\AA~luminosity, observation ID, exposure time, observation 
date and off-axis angle for each target is given in Table \ref{t:params}, 
where objects are identified by their J2000 coordinates and shortened to 
\textit{hhmm+ddmm} notation elsewhere. 
We obtain the radio fluxes of our sample from the FIRST \citep{1998AJ....115.1693C} and NVSS \citep{1995ApJ...450..559B} radio catalog.
By  assuming a power-law $F_{\nu}\propto \nu^{\alpha}$ with the spectral index $\alpha=-1$ at 1.4 GHz and comparing their rest-frame luminosity $\nu L_{\nu}\rm{(1.4GHz)}$
with $\oiii\lambda$5007\AA~luminosity, 6 of them are classified as radio loud (RL) sources \citep{1999AJ....118.1169X,zakamska04},
which are 0812+4018, 0834+5534, 1119+6004, 1347+1217, 1411+5212, and 1449+4221.
Some sources were also studied and 
published in other papers, and they are marked in the last column of Table 
\ref{t:params}. 9 objects have multiple observations, and the number of 
total \ch\ and \xmm\ observations for the whole sample is 85. In 52 of them, 
the sources in our sample are the targets of observations.

The data pipeline is done by \texttt{XAssist}\footnote{version 0.9993, 
\url{http://xassist.pha.jhu.edu}}, which is a software package for automatic 
analysis of X-ray astrophysics data. \texttt{XAssist} generates the light curves and 
can filter the raw data for flaring by its default parameter setting. However, we 
also checked the light curve and filtered the flaring of each observation manually.
Point sources with sufficient photons 
are detected by \texttt{XAssist} automatically. In cases where 
sources are not detected due to insufficient counts, user-specified region 
files which contain the source coordinates are supplied as input to 
\texttt{XAssist}. CIAO (ver. 4.3) and XMMSAS (ver. 10.0.0) were called in 
processing \ch\ and \xmmn\ data, respectively. The size of each point source 
extraction region is set by fitting an elliptical Gaussian function to a 
``stamp'' image for each source, which typically results in a region size of 
2$\arcsec$ (\ch) and 18$\arcsec$ (\xmmn) for on-axis sources. Depending on 
how large the off-axis angles are, the region sizes of \ch\ sources vary from 
about 4$\arcsec$ to 9$\arcsec$, and those of \xmmn\ sources vary from about 
20$\arcsec$ to 40$\arcsec$. 
The fraction of energy encircled in these extraction regions from PSF integration is above 80\%
 \citep{2004SPIE.5165..423A,2011A&A...534A..34R}. 
Background regions are set as annuli centered on 
the sources, but if the source is located in a crowded region or on the edge 
of the detector, another circular region in the field was chosen manually 
for background extraction.

\section{Spectral analysis\label{spec}}

We extract the spectra in the energy range of 0.3-8 keV for the \ch\ observations.
For the \xmm\ ones, we used the 0.3-10 keV regime. Although the 8-10 keV emission
of \xmm\ data might be dominated by background and spurious spectral lines, 
the spectral results are nearly the same as if the 8-10 keV data are removed for the 
weak X-ray sources.
X-ray spectral fitting is 
performed with \texttt{XSPEC (ver.12)}. The spectra are grouped to one count 
per bin, and the $C$-statistic \citep{cash79} is used in fitting the spectra. 
Although the $C$-statistic is devised for unbinned spectra, $C$-statistic 
fitting in \texttt{XSPEC} performs better if the spectra are binned to at 
least one count per bin \citep{teng05}. For those sources with more than 200 
photon counts collected, we group their spectra to 10  (total counts less than 500)
or 20 (total counts more than 500) counts per bin, and use 
$\chi^{2}$ statistic in the spectral fitting. X-ray photons are collected by 
three detectors on \xmmn\, i.e., PN, MOS1 and MOS2. The two MOS spectra are 
combined and fitted simultaneously with PN spectra in \texttt{XSPEC}, and all 
parameters are tied together except for a constant multiplicative factor
to account for relative flux calibration differences among the detectors. 
Five \xmmn\ sources have counts detected in only one or two of the three 
detectors, which are noted in the second column in Table \ref{t:fits}.  
Errors are calculated at 90\% significance, i.e., $\Delta \chi^2$ or $\Delta 
C$ =2.7 for one parameter of interest \citep{1976ApJ...210..642A}. 

The X-ray spectra of obscured (type 2) AGN are complicated and usually 
consisted of multiple components: power-law, thermal, scattering, reflection,
and emission lines \citep[see][]{turner97,risaliti02,ptak06,lamassa09}. 
Thus, no single model could fit the spectra well in all cases. 
We carry out the spectral fit with \texttt{XSPEC} using several spectral models:

(1) Single-absorber power-law: Initially, the spectrum is fitted as a power-law 
continuum absorbed by the Galactic column density ($N_{\rm{H,G}}$) and an 
intrinsic redshifted absorption column density ($N_{\rm{H}}$). This model 
results in three free parameters: the column density $N_{\rm H}$, the photon 
index $\Gamma$, and the power-law normalization. The Galactic neutral 
hydrogen column density $N_{\rm{H,G}}$ is a fixed parameter \citep{dickey90}, 
which is calculated from HEAsoft $N_{\rm{H}}$ tool. However, in some cases, 
we fixed the photon index at $\Gamma=1.7$ (which is a typical value for AGN, 
\citealt{nandra05}) if it is unconstrained, i.e., the errors exceeded 
reasonable bounds. 

(2) Double-absorber power-law: In some cases, a single absorbed power-law cannot
model the data accurately and a two-absorber model could be an approximation to the case of X-ray photons being
scattered into the line of sight \citep{turner97,ptak06,lamassa09}.
We applied this model on 17 sources and considered this approach to be the best-fitting 
model. The photon indices of both power-law components are tied 
together when fitting the spectra. However,  tying the photon indices in the 
case of SDSS~J1034+6001 results in a very large $\chi^2$, and we thus use two 
different indices in fitting its spectrum. For those sources which have very small values 
for  $N_{\rm H,1}$ (lower than $N_{\rm{H,G}}$) during spectral fitting, 
we then fixed their values to $N_{\rm{H,G}}$.

(3) Absorbed power-law plus Gaussian Fe K$\alpha$ line: 
Eleven objects show visually-detected Fe K$\alpha$ emission lines, and a Gaussian 
component was added to the best-fitting power-law continuum. We initially fixed the line 
energy $E_{\rm{line}}$ at 6.4 keV (in the source rest frame) and the line width 
($\sigma$) at 0.01 keV ($\sim 10\%$ of the instrumental line resolution for 
\ch\ and \xmmn). In \texttt{XSPEC}, we first \texttt{ignore} the photon 
counts in the energy range of 5-7 keV to get the power-law index of the 
continuum, and then \texttt{notice} them to fit the emission line around 6.4 
keV. The line energy of 0834+5534 is around 6.7 keV instead of 6.4 keV.

We list the photon counts, the column densities and the 
photon indices of the best-fitting spectral fits for 54 sources which have enough photon counts
to result in a moderate-quality spectral fit in Table \ref{t:fits}, as well as the 
derived observed and intrinsic (de-absorbed) 2-10 keV luminosities and the ratios of the X-
ray to $\oiii$ luminosity. For the cases with double power-law fits, we also list the ratio
of the normalization of both power-law components.
Some quasars have very small column densities in 
the spectral fits, and we use the upper limit instead in Table \ref{t:fits}. 
The spectral plots of each quasar are shown in Figure \ref{f:all}
\notetoeditor{Figure 1 is online-only figures. Please make the first 6 
subfigures of Figures 1 appear in print.}.
For those with multiple observations in either \ch\ or \xmm\ or both, we 
also report in Table \ref{t:fits} the column density, photon index and $\chi^2$ from 
the simultaneous fits of all spectral data, and we use these values in following discussions.  
Discrepancies between each individual observation are discussed in Appendix \ref{appendixb}.

There are 17 sources whose observations are dominated by background. 
The photon counts are too low to constrain 
the spectral parameters in spectral fitting. Therefore, we calculate the 
upper limit of the 2-10 keV flux at a 3-$\sigma$ level. We assume that their 
spectra are an absorbed power-law with $\Gamma=1.7$ and $N_{\rm 
H}=10^{23}~\rm{cm}^{-2}$, which is close to the mean value of the column densities given 
in Table \ref{t:fits} (see Section \ref{gamma}) \footnote{The low photon counts of 
0028$-$0014 and 1606+2725 might be due to their short effective exposure time rather than 
heavy absorption. However, we use the same assumption as the other 15 sources to estimate 
their upper limit fluxes.}
. The 3-$\sigma$ upper limit of the 2-10 keV photon 
count rates are calculated by using the Bayesian statistical method by 
\citet{kraft91}. We determined the count rate to flux conversion 
coefficient using \texttt{XSPEC}, and multiply it by the count rate upper 
limit to calculate the 2-10 keV flux upper limit. The detected counts, the 
source count upper limits, and the associated upper limits on the count 
rates, fluxes and luminosities are listed in Table \ref{t:upper}.
Table \ref{t:fe} lists the Gaussian fit parameters of the iron lines as well as the equivalent width 
and line luminosity. The change of $\chi^2$ if we remove the 
Gaussian component from the spectral fit is also listed in Table \ref{t:fe} to show how significant this 
emission line is.

\section{Results and discussions\label{results}}

\subsection{Column density and photon index distribution\label{gamma}} 

Of our 71 quasars, at least crude spectral fitting is possible for 54. For 
these, we find that the mean power-law index is $\Gamma=1.87\pm 0.65$
using the best-fitting results in Table \ref{t:fits} (those with 
photon indices fixed at 1.7 are excluded), 
where the error bar is the standard deviation of the power-law indices of 
the sample neglecting the individual fitting errors. 
In the case that there are multiple observations for one 
object, we use the values of the simultaneous joint fit instead. 
Multiple observations may give different fluxes or observed luminosities due to
AGN variability. However, the spectral shape between different observations does 
not change significantly (see Figure \ref{f:multi}). Thus, it is safe for us to use
the photon index derived from the simultaneous joint fit.
The 6 sources also claimed as RL sources have a mean photon index of 2.14
compared to 1.83 for the remainder of the sample. Therefore, their presence 
does to affect the statistical result of the photon index distribution.
The mean value of our sample is consistent with the result from a sample of 
type 2 AGNs in the {\it SWIFT}-BAT survey, which finds the mean value of 
photon index of the continuum power-law in the energy regime 15-195 keV is 
$\Gamma=1.90\pm 0.27$ \citep{2011ApJ...728...58B}. It is also roughly consistent with
that found in a sample of obscured AGNs selected by 
\textit{INTEGRAL}, which is $\Gamma=1.68\pm 0.30$. 
\citep{2012MNRAS.420.2087D}. However, if we 
use only the results in Table \ref{t:fits} for double-absorber power-law fits, 
it becomes larger, i.e., $\Gamma=2.14\pm 0.60$. This distribution is 
much like the one found in the best fits of a sample of local Seyfert 2s 
studied by \citet{lamassa09}, where more than half of the objects have
double-absorbed power-laws as their best-fitting model. 
Since the soft X-ray with steep slope could be biasing the spectral fit with power-law slopes tied, i.e., 
the slope of AGN only is flatter than the slope of AGN plus star formation, 
this might result in the larger index of double-absorber power-law.
We show the comparison between our best-fitting results and their samples in 
Figure \ref{f:gamma}, where we use different bins for the sample of
\cite{2011ApJ...728...58B} for display purpose.

By excluding those with upper limits or fixed at $N_{\rm H,G}$ for their column densities in spectral fits,
we find the mean $N_{\rm{H}}$ of our sample is $\log N_{\rm{H}}=22.9\pm0.9$ cm$^{-
2}$ using $N_{\rm H,1}$ for single power-law and $N_{\rm H,2}$ for
double power-law fits from 
the best-fitting models listed in Table \ref{t:fits}.  The $N_H$ distribution is consistent with those Seyfert 2s as shown in 
Figure \ref{f:nh}. We will discuss the possible luminosity dependence of 
obscuration in the following sections.

\subsection{The \lx/\loiii\ ratio as an indicator of obscuration}

As the $\oiii$$\lambda$5007 line emission originates in the narrow line 
region and so is not affected by the circumnuclear obscuration, 
the ratio between the observed hard X-ray (2-10 keV) and $\oiii$ line 
luminosity could be used as an indicator of the obscuration of the 
hard X-ray emission \citep{mulchaey94,heckman05, 2006A&A...455..173P,lamastra09,lamassa09,2010ApJ...722..212T}.
In Figure \ref{f:lxlo}, we plot a histogram of the \lx/\loiii\ ratios for our 
sample listed in Table \ref{t:fits}. We also show the observed distributions for type 1 (dashed blue 
line) and type 2 AGNs (dot-dashed red line) \citep{heckman05}. The X-ray to 
$\oiii$ luminosity ratio of our sample agrees well with that of type 2
AGNs from \cite{heckman05} with a KS test $P=0.645$, indicating that this sample is also 
likely experiencing obscuration. However, the fitted obscuring column 
densities inferred from the single-absorber power-law spectral fits are often 
too low to be consistent with the \lx/\loiii\ ratios of type 2 quasars, e.g., 
the single-absorber model likely underestimates the amount of X-ray 
obscuration in our sample. Thus, we estimate their obscuration in the 
following subsection by using the X-ray to \oiii\ ratios.

\subsection{Estimation of the absorbing column density \label{simulation}}

Compared with the local Type 1 AGNs, the derived observed \lx/\loiii\ ratio in Table \ref{t:fits}
implies that the targets in our sample are more highly obscured 
than would be implied by the fitted column densities $N_{\rm{H}}$ from our 
spectral models, i.e., the column density is underestimated in our spectral 
fits for at least half of the whole sample.
We therefore use the correlation between the hard X-ray and $\oiii$ 
luminosity for both type 1 and 2 AGNs \citep{heckman05} to more realistically 
estimate the absorbing column densities of our targets \citep{lamassa09}. We 
employ a Monte Carlo approach to take the dispersion in the Sy 1 \lx/\loiii\ 
distribution into account. First we generate 1000 random numbers which 
follow a Gaussian distribution with the same mean and dispersion as the 
$L_{2-10\ \rm keV}/L_{\rm [O III]}$ distribution of unobscured (type 1) AGNs 
in \cite{heckman05}. For each AGN in our sample, the simulated
  unabsorbed 2-10 keV X-ray luminosities are
computed by multiplying the observed $\oiii$ luminosity by the
random draws from the Sy 1 $L_{2-10\ \rm keV}/L_{\rm [O III]}$
distribution. The difference between these simulated unobscured X-ray luminosities
and the observed value is considered to be due to absorption. In order
to assess how much absorption is consistent with the difference
between the simulated and observed X-ray luminosities, we
tabulated the expected fluxes and count rates for a partial covering 
model with covering fraction of 0.99 and photon index fixed at 1.7 and 
column densities varying from 0 to $10^{25}$ cm$^{-2}$. We then 
interpolated the effective column density $N_{\rm{H,sim}}$ that predicts a 
model count rate consistent with the observed count rate for each
AGN.

We compare the results from these simulations and the absorbed
power-law spectral fits in Figure \ref{f:nhsim}. The fitted $N_{\rm 
H}$ values from the single-absorber model (black plus symbols) are 
systematically lower than the simulated column densities, while $N_{\rm H,2}$ 
from double-absorber model (red asterisks) are more consistent with the 
simulated column densities, showing that not surprisingly more complex 
spectral models do a better job of recovering the intrinsic column density 
implied by the attenuated X-ray flux relative to the $\oiii$ emission.

Additionally, we used the \textit{plcabs} model in \texttt{XSPEC} 
\citep{yaqoob97} to fit the spectra in order to approximately take Compton 
scattering into account.  This model assumes a spherical covering which is not 
likely to be the case but is nevertheless an improvement over fitting
with absorption models that do not include scattering.  In future work
we will consider more advanced absorption models such as MyTorus for
sources with high enough signal-to-noise to warrant more advanced fitting. 
The results from fitting with both the simple partial covering model and \textit{plcabs} are shown in 
Table \ref{t:sim}, where the lower limits for the simulated $N_{\rm H}$ are 
derived for non-detections based on the upper limits for the photon count rates in Table 
\ref{t:upper}. As shown in Table \ref{t:sim} about half of the sources have a 
fitted column density $N_{\rm{H,plcabs}}$ much lower than the simulated 
$N_{\rm{H,sim}}$. This indicates that direct spectral fitting still 
underpredicts the column density even by introducing Compton scattering in
some cases, which reaffirms the necessity of using \lx/\loiii\ ratio as
an indicator of intrinsic obscuration.
In summary these results imply that high signal-to-noise 
broadband spectra fitted with more complex (and realistic) models are more 
likely to recover the true (higher) column densities than simple power-law 
fits. This is also seen in lower luminosity Seyfert 2 galaxies 
\citep{lamassa09, rigby09, melendez09}. 

\subsection{Iron Line Emission}
By visual examination of the spectra, iron emission line is found in 11 of the type 2 quasars.
The line energy and equivalent width (EW), both in rest frame, are listed in Table \ref{t:fe},
as well as the line luminosity, $\chi^2$ and degrees of freedom in spectral fitting.
For the rest of the sample which do not show a significant Fe K$\alpha$ 
component in their individual spectra, we grouped them according to their 
observed \lx/\loiii\ ratio, and then applied a `spectral' stacking procedure
or also referred as simultaneous spectral fitting. 
In Table \ref{t:stack}, we show the four bins of the X-ray to $\oiii$ 
luminosity ratio that are used to group the sources, 
and we exclude those with photon counts fewer than 10 in 
the 2-10 keV band. We load the spectra of the objects in the same bin into 
\texttt{XSPEC} and only fit their spectra in the 3-8 keV range to minimize 
the impact of the spectral complexity discussed above. We assume that they 
have approximately the same properties for the power-law continuum and iron 
emission line. The intrinsic line width ($\sigma$) in the Gaussian component 
is fixed at 0.01 keV (i.e., unresolved for CCD spectra), and the photon 
indices of the continuum power-law is fixed at 1.7. The spectrum of each 
object is not physically shifted to account for redshift since the redshift 
is instead taken into account in the spectral model.  In each group, the normalization of 
the power-law component and parameters of the Gaussian 
component for each source are tied together between the fits. As we assume the sources 
in the same group suffer similar obscuration, tying the parameters can ensure
that the sources with similar \lx/\loiii\ ratio have the same EW of iron line. 
But their relative intensity (both continuum and emission line)  
for each source is allowed to be free, which is controlled by a constant factor during fitting. 
The line energy and equivalent width of iron line of each bin are shown in Table \ref{t:stack}.

We show the correlation between the (effective average) Fe K$\alpha$ EW
and the ratio of hard X-ray and $\oiii$ luminosities (\lx/\loiii) in 
Figure \ref{f:felx}.  This includes the stacking procedure along with the 11 
quasars with prominent iron lines in Table \ref{t:fe} (black plus symbols 
with error bar), the four groups classified by their \lx/\loiii\ ratio in 
Table \ref{t:stack} (blue plus symbols without error bar), and the sample of 
type 2 Seyfert galaxies from \cite{lamassa09} (red asterisk with error bar). 
Two objects (SDSS~J1218+4706 and SDSS~J1238+0927) are included in both our sample and 
theirs, and we use the EW and luminosity in Table \ref{t:fe} to make the 
plots as both papers give similar results. In order to fit the correlation
by taking the upper limits into account, we use survival analysis 
\texttt{ASURV} Rev 1.2, which implements the method presented in 
\cite{isobe90} and \cite{lavalley92} to investigate the correlation between 
these two parameters ($\log$ EW in units of eV and \lx/\loiii), which uses the 
bivariate data algorithm by \cite{isobe86}. The correlation coefficient found 
in the survival analysis is $-0.52\pm0.10$ with a $>3\sigma$ significance.

We also investigate the correlation between the iron emission line luminosity 
and the $\oiii$ luminosity by applying survival analysis. This is shown in 
Figure \ref{f:feloiii}, which includes the 11 individual objects listed in 
Table \ref{t:fe} (symbols in black), the sample from \cite{lamassa09} 
(symbols in red) and those in our sample with no visually-detected iron lines 
(symbols in blue). For those not listed in Table \ref{t:fe}, we grouped them 
in bins defined by their \oiii\ luminosities. The iron line luminosity in 
each bin is calculated as the mean of \loiii\ by multiplying the ratio of 
$<f_{\rm Fe}>/<f_{\rm [OIII]}>$, where $<f_{\rm Fe}>$ and $<f_{\rm [OIII]}>$ 
are the means of iron line and \oiii\ fluxes in each bin respectively. The 
mean values of iron line luminosity in \loiii\ bins are listed in 
Table \ref{t:loiii}, where the error of $L_{\rm{Fe}}$ is calculated using 
error propagation of $\delta f_{\rm Fe}$ and $\delta f_{\rm [OIII]}$. The 
slope of the linear regression fit is $1.13\pm0.15$, with the significance 
of correlation greater than 99.99\%. 
Compared with the value 
of 1 with a scatter of 0.5 dex given by \cite{ptak03} and $0.7\pm0.3$ by \cite{lamassa09}, it implies 
that the Fe K$\alpha$ line luminosity is roughly tracking the intrinsic AGN 
luminosity in a similar fashion to lower luminosity obscured AGN.

\subsection{Luminosity Dependence of Obscuration\label{nh}}

\cite{lamassa11} studied a sample of 45 type 2 Seyfert galaxies selected 
based on their mid-infrared continuum and [OIII]$\lambda$5007 and emission 
line fluxes. They found that the observed hard X-ray to $\oiii$ flux ratios 
are one order of magnitude lower on average than that of type 1 Seyfert 
galaxies (in agreement with \citealt{heckman05}), and they show a continuum of 
inferred X-ray obscuration without a clear separation into Compton-thin and 
Compton-thick populations. Here we similarly find that there is no strong 
break in the distributions of either the fitted $N_{\rm H}$ distribution or 
the \lx/\loiii\ ratio for high luminosity type 2 AGN 
(Figure \ref{f:nh} and \ref{f:lxlo}). 
We also find that the correlation between the Fe K$\alpha$ and $\oiii$ luminosities 
is evidently the same between this sample of type 2 quasars and type 2 
Seyfert galaxies. Finally, Figure \ref{f:felx} shows that the correlation 
between the EW of iron line and the \lx/\loiii\ ratio is also the same for 
both the low luminosity (Seyfert) and high luminosity (quasar) type 2 AGN. 
Taken together, these results show that low and high luminosity optically-selected 
type 2 AGNs have similar properties with respect to 
their X-ray obscuration. 

We examine the possible luminosity dependence of obscuration more directly in 
Figure \ref{f:nh2loiii}, in which we plot the column density of the second 
absorber versus the observed \oiii\ luminosity for those AGN having double-absorber 
power-law fits in Table \ref{t:fits}. We also add the corresponding 
data for the type 2 Seyferts from \cite{lamassa09}. There is no tendency 
for the column density to be correlated with the \oiii\ luminosity (over a 
range of more than three orders-of-magnitude in luminosity). Finally, in 
Figure \ref{f:lxlo2} we plot the \oiii\ luminosity vs. the hard X-ray luminosity for the 
combination of our type 2 quasar sample and the \citeauthor{lamassa09} type 2 Seyfert 
sample. Using survival analysis to account for the objects with upper limits 
on the X-ray luminosity we find a best-fit slope in the log-log plot of 
0.88$\pm$0.11 (consistent with no significant luminosity-dependent X-ray 
obscuration), with significance of correlation $>99.99\%$.
In fact, type 1 AGNs show a systematic decrease in the ratio of 
hard X-ray to bolometric luminosity at increasing bolometric luminosity 
\citep[e.g.,][]{marconi04, 2007MNRAS.381.1235V,
2009MNRAS.399.1553V, 2010A&A...512A..34L}. If the \oiii\ 
luminosity is proportional to the bolometric luminosity, and if the amount of 
X-ray obscuration is independent on AGN luminosity, then the relationship in 
\cite{marconi04} would imply a slope of $\sim$0.8. This is fully consistent 
with the fitted slope in Figure \ref{f:lxlo2}. Recently,
\cite{jin12} reported a nearly linear correlation between 
\loiii\ and $L_{\rm 2-10 keV}$ of a sample of type 1 AGNs selected from the 
cross-correlation of the 2XMMi and SDSS DR7 catalogs. We show the correlation
with the slope found by them in Figure \ref{f:lxlo2} with $1\sigma$ deviation
of our sample, where the line is shifted 1.26 dex downward to line up with
the sample in this paper. This offset between the type 1 sample by
\cite{jin12} and our type 2 sample is consistent with that reported by 
\cite{heckman05}, indicating that $L_{\rm X}/L_{\rm OIII}$ ratio is still 
a good indicator of intrinsic obscuration for high-luminosity AGNs.

Additionally, we compare the ratio of their X-ray and \oiii\ luminosity with 
their geometric means in Figure \ref{f:lxlo_geo}. There appears to be a slight
correlation (slope $0.24\pm 0.09$ in log-log scale) between the two 
quantities as shown in the upper panel of Figure \ref{f:lxlo_geo}. 
However, if we exclude those highly-obscured sources with 
$L_{\rm X}/L_{\rm OIII} < 1$, this correlation becomes negligible, i.e., the 
slope is nearly zero (see the lower panel of Figure \ref{f:lxlo_geo}). 
Comparing both cases, we find that the ``correlation" in the top panel of 
$L_{\rm X}/L_{\rm OIII}$ vs. $(L_{\rm X}L_{\rm OIII})^{1/2}$ is driven by
the highly-obscured AGNs at lower luminosity.

\subsection{The Fraction of Compton-thick AGN\label{ct}} 

In order to explain the X-ray background (XRB) spectrum above 10 keV, 
\cite{gch07} predict that the population of Compton-thick AGN is as numerous 
as that of Compton-thin ones in their synthesis model of XRB fitting.  

In Figure \ref{f:lxlonh}, we plot the column densities we derived from the 
simulations described in Section \ref{simulation} versus the \lx/\loiii\ ratio. Since 
$N_{\rm H,sim}$ is derived from the difference between the typical 
Seyfert 1 \lx/\loiii\ value and our observed \lx/\loiii, it is not surprising 
that we find that the \lx/\loiii\ ratio decreases as the simulated $N_{\rm 
H,sim}$ increases. We designate a source as a Compton-thick candidate if
the $1\sigma$ confidence interval of simulated column density exceeds
$1.6\times10^{24}$ cm$^{-2}$ in Figure \ref{f:lxlonh}.
In addition, sources with an iron line EW larger than 1 keV in Table \ref{t:fe} 
are also considered to be Compton-thick, although the errors are often
large. Also note that in some cases, there is a possibility that the
AGN is Compton-thick even thought the Fe-K emission has a low EW
(e.g., Mkn 231). By also including the three sources 
which have no hard X-ray photons detected, we find 39 quasars out of 71 
($55\pm9\%$) are classified as Compton-thick. We flagged them in Table 
\ref{t:fits} and \ref{t:upper}. Of course, the Compton-thick fraction calculated in this way
 has large uncertainty due to the inaccuracy of simulated obscuration. Taken the lower 
 error bars of $N_{H, sim}$ into account, there are 30 sources with 
 $N_{\rm H, sim}-\sigma_{N_{\rm H,sim}}>10^{23.5}$  cm$^{-2}$, which is still a significant
 fraction of heavily obscured sources.

This selection is basically equivalent to the approach based on $L_{\rm X}/L_{\rm OIII}$
in \cite{vignali10}. \cite{lamassa11} found that a majority of Compton-thick AGNs selected based
on various obscuration diagnostics have ratios of 2-10 keV flux to intrinsic flux an order of
magnitude lower than the mean values for Seyfert 1s.  If we adopt the mean 
$L_{\rm X}/L_{\rm OIII}$ value of type 1 Seyferts found in \cite{heckman05}, we
find that the sources marked as Compton-thick in Table \ref{t:fits} agree with the
conclusion of flux ratio in \cite{lamassa11}, except a few outliers.
 
\subsection{Sample completeness and selection bias}

As stated above, in a sample of 25 obscured quasars optically selected from 
SDSS, V10 estimated the intrinsic X-ray luminosity from the observed $\oiii$ 
emission line flux using \citet{mulchaey94}, and compared it with the 
observed X-ray luminosities, i.e., similar to our simulation procedure 
although our simulations take the dispersion in the Seyfert 1 distribution into 
account. V10 conclude that a quasar could be identified as Compton-thick if the 
ratio between the observed and predicted X-ray luminosities is less than 
0.01 and find the fraction of Compton-thick AGN to be 65 per cent.  
However, they point out that \oiii-based selection results in an Eddington bias that 
would naively lower the observed \lx/\loiii\ ratios and  estimate that the 
true fraction is likely closer to 50\% on the basis of the observed $L_{\rm 
X}/L_{\rm MIR}$ values for their sample.

The V10 sample is selected from the catalog of 291 type 2 quasars in Z03, 
with $L_{\rm [O\ III]}>10^{9.28}L_{\odot}$ (note that the \oiii\ luminosities 
used by V10 are from Z03, which are slightly different from those given 
by R08 due to a different \oiii\ line fitting procedure). This sample had 
complete X-ray coverage. However, the R08 catalog is significantly larger, 
with 887 type 2 quasars selected by applying the same criteria to newer and 
more extensive SDSS data. This increase in sample size, plus the larger range 
in \loiii\ that we have probed means that our sample is not complete with 
respect to the optical selection. 
Also, as discussed in V10, the selection based on \oiii\ line may miss some type 2
AGNs due to extinction.
Thus, it is necessary to discuss how the 
completeness may affect our estimation of the fraction of Compton-thick AGNs. 
In Figure \ref{f:completeness}, we show the completeness of our sample in the 
catalog of R08, which is the number of AGNs in our sample above a given 
\oiii\ luminosity divided by the number of AGNs in R08 sample above the same 
\oiii\ luminosity. Although our sample only covers a small fraction 
($\sim8\%$) of the parent sample in \cite{reyes08} over most of the \oiii\ 
luminosity range, the completeness rises rapidly at higher luminosities, 
reaching over $>20\%$ in the luminosity range studied by V10  ($L_{\rm [O\ 
III]}>10^{9.10}L_{\odot}$ according to the new measurement of \oiii\ 
luminosity by \citealt{reyes08}). 

If we limit the \oiii\ luminosity range of our sample to that in V10, the 
Compton-thick fraction becomes $56\%$ (19 out of 34) with $L_{\rm [O\ 
III]}>10^{9.10}L_{\odot}$, consistent with the fraction reported in V10. 
While with \oiii\ luminosity above $10^{9.50}L_{\odot}$, the Compton-thick 
fraction is $53\%$ (8 out of 15). 

Although 45 out of the total 72 sources are on-axis targets, 
only 13 quasars in our sample were initially targeted 
observations by \ch\ and \xmmn\ and were not obviously selected independently 
of their X-ray properties. The others are either serendipitous objects in the 
field of view (27) or were observed in X-rays based on their \oiii\ luminosities (32). 
Thus, the majority of our sample were not observed 
in X-rays based on their known X-ray properties. From this point of view, we 
can safely claim that our sample is not X-ray biased.

\section{Summary\label{con}}

We have presented the hard (2-10 keV) X-ray spectral properties of 71 type 2 
quasars in the redshift range of $z\sim0.05-0.73$ from \ch\ and \xmmn\ 
archival data, which are selected based on their $\oiii$$\lambda$5007 
emission line luminosity. This is the largest sample of optically selected obscured quasars 
studied in X-rays to date. Their observed $\oiii$ luminosities range from 
$10^8$-$10^{10.3}~L_{\odot}$. 

Of these 71 objects, 17 have limited photons detected, and we gave the 3-$\sigma$ 
upper limits to their X-ray fluxes. For the remainder, we have 
fitted their X-ray spectra by assuming a single absorbed power-law to probe 
their spectral slope and circumnuclear obscuration. 
We use a more complicated model (double-absorber power-law) to re-do 
the spectral fits on 17 sources. We also fit the Fe 
K$\alpha$ fluorescent emission line in individual sources. For the others, we 
grouped them in four bins according to their observed \lx/\loiii\ ratios and 
\loiii\, and jointly fit their spectra to investigate the Fe K$\alpha$ 
feature. We also used a more physically realistic model to simulate the X-ray 
spectrum, which included partial covering by the absorber and the effects of 
Compton scattering. Our main results are summarized as follows:

\begin{enumerate}
\item For the 54 sources fitted with absorbed power-law we find the 
average value for the power-law index is $<\Gamma>=1.87\pm0.74$. The average 
column density of our sample from the direct spectral fit is $\log N_{\rm 
H}=22.9\pm0.9$ cm$^{-2}$. 

\item The distribution of the \lx/\loiii\ ratio of our type 2 quasar sample 
agrees with that of local lower luminosity type 2 Seyferts studied 
previously, indicating that they are experiencing similar amounts of X-ray 
obscuration. Based on the small ratios of \lx/\loiii, we find that the 
single-absorber power-law model underestimates the intrinsic X-ray 
obscuration. The double-absorber power-law model we applied to the 17 
brightest sources also gave a higher column density than the single-absorber 
model.

\item We constructed a more physically realistic model with partial covering 
of the central source and Compton scattering to simulate the intrinsic column 
densities that produced the observed low \lx/\loiii\ ratio. We find that 
about half of our sample have simulated column densities one order of 
magnitude higher than from their single power-law spectral fits, but a 
significantly better agreement with the double power-law model results.

\item We investigated the Fe K$\alpha$ features directly detected in 11 
individual sources and the rest in groups by stacking (jointly fitting) their 
spectra. The anti-correlation between the iron line equivalent width and the 
\lx/\loiii\ ratio confirms the relationship studied previously 
\citep{krolik87,bassani99,lamassa09}. Also, we find that the iron line 
luminosity correlates well with the $\oiii$ line luminosity, extending the 
relation seen in type 2 Seyferts to higher luminosities. These correlations 
illustrate that the weak observed hard X-ray emission is due to the heavy 
absorption around the central SMBH, not due to intrinsically weak X-ray 
emission. The consistency of these correlations with those found in low-luminosity 
Seyfert galaxies supports the standard model of AGN at the high 
luminosity end.

\item By combining our analysis with results for type 2 Seyferts from \cite{
lamassa09,lamassa11} we find no dependence of the simulated absorbing column 
densities on AGN luminosity. We also find a nearly linear relationship 
between the [OIII] and X-ray luminosities. These results show that the amount 
of X-ray obscuration does not depend significantly on AGN luminosity (over a 
range in luminosity of over three orders-of-magnitude).

\item Based on the observed \lx/\loiii\ ratio and the simulated column 
densities, we find that about half of the total 71 quasars would be 
classified as Compton-thick AGNs. When limiting the \loiii\ range to higher 
values, the Compton-thick fraction does not change significantly. However, 
more accurate quantification of the Compton-thick fraction and its dependence 
on intrinsic luminosity requires a larger sample.

\end{enumerate}


\acknowledgments

We thank the anonymous referee for helpful comments and suggestions.
We also thank Tahir Yaqoob for the discussion on the issues of Compton-thick torus.

\appendix

\section*{Appendix}

\section{Objects studied in other literature \label{appendixa}}

35 quasars in our sample were also found in papers of X-ray studies of Type 2 
AGN (V04; V06; V10; LM09; P06 and L09), which are flagged in the last column of 
Table \ref{t:params}. There are 17 objects studied in V04, but only SDSS~J1226+0131 
has \xmm\ data and others are observed by {\it ROSAT}. Two objects (SDSS~J0115+0015 
and SDSS~J0243+0006) in P06 were included in Z03, but the $\oiii$ luminosity cut 
excludes them in R08. Therefore, we remove these two objects in this paper. 

{\it Objects with limited photon counts}. SDSS~J0120$-$0050, SDSS~J0134+0014, SDSS~J0319$-$0058, SDSS~J0737+4021, SDSS~J1027+0032, SDSS~J1446+0113, SDSS~J1517+0331 and SDSS~J2358$-$0022 have their 
X-ray luminosity given as a 3-$\sigma$ upper limit in our work (see Table 
\ref{t:upper}). However, the de-absorbed X-ray luminosity of these sources in 
V06 and V10 are not listed as upper limits. The luminosities are based on 
directly converting from their observed 2-8 keV count rates, and are about 
one order of magnitude lower than our upper limits. 

SDSS~J0149$-$0048, SDSS~J0815+4304, SDSS~J0842+3625, SDSS~J0921+4531 and SDSS~J1157+6003 have upper limits 
on the observed flux and derived X-ray luminosity given in our work, V06 and 
V10. However, we find that our values are systematically one order of 
magnitude larger than those in V04, V06, V10. This difference is due
to our assumption of an intrinsic column density of $10^{23}$ cm$^{-2}$ 
in converting the source count rates to flux, while only Galactic absorption 
was assumed by them.

{\it SDSS~J0050-0039}. The spectral parameters given by V06 are $N_{\rm 
H}=3.75\times 10^{23}$ cm$^{-2}$ and $\Gamma=1.78$, and the derived de-absorbed 2-10 keV luminosity is $7.2\times 10^{44}$ erg s$^{-1}$. These 
values are consistent with our analysis of the same \ch\ observation (Obs ID: 
5694), and we also derive the observed 2-10 keV luminosity of $1.8\times 
10^{44}$ erg s$^{-1}$. 

{\it SDSS~J0123+0044}. This object has enough photons to constrain the spectral 
parameters. Photon index as a free parameter in V10's initial spectral fitting resulted in
a very flat spectrum, and they then fixed it to 2 and derived derived the column 
density of $N_{\rm H}=1.44\times 10^{23}$ cm$^{-2}$, which is twice of 
our value. However, we did not fix the photon index and got its value of $\Gamma=0.69$.

{\it SDSS~J0157+0053}. The \ch\ observation (Obs ID:7750) is studied by both V10 
and us. The de-absorbed X-ray luminosity of this \ch\ observation from our 
work is one order of magnitude larger than that given by them. However, we 
also found an \xmm\ observation available, which has many more photon counts 
than the \ch\ data to constrain the spectral parameters. The result of 
multiple observations is shown in Appendix \ref{appendixb}. 

{\it SDSS~J0210-1001}. P06 presented the spectral properties of the object by 
analyzing the \xmm\ observation (Obs ID: 0204340201), which gives a column 
density of $N_{\rm H}=2.3\times 10^{22}$ cm$^{-2}$ and a flat photon index of 
$\Gamma=0.46$. V06 re-analyzed the data but only gave the de-absorbed 2-10 
keV luminosity, which is close to the value from P06. We have similar 
results in this paper.

{\it SDSS~J0801+4412}. We obtain similar spectral parameters and flux for this 
object as P06 did. The column density given by V06 is $N_{\rm H}=4.29\times 
10^{23}$ cm$^{-2}$, while it is $4.08\times 10^{23}$ cm$^{-2}$ in our work. 

{\it SDSS~J0812+4018}. The best-fit photon index and absorption of SDSS~J0812+4018 in V10 
are $\Gamma=2.6$ and $N_{\rm H}=2.14\times 10^{22}$ cm$^{-2}$. Our results 
are $\Gamma=1.91$ and $N_{\rm H}=9.3\times 10^{21}$ cm$^{-2}$, which has a 
flatter spectral slope and slightly smaller obscuration.

{\it SDSS~J0920+4531}. Neither V10 nor our work is able to constrain the column 
density from the spectral fit. They fixed the photon index at $\Gamma=2$ and 
our value is $\Gamma=1.38$, and our value of the derived X-ray luminosity is 
twice as large as theirs.

{\it SDSS~J1039+6430}. Very limited photons are detected, the spectral fit by 
both V10 and us fixed the photon index. V10 also fixed the column density at
the Galactic value, while we derived an upper limit for it.  Our results are similar to the values in V10.

{\it SDSS~J1153+0326}. V06 fitted the spectrum firstly by a power-law and Galactic 
absorption only, and they got a flat photon index of $\Gamma=0.56$. This is 
consistent with our result in Table \ref{t:fits}. They then fixed the index 
at $\Gamma=2$ and got an absorption of  $N_{\rm H}=1.54\times 10^{22}$ cm$^{-
2}$. 

{\it SDSS~J1218+4706}. Our spectral fit results are very similar to those from L09. 
Both works performed the double-absorber power-law model in the spectral 
fitting.

{\it SDSS~J1226+0131}. The \xmm\ observation (Obs ID: 0110990201) is studied by 
both V04 and P06. The best-fitting spectrum of SDSS~J1226+0131 in V04 gives a flat 
photon index of $\Gamma=1.3$ and column density $N_{\rm H}=1.26\times 
10^{22}$ cm$^{-2}$. In P06, the simple power-law model fitting gives 
$\Gamma=1.41$ and $N_{\rm H}=2.0\times 10^{22}$ cm$^{-2}$. Our $N_{\rm H}$ value
are close to their results. The observed hard X-ray luminosity is consistent with 
the two papers.  

{\it SDSS~J1228+0050}. The column density from the spectral fit by V10 is $N_{\rm 
H}=1.52\times 10^{23}$ cm$^{-2}$, which is very close to our value of $N_{\rm 
H}=1.32\times 10^{23}$ cm$^{-2}$. The photon index given by both works is 
slightly different: $\Gamma=1.9$ in their paper and 1.55 in ours, but  they are
consistent if considering uncertainty.

{\it SDSS~J1232+0206}.  P06 fixed both photon index and column density (
$\Gamma=1.7$ and $N_{\rm H}=1.0\times 10^{23}$ cm$^{-2}$) in the spectral fitting. 
We got $\Gamma=2.11$ and $N_{\rm H}=7.45\times 10^{22}$ 
cm$^{-2}$. Our derived flux value is consistent with P06 within a factor of 
two. 

{\it SDSS~J1238+0927}. Our spectral fit results are very similar to those from L09. 
Both works performed the double-absorber power-law model in the spectral 
fitting.

{\it SDSS~J1641+3858}. The spectral properties obtained by P06 are very close to 
the values in our paper. V06 got a column density slightly higher but
still consistent with our value.

{\it SDSS~J2358-0009}. This object was considered to be a serendipitous source with 
a large off-axis angle in the \ch\ observation (Obs ID: 5699). Only upper limits of flux and luminosity were 
given in V06 due to the very limited photon counts. This data set 
is ruled out for this object by the search radius described in Section 
\ref{pipeline}. Instead, we found that it is covered by two \xmm\ observation 
(see Table \ref{t:params}). We performed a moderate-quality spectral fit by 
using the \xmm\ data.

\section{Objects with multiple observations \label{appendixb}}

{\it SDSS~J0056+0032}. It was observed by \xmm\ (Obs ID: 0303110401) and \ch\ (Obs 
ID: 7746) in 2005 and 2008, respectively. The \xmm\ observation had 59 total photons detected, 
which allows us to perform a moderate quality spectral fit. The \ch\ 
observation detected only 6 photons, and is not sufficient for spectral fit. 
Thus, we do not report the spectral results of the \ch\ observation in Table \ref{t:fits} and 
adopt photon index, column density and observed
X-ray luminosity from \xmm\ data in discussion.

{\it SDSS~J0157-0053}. The \ch\ observation (Obs ID: 7750) has 23 photons detected, 
which allows a moderate quality spectral fit. The photon index is $\Gamma=-
0.47$ for this \ch\ observation in the single-absorber power-law model, and results
in large data-to-model ratio. Thus, the double-absorber power-law model is used in 
the spectral fit instead. \xmm\ observation (Obs ID: 0303110101) 
detected $\sim 500$ photons and the spectral fit gives $\Gamma=1.64$. 
Due to the insufficient photon counts in \ch\ observation, we use the spectral 
properties and derived flux from the \xmm\ observation in the sample statistics.

\begin{figure}
\centering
\begin{tabular}{ccc}
\epsfig{file=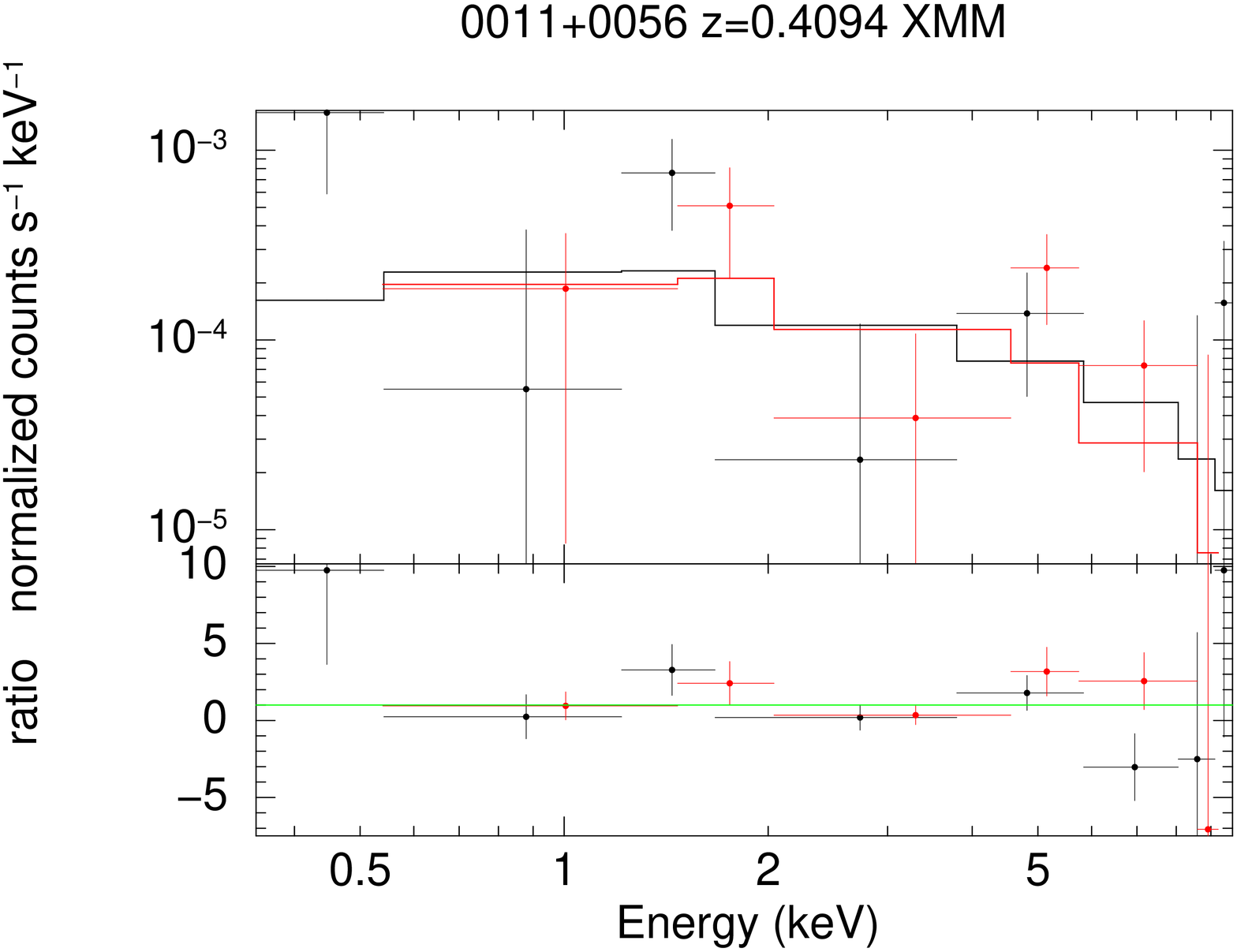,width=0.25\linewidth,angle=-90,clip=} &
\epsfig{file=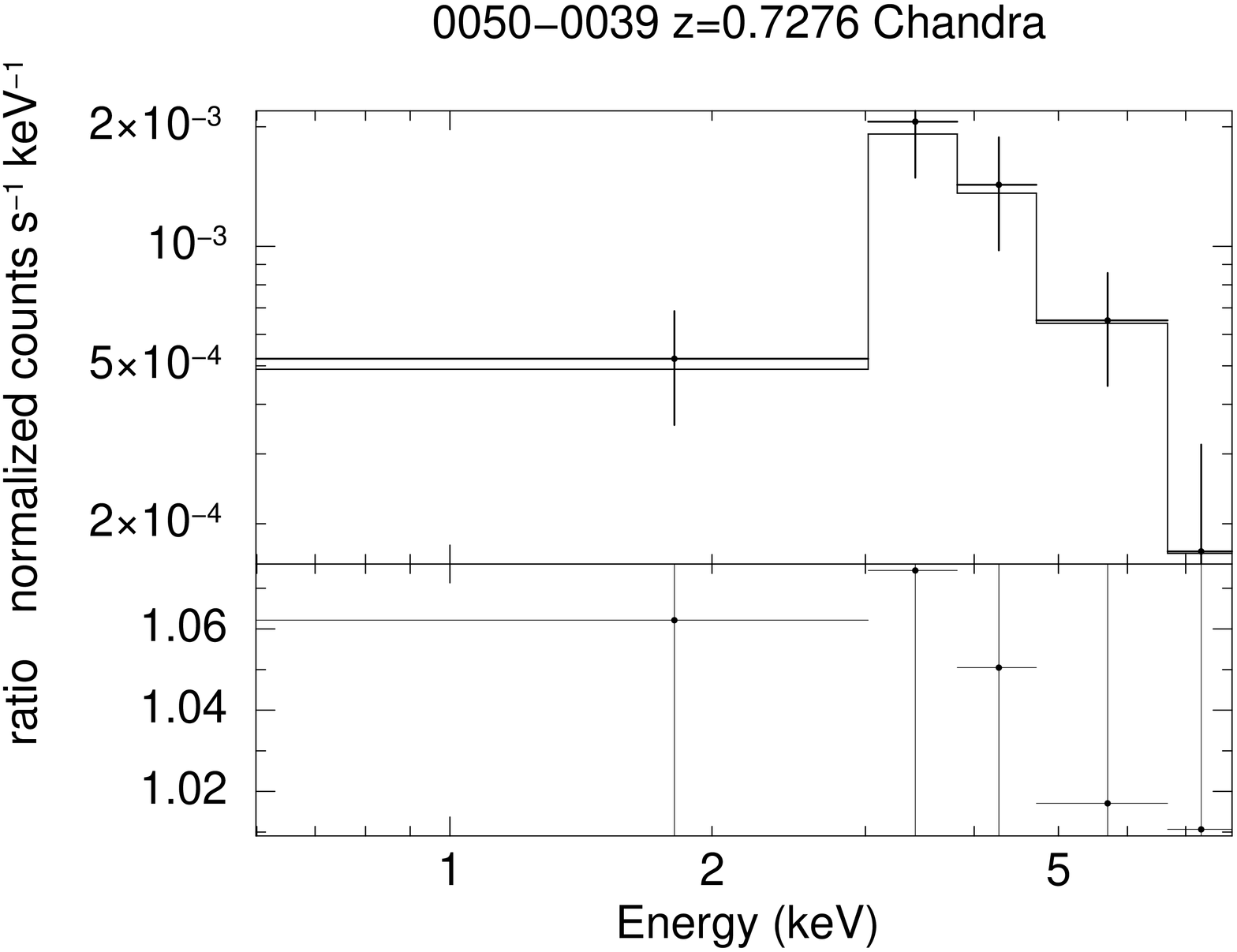,width=0.25\linewidth,angle=-90,clip=} &
\epsfig{file=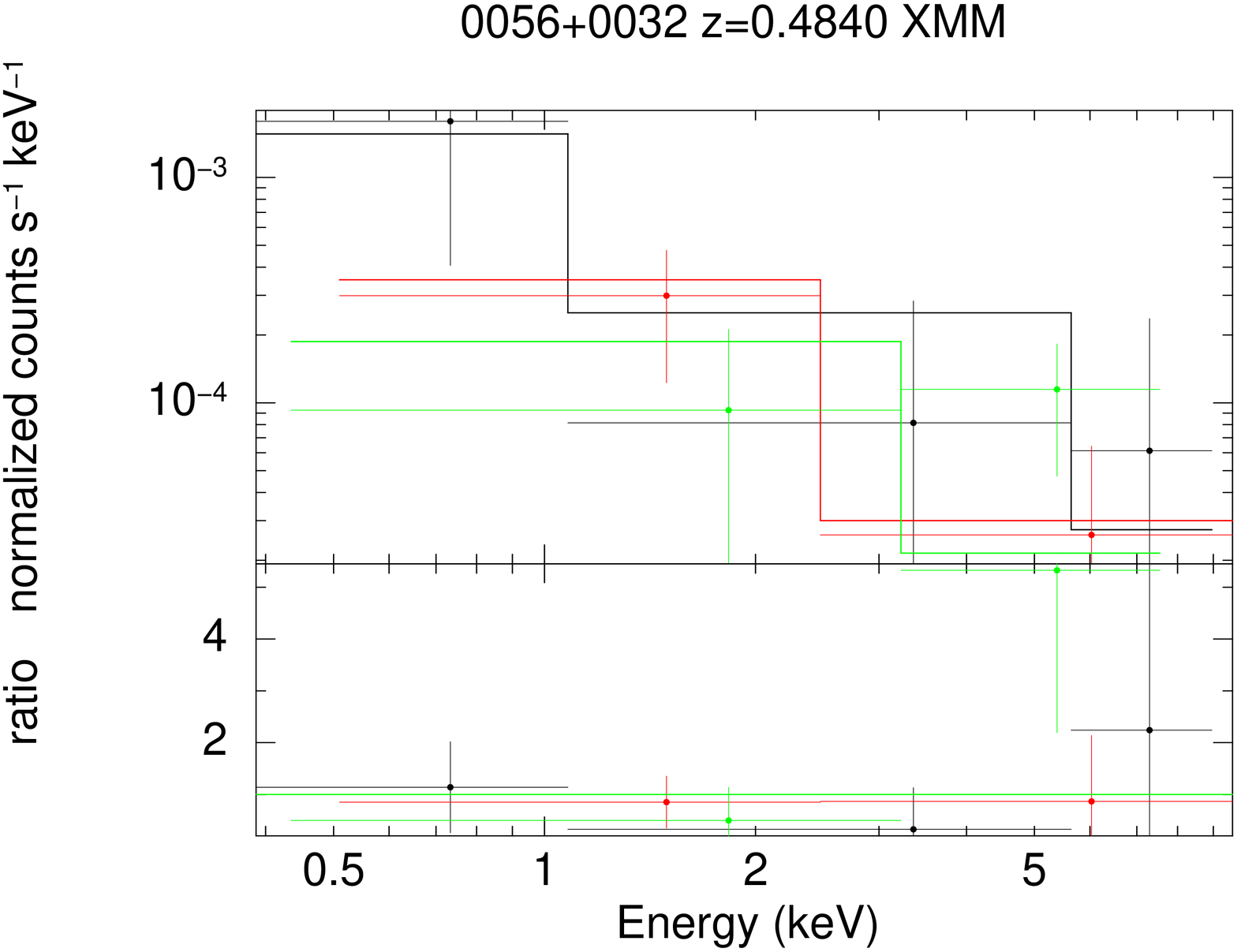,width=0.25\linewidth,angle=-90,clip=} \\
\epsfig{file=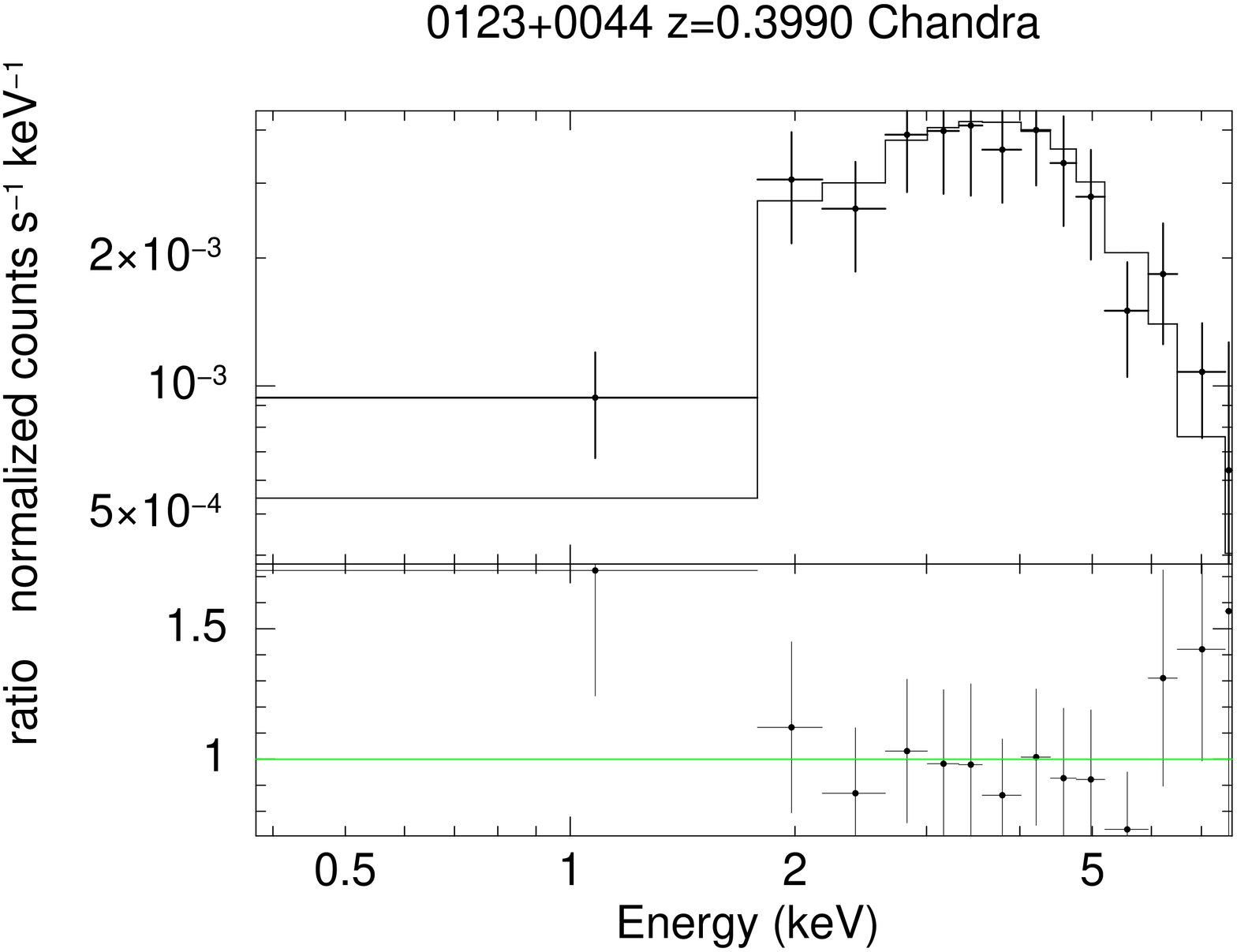,width=0.25\linewidth,angle=-90,clip=} &
\epsfig{file=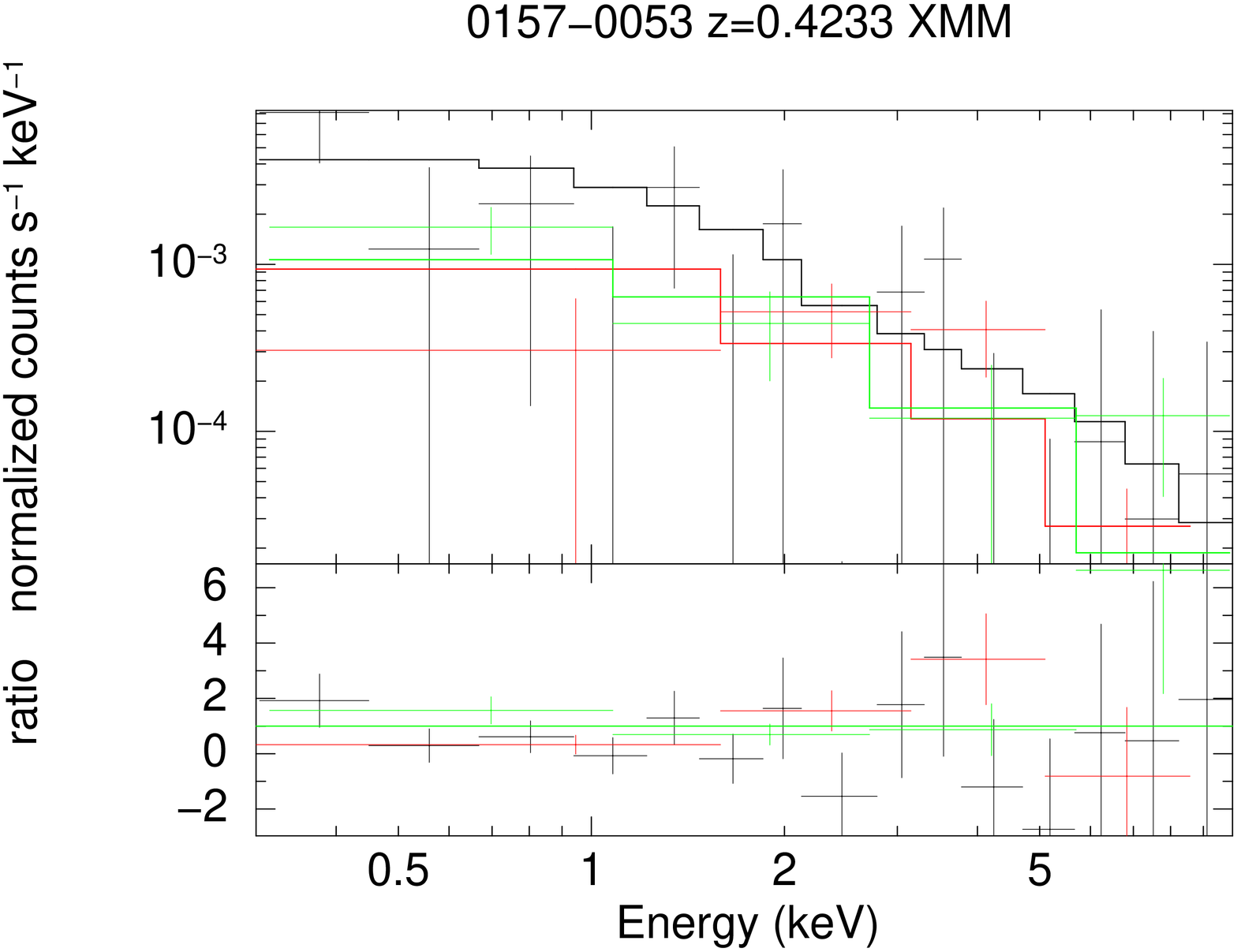,width=0.25\linewidth,angle=-90,clip=} &
\epsfig{file=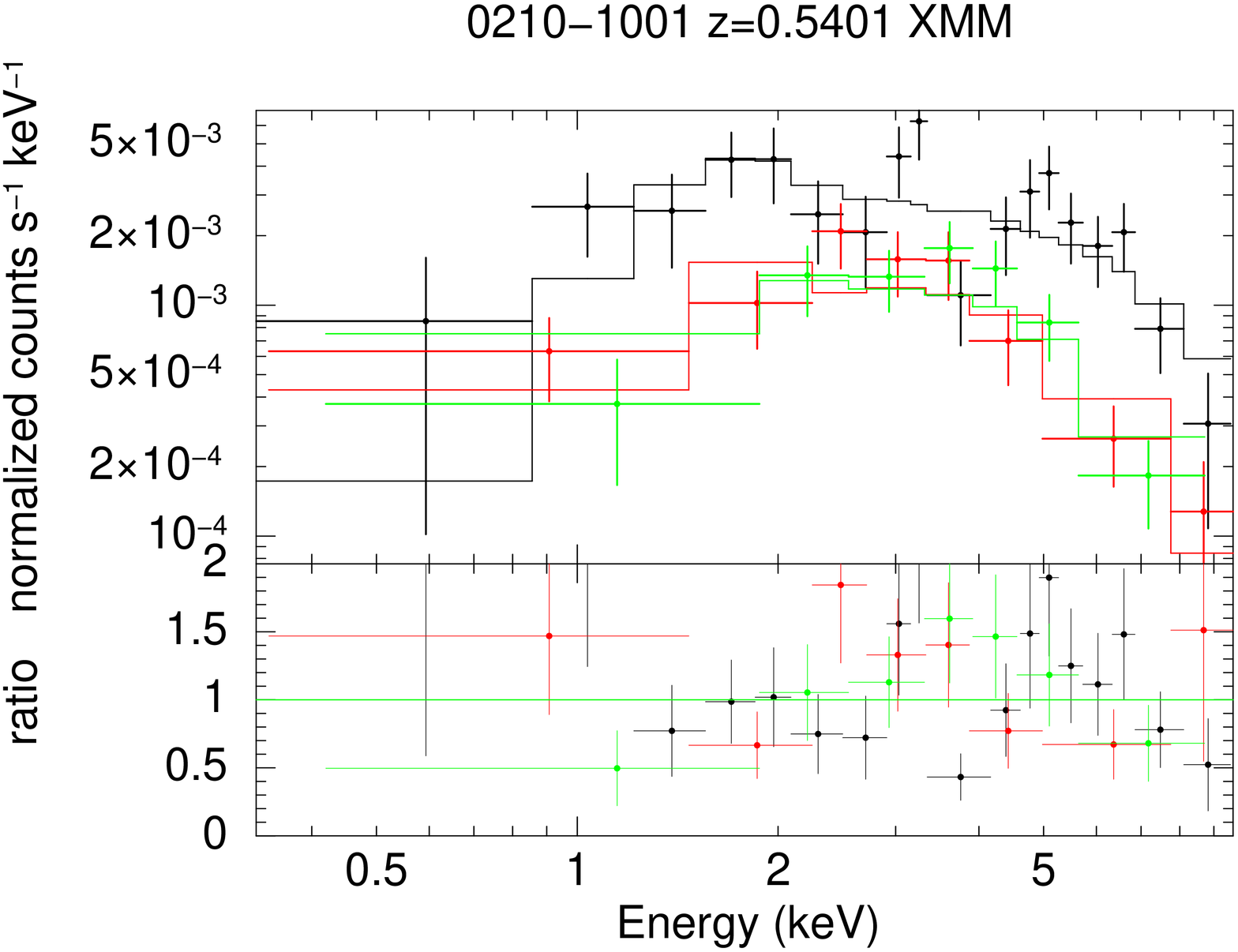,width=0.25\linewidth,angle=-90,clip=} \\
\epsfig{file=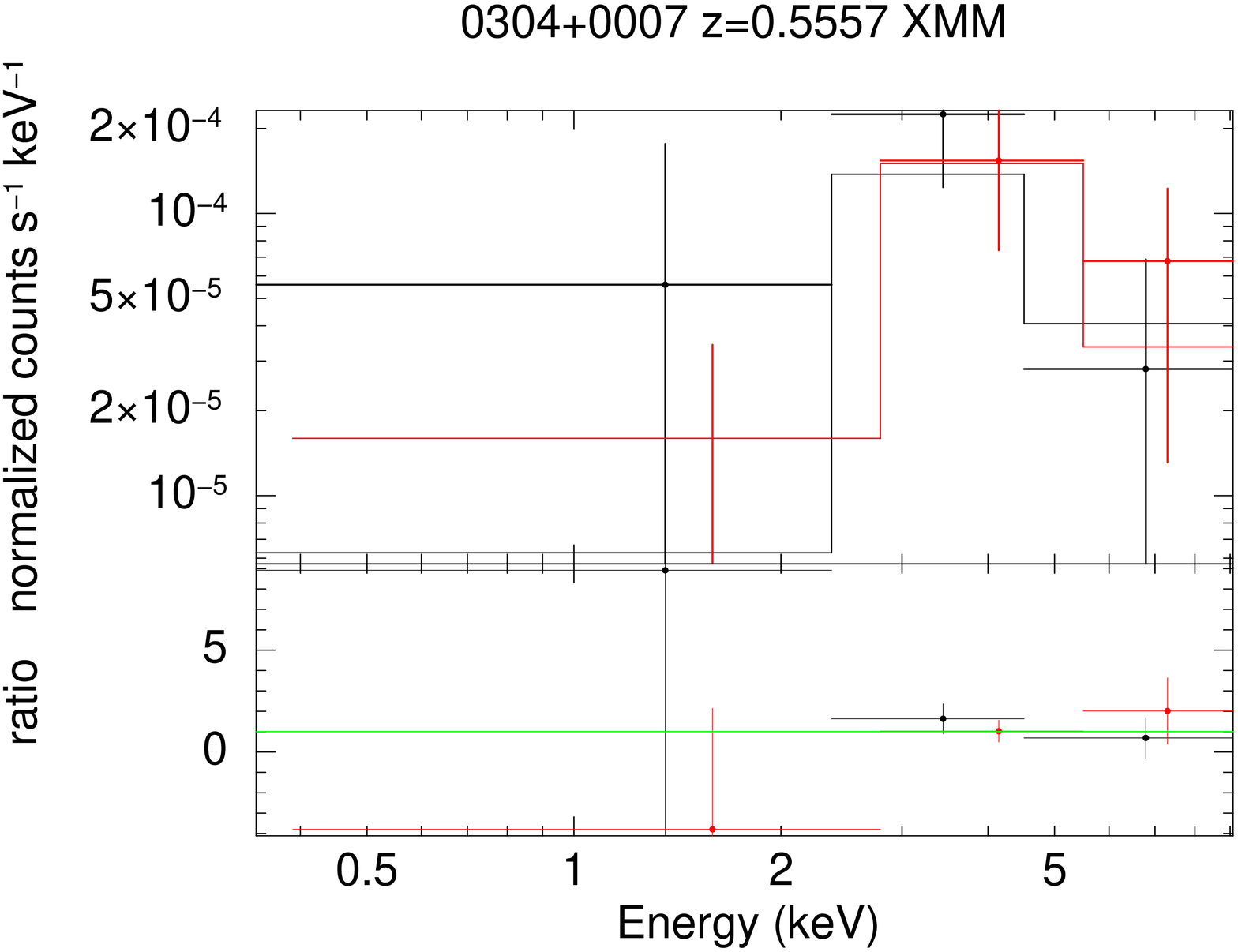,width=0.25\linewidth,angle=-90,clip=} & 
\epsfig{file=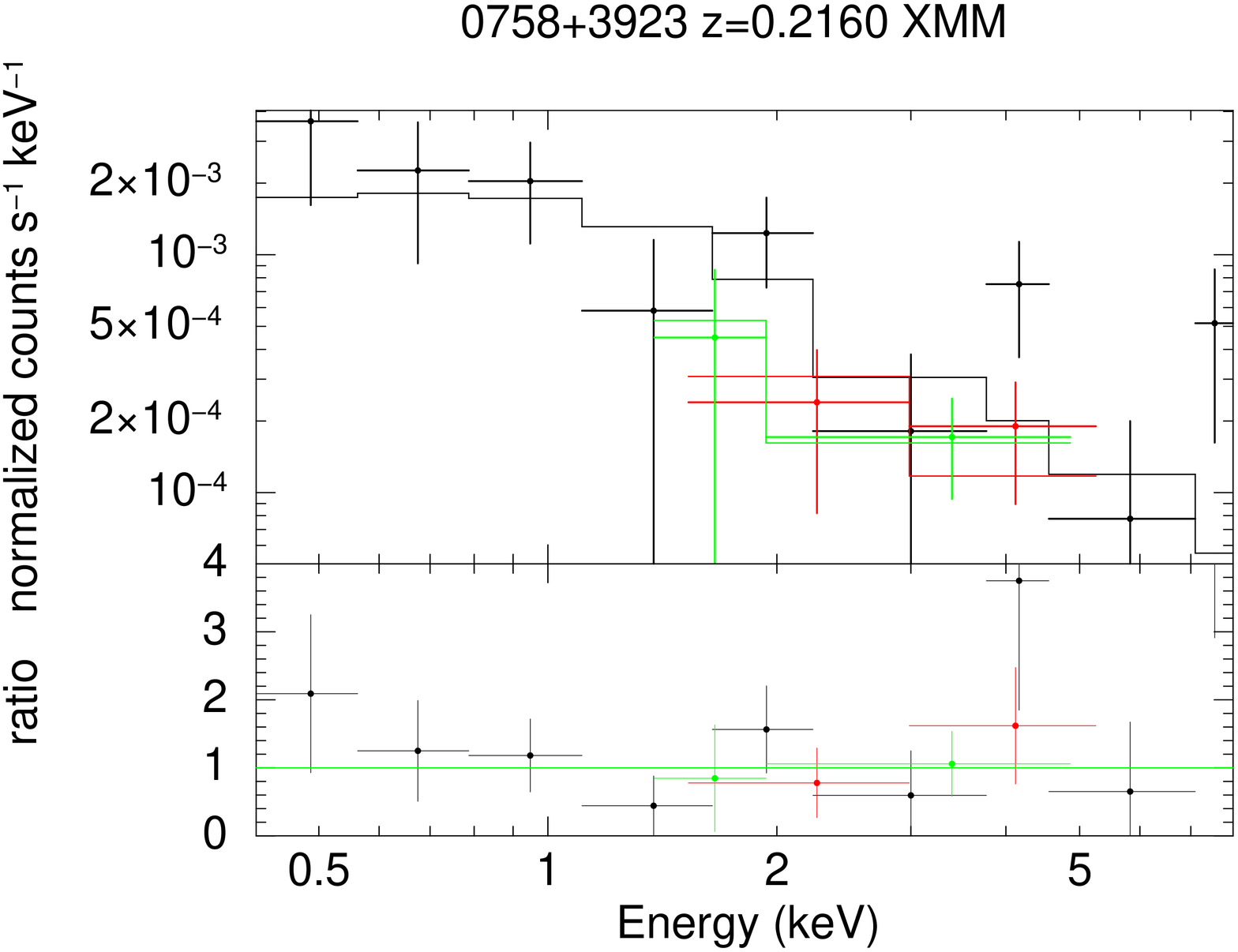,width=0.25\linewidth,angle=-90,clip=} &
\epsfig{file=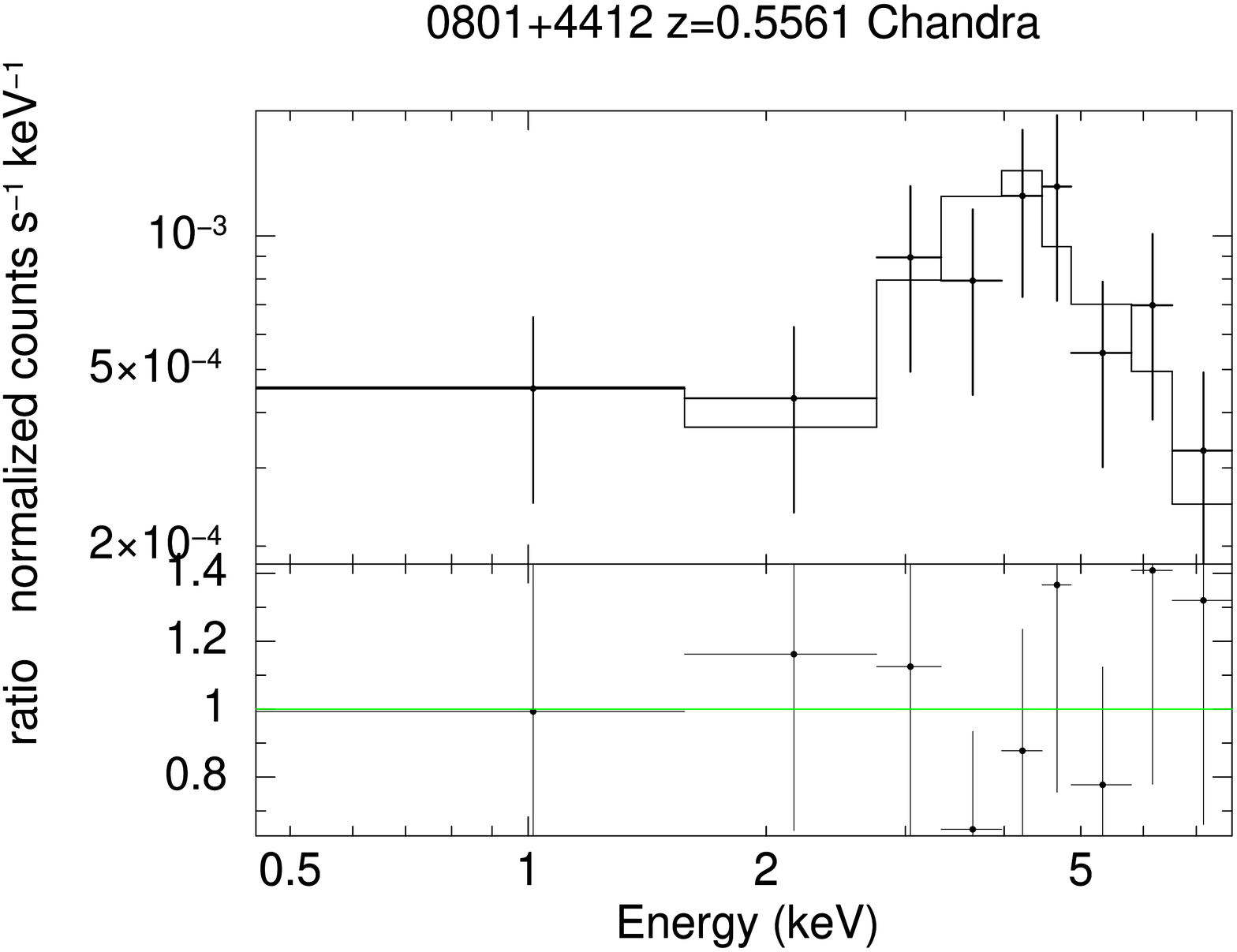,width=0.25\linewidth,angle=-90,clip=} \\
\epsfig{file=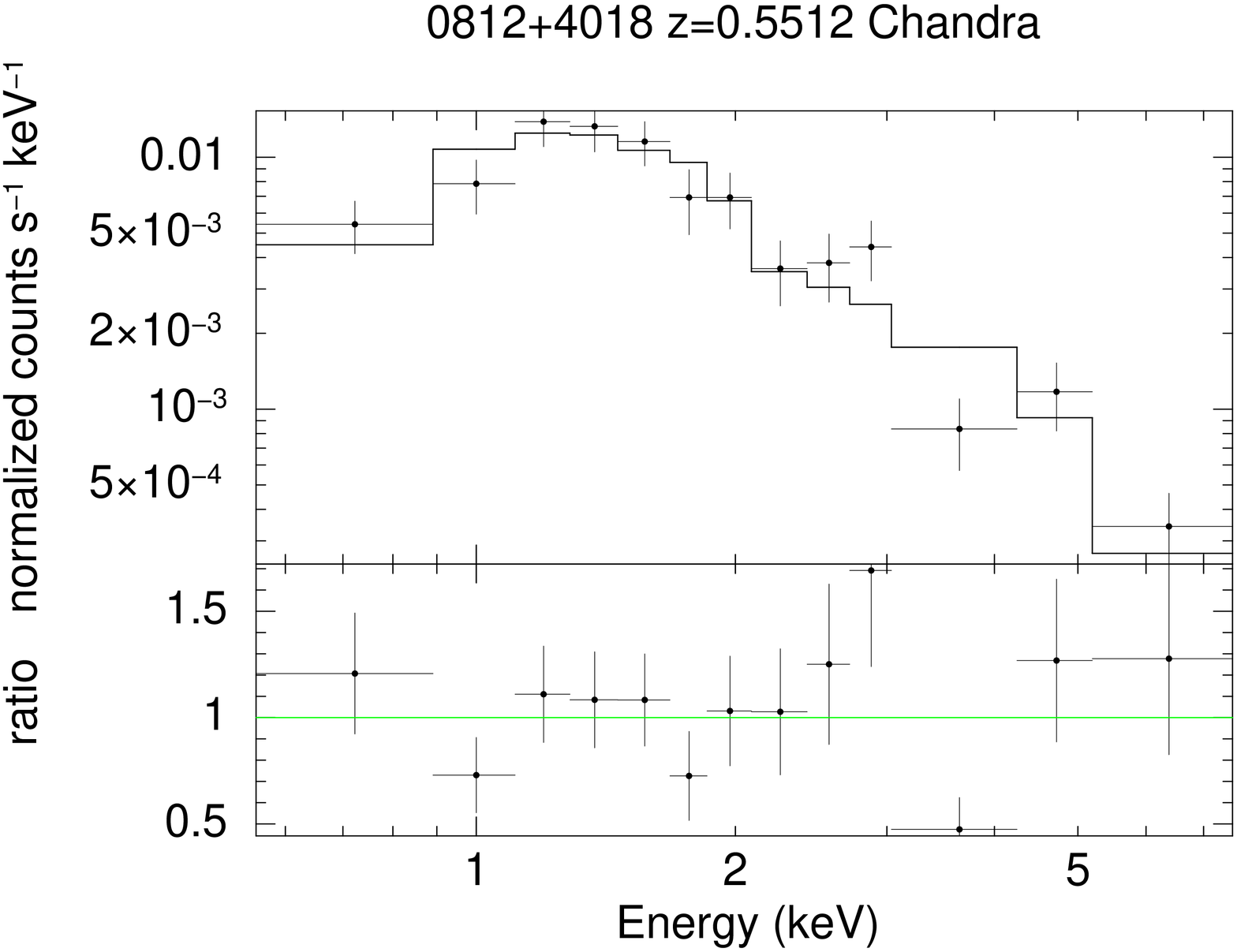,width=0.25\linewidth,angle=-90,clip=} &
\epsfig{file=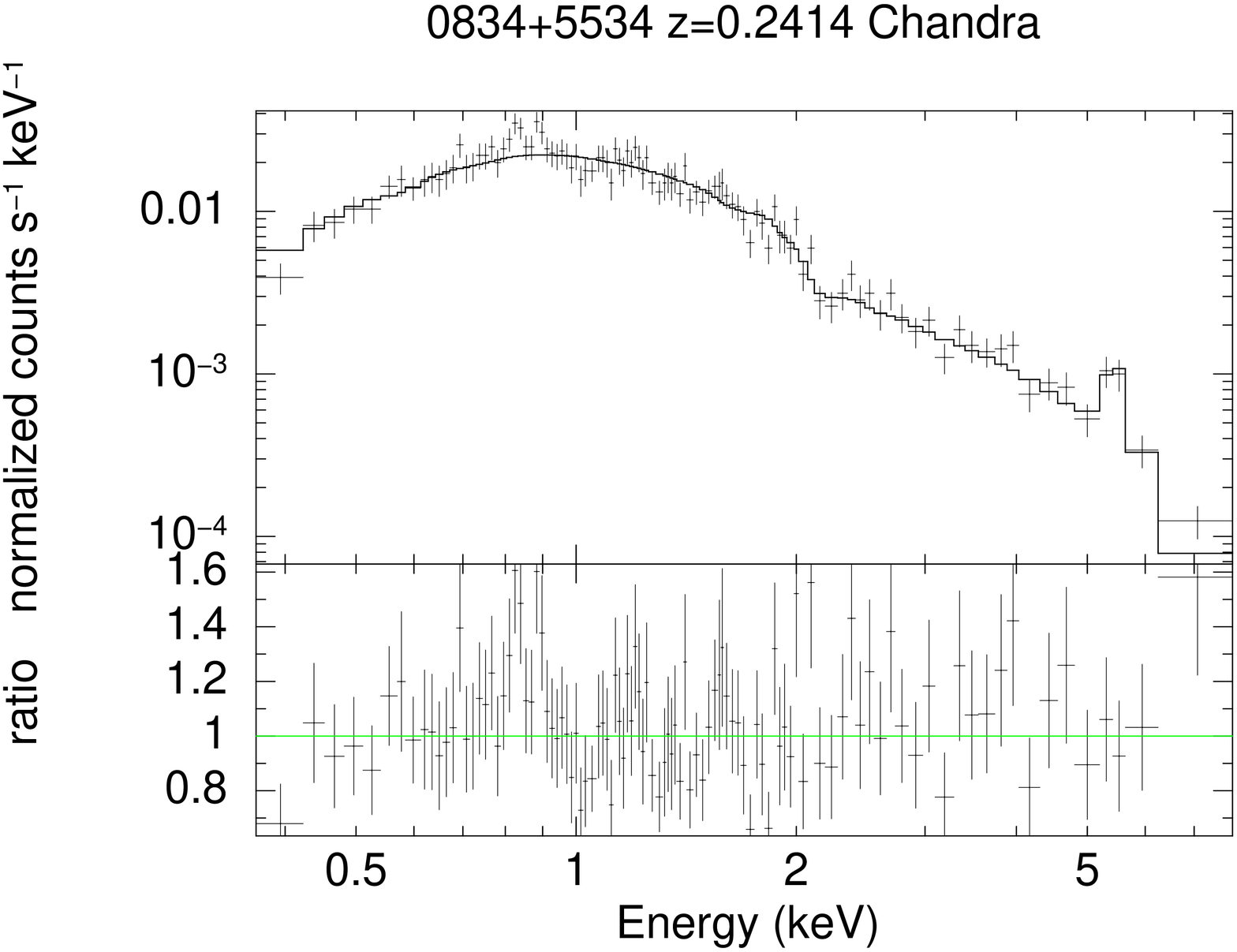,width=0.25\linewidth,angle=-90,clip=} &
\epsfig{file=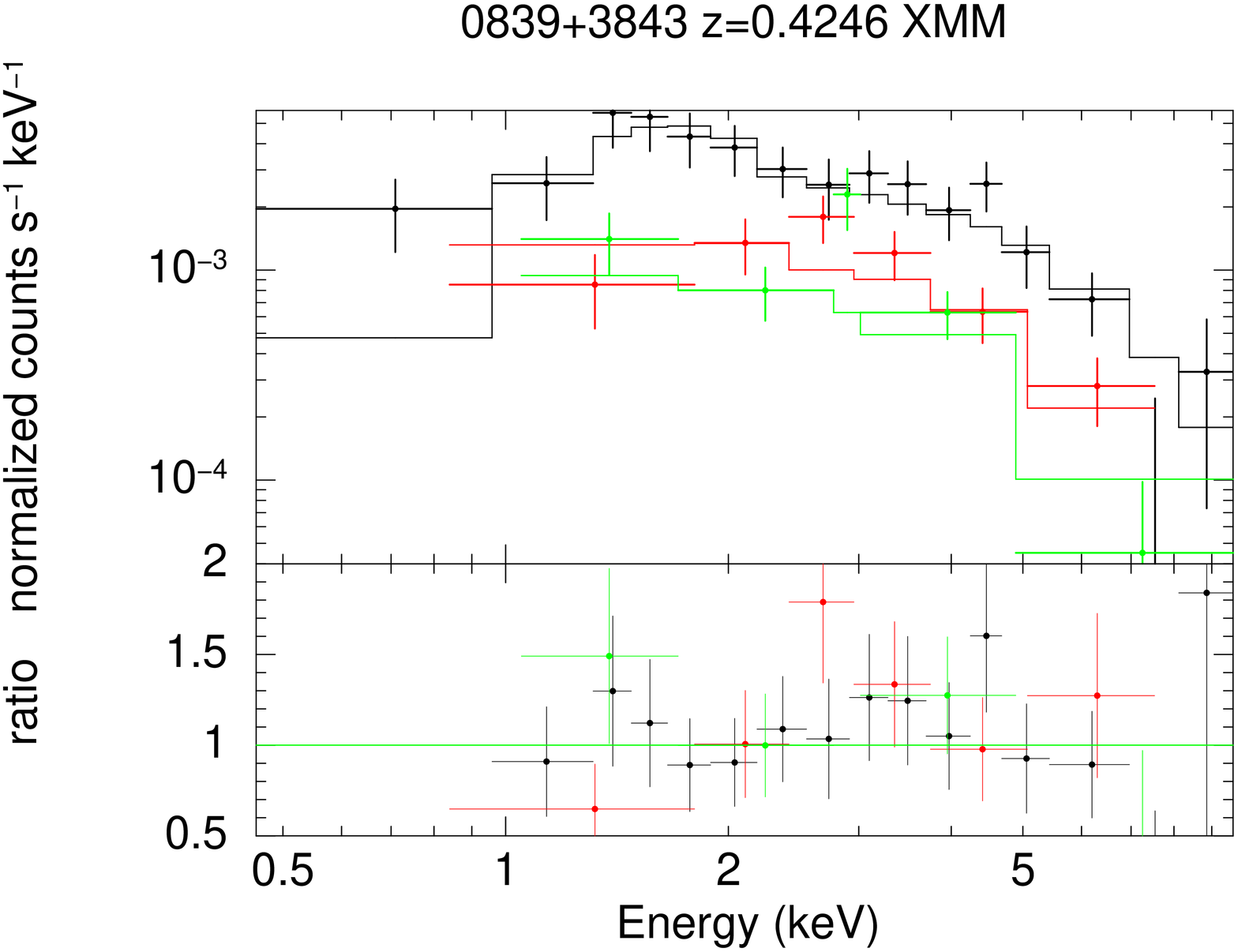,width=0.25\linewidth,angle=-90,clip=} \\
\end{tabular}
\caption{Spectral plots of the best fits of each source. 
The ratio of the data divided by the folded model is shown in the bottom panels. The spectral 
data in some plots are rebinned for display purpose. (A color version and the 
complete figure set (54 images) are available in the online journal.) 
\label{f:all}}
\end{figure}

{\it SDSS~J0758+3923}. There are two \xmm\ observations available for this object 
with Obs ID: 0406740101 and 0305990101. No significant flux variability is 
observed. The spectral fit parameters for both individual and combined 
observations are listed in Table \ref{t:fits}. 
We use the luminosity information from the observation with longer exposure time.
The spectral plot of \xmm-
0406740101 is shown in Figure \ref{f:all}, and Figure \ref{f:multi} shows the 
simultaneous spectral fit for multi-observations.

{\it SDSS~J0834+5534}. Also know as 4C 55.16. 
Two \ch\ observations (Obs ID: 1645 and 4940) and one \xmm\ 
observation (Obs ID: 0143653901) are found to cover 0834+5534. The \xmm\ 
imaging shows a point-like morphology of this object, but it is extended in 
the \ch\ observation. The radii of extraction circles on \ch\ and \xmm\ 
images are 2.5\arcsec and 38\arcsec, respectively. The 2-10 keV flux measured 
from \xmm\ data is one order of magnitude higher than that from \ch\ 
observations (see Table \ref{t:fits}).
Since it is radio-loud, the extended emission is probably due to the jets.
Therefore, we use the results of the 2.5\arcsec extraction region in \ch\
data. A simultaneous spectral fit of both \ch\ observations is shown in Figure 
\ref{f:multi}.

{\it SDSS~J0900+2053}. Two \ch\ observations (Obs ID: 10463 and 7897) and one \xmm\ 
observation (Obs ID: 0402250701) are found to cover 0900+2053. 
The \ch\ observations show an extended morphology in X-ray emission.
The star formation rate of the galaxy is 12.5 $M_{\odot}~{\rm yr}^{-1}$
given by the MPA/JHU DR7 of SDSS \footnote{http://www.mpa-garching.mpg.de/SDSS/DR7/}. 
We extracted the spectra from concentric 
regions with radii of 2.5\arcsec, 10\arcsec\
and 20\arcsec. The soft X-ray fluxes of the two larger regions are 7 and 10
times of that in the 2.5\arcsec\ region, while the hard X-ray fluxes of the
two larger regions are only 2 and 3 times of that in the smallest region.
Thus, the extended emission is dominated by soft X-ray photons from star 
formation. We use the 2.5\arcsec\ region to estimate the quasar emission
in this paper. Simultaneous spectral fit of both \ch\ 
observations is shown in Figure \ref{f:multi}.  

{\it SDSS~J0913+4056}. This is a hyperluminous infrared galaxy.
Two \ch\ observations (Obs ID: 10445 and 509) and one \xmm\ 
observation (Obs ID: 0147671001) are found to cover SDSS~J0913+4056. 
Like SDSS~J0900+2053, soft X-ray photons dominates the extended emission, and we
use the 2.5\arcsec region for the spectral analysis of quasar emission.
A simultaneous spectral fit of both \ch\ observations is shown in Figure 
\ref{f:multi}. The spectral parameters from our fits are consistent with the original papers
which studied these three observations \citep{iwasawa01,piconcelli07,vignali11}.
However, they came to different conclusions whether it is Compton-thin or Compton-thick.

{\it SDSS~J1227+1248}. Three \ch\ observations (Obs ID: 5912, 9509 and 9510) and 
one \xmm\ observation (Obs ID: 0210270101) have SDSS~J1227+1248 covered in the 
field of view. The simultaneous fit of three \ch\ data sets is shown in 
Figure \ref{f:multi}. However, we only use the \xmm\ observation in the 
double power-law spectral fit to derive the spectral properties.

{\it SDSS~J1311+2728}. This object is observed by \xmm\ (Obs ID: 0021740201) and 
\ch\ (Obs ID: 12735) with exposure times of 44 ks and 8 ks, respectively. The 
\xmm\ observation has 588 total X-ray photons detected, while only 19 photons 
are captured by \ch. Therefore, the spectral properties of SDSS~J1311+2728 
presented in this paper are from the \xmm\ observation. 

{\it SDSS~J2358-0009}. This object is observed by two \xmm\ observations (Obs ID: 
0303110301 and 0303110801). The simultaneous fit of both observations is 
shown in Figure \ref{f:multi}.

\clearpage

\begin{landscape}
\begin{deluxetable}{lcccccccc}
\tablewidth{0pt}
\tabletypesize{\scriptsize}
\setlength{\tabcolsep}{1pt}
\tablecaption{SDSS type 2 AGN observed with \ch\ or \xmmn or both \label{t:params}}
\tablehead{Source ID & Galactic $N_{\rm{H,G}}$ & z & $\log$(\loiii/$L_{\odot})$ & 
Observation & Exposure &
  Date & off-axis & ref.\\ 
J2000 coordinates & ($\times 10^{20}$ cm$^{-2}$) & & & ID & (ks) & mm/dd/yy & 
angle (\arcmin) &\\
(1) & (2) & (3) & (4) & (5) & (6) & (7) & (8) & (9)} 
\startdata
SDSS~J001111.97+005626.3 & 2.89 & 0.4094 & 8.67 & \xmm-0403760301 & 19.9 (P) 
25.1 (M1) 25.1 (M2) & 08/07/07 & 4.8 \\

SDSS~J002852.86$-$001433.5 & 2.66 & 0.3103 & 8.08 & \xmm-0403160101 & 0.84 
(P) 1.4 (M1) 1.5 (M2) & 06/29/07 & 7.9 \\ 

SDSS~J005009.81$-$003900.6 & 2.57 & 0.7276 & 10.06 & \ch-5694 & 8.0 & 
08/28/05 & & $b$ \\

SDSS~J005621.72+003235.8 & 2.86 & 0.4840 & 9.25 & \xmm-0303110401 & 8.7 (P) 
11.4 (M1) 11.4 (M2) & 07/16/05 & \\

& & & & \ch-7746 & 9.9 & 02/08/08 & & $c$ \\

SDSS~J012032.21$-$005502.0 & 3.69 & 0.6010 & 8.85 & \ch-7747 & 10.2 & 
02/18/07 & & $c$ \\ 

SDSS~J012341.47+004435.9 & 3.24 & 0.3990 & 9.14 & \ch-6802 & 10.0 & 02/07/06 
& & $c$ \\

SDSS~J013416.34+001413.6 & 2.91 & 0.5559 & 9.53 & \ch-7748 & 10.0 & 09/10/07 
& & $c$ \\

SDSS~J014932.53$-$004803.7 & 2.85 & 0.5669 & 9.29 & \ch-7749 & 10.1 & 
08/30/07 & & $c$ \\

SDSS~J015716.92$-$005304.8 & 2.58 & 0.4223 & 9.19 & \ch-7750 & 9.7 & 06/18/07 
& & $c$ \\

& & & & \xmm-0303110101 & 9.9 (P) 12.7 (M1) 12.7 (M2) & 07/14/05 & \\

SDSS~J021047.01$-$100152.9 & 2.17 & 0.5401 & 9.87 & \xmm-0204340201 & 9.1 (P) 
11.6 (M1) 11.6 (M2) & 01/12/04 & & $b,e$ \\

SDSS~J030425.69+000740.9 & 7.05 & 0.5557 & 9.26 & \xmm-0203160201 & 15.4 (P) 
14.9 (M1) 14.9 (M2) & 07/19/04 &  8.1 \\ 

SDSS~J031950.54$-$005850.6 & 6.05 & 0.6261 & 9.59 & \ch-5695 & 11.6 & 
03/10/05 & & $b$ \\


SDSS~J073745.88+402146.5 & 6.18 & 0.6142 & 9.31 & \ch-7751 & 9.5 & 02/03/07 
& & $c$ \\

SDSS~J075820.98+392336.0 & 5.22 & 0.2160 & 9.02 & \xmm-0406740101 & 10.89 (P) 
14.22 (M1) 14.24 (M2) & 10/22/06 & 4.1 \\

& & & & \xmm-0305990101 & 2.0 (P) 7.9 (M1) 7.9 (M2) & 04/18/06 & 6.1 \\

SDSS~J080154.24+441233.9 & 4.79 & 0.5561 & 9.64 & \ch-5248 & 9.9 & 11/27/03 
& & $b,e$ \\

SDSS~J081253.10+401859.9 & 5.16 & 0.5512 & 9.39 & \ch-6801 & 10.0 & 12/11/05 
& & $c$ \\

SDSS~J081507.42+430427.2 & 5.02 & 0.5099 & 9.44 & \ch-5696 & 8.3 & 12/27/05 
& & $b$ \\

SDSS~J083454.89+553421.1 & 4.14 & 0.2414 & 8.69 & \ch-1645 & 9.0 & 10/17/01 
& \\

 & & & & \ch-4940 & 96.0 & 01/03/04 & \\

& & & & \xmm-0143653901 & 6.3 (P) 9.6 (M1) 9.6 (M2) & 10/09/03 & 13.1 \\

SDSS~J083945.98+384319.0 & 3.55 & 0.4246 & 8.60 & \xmm-0502060201 & 15.4 (P) 
18.7 (M1) 18.7 (M2) & 10/16/07 & 10.8 & $f$ \\

SDSS~J084041.08+383819.8 & 3.45 & 0.3132 & 8.45 & \xmm-0502060201 & 15.4 (P) 
18.8 (M1) 18.8 (M2) & 10/16/07 & & $f$\\

SDSS~J084234.94+362503.1 & 3.41 & 0.5615 & 10.02 & \ch-532 & 19.7 & 10/21/99 
& 5.4 & $b,e$ \\

SDSS~J085331.39+175347.3 & 2.94 & 0.1865 & 8.92 & \xmm-0305480301 & 23.3 (P) 
68.6 (M1) 68.4 (M2) & 10/28/05 & 11.4 \\

SDSS~J085554.47+370900.4 & 2.93 & 0.3567 & 8.84 & \ch-6807 & 10.5 & 02/17/06 
& 4.93 \\

SDSS~J090037.09+205340.2 & 3.39 & 0.2357 & 8.98 & \ch-10463 & 41.2 & 
02/24/09 & \\

& & & & \ch-7897 & 9.1 & 12/23/06 & 1.3 \\

& & & & \xmm-0402250701 & 9.9 (P) 15.7 (M1) 15.7 (M2) & 04/13/07 & \\

SDSS~J091345.48+405628.2 & 1.82 & 0.4409 & 10.33 & \ch-509 & 9.2 & 11/03/99 
& \\

& & & & \ch-10445 & 76.2 & 01/06/09 & \\

& & & & \xmm-0147671001 & 10.2 (P) 13.5 (M1) 13.5 (M2) & 04/24/03 & 1.1 \\

SDSS~J092014.10+453157.3 & 1.51 & 0.4025 & 9.15 & \ch-6803 & 10.2 & 03/05/06 
& & $c$ \\

SDSS~J092152.45+515348.1 & 1.42 & 0.5877 & 9.41 & \ch-7752 & 10.2 & 09/27/07 
& & $c$ \\

SDSS~J092318.06+010144.8 & 3.32 & 0.3873 & 8.77 & \xmm-0551201001 & 23.1 (P) 
26.7 (M1) & 11/06/08 &&  $f$\\

SDSS~J092438.24+302837.1 & 1.94 & 0.2727 & 8.80 & \xmm-0553440601 & 4.4 (P) 
6.5 (M1) & 11/22/08 & 10.3 \\

SDSS~J093952.74+355358.0 & 1.43 & 0.1366 & 8.75 & \xmm-0021740101 & 26.6 (P) 
33.9 (M1) 33.9 (M2) & 10/27/01 & \\

SDSS~J094506.39+035551.1 & 3.71 & 0.1559 & 8.60 & \xmm-0201290301 & 24.9 (P) 
37.0 (M1) 37.0 (M2) & 05/19/04 & 10.0 \\

SDSS~J100327.93+554153.9 & 0.775 & 0.1460 & 8.24 & \xmm-0110930201 & 17.1 (P) 
24.5 (M1) 24.5 (M2) & 04/13/01 & 13.2 \\

SDSS~J102229.00+192939.0 & 2.36 & 0.4063 & 9.13 & \ch-4907 & 7.3 & 03/31/05 
& \\

SDSS~J102746.03+003205.0 & 4.47 & 0.6137 & 9.46 & \ch-7883 & 10.0 & 01/13/07 
& & $c$ \\

SDSS~J103408.59+600152.2 & 0.69 & 0.0511 & 8.81 & \xmm-0306050701 & 8.8 (P) 
11.4 (M1) 11.4 (M2) & 04/04/05 & 1.2 \\

SDSS~J103456.40+393940.0 & 1.47 & 0.1507 & 8.91 & \xmm-0506440101 & 11.9 (P) 
15.0 (M1) 15.0 (M2) & 05/01/02 & 4.6 \\

SDSS~J103951.49+643004.2 & 1.18 & 0.4018 & 9.43 & \ch-7753 & 10.0 & 02/04/07 
& & $c$ \\

SDSS~J104426.70+063753.8 & 2.82 & 0.2104 & 8.16 & \xmm-0405240901 & 24.0 (P) 
31.0 (M1) 31.0 (M2) & 06/05/07 & 5.5 \\

SDSS~J110621.96+035747.1 & 4.58 & 0.2424 & 9.01 & \ch-6806 & 10.2 & 02/02/06 
& \\

SDSS~J111907.01+600430.8 & 0.71 & 0.2642 & 8.28 & \xmm-0502780201 & 9.6 (P) 
13.5 (M1) 13.5 (M2) & 05/20/07 & \\

SDSS~J113153.75+310639.7 & 1.96 & 0.3727 & 8.52 & \xmm-0102040201 & 17.2 
(M1) 23.3 (M2) & 11/22/00 & 12.1 \\

SDSS~J114544.99+024126.9 & 2.21 & 0.1283 & 8.19 & \xmm-0551022701 & 13.8 (P) 
& 06/15/08 & 8.0 \\

SDSS~J115138.24+004946.4 & 2.26 & 0.1951 & 8.40 & \ch-7735 & 4.7 & 07/09/07 
& \\ 

SDSS~J115314.36+032658.6 & 1.89 & 0.5748 & 9.64 & \ch-5697 & 8.3 & 04/10/05 
& & $b$ \\

SDSS~J115718.35+600345.6 & 1.65 & 0.4903 & 9.61 & \ch-5698 & 7.1 & 06/06/06 
& & $b$ \\

SDSS~J121839.40+470627.7 & 1.17 & 0.0939 & 8.56 & \xmm-0203270201 & 14.2 (P) 
33.3 (M1) 35.0 (M2) & 06/01/04 & 6.0 & $d$ \\

SDSS~J122656.40+013124.3 & 1.84 & 0.7321 & 9.8 & \xmm-0110990201 & 21.3 (P) 
28.6 (M1) 28.6 (M2) & 06/23/01 & 5.0 & $a,e$ \\

SDSS~J122709.84+124854.5 & 2.64 & 0.1945 & 8.5 & \xmm-0210270101 & 22.0 (P) 
26.2 (M1) 26.2 (M2) & 12/19/04 & 3.8 \\

& & & & \ch-5912 & 32.6 & 03/09/05 & 4.2 \\

& & & & \ch-9509 & 25.8 & 04/14/08 & 6.7 \\

& & & & \ch-9510 & 25.2 & 04/14/08 & 7.5 \\

SDSS~J122845.74+005018.7 & 1.88 & 0.5750 & 9.28 & \ch-7754 & 9.5 & 03/12/07 
& & $c$ \\

SDSS~J123215.81+020610.0 & 1.80 & 0.4807 & 9.62 & \ch-4911 & 9.7 & 04/21/05 
& & $b,e$ \\

SDSS~J123843.02+092744.0 & 1.87 & 0.0829 & 8.51 & \xmm-0504100601 & 17.4 (P) 
21.3 (M1) 21.3 (M2) & 12/09/07 & 1.7 & $d$ \\

SDSS~J124302.48+122022.8 & 2.34 & 0.4857 & 9.09 & \ch-11322 & 10.6 & 
02/28/10 & 3.4 \\

SDSS~J124337.34$-$023200.2 & 2.03 & 0.2814 & 8.88 & \ch-6805 & 10.2 & 
04/25/06 & \\

SDSS~J130128.76$-$005804.3 & 1.59 & 0.2455 & 9.12 & \ch-6804 & 10.2 & 
05/30/06 & \\

SDSS~J131104.36+272813.4 & 0.98 & 0.2398 & 8.46 & \xmm-0021740201 & 40.3 (P) 
43.7 (M1) 43.7 (M2) & 12/12/02 & \\ 

& & & & \ch-12735 & 8.0 & 11/17/10 & \\

SDSS~J132419.88+053704.6 & 2.26 & 0.2027 & 8.49 & \xmm-0200660301 & 10.7 (P) 
10.0 (M1) 10.2 (M2) & 07/11/04 & 1.7 \\

SDSS~J132946.20+114009.3 & 1.93 & 0.5596 & 9.36 & \xmm-0041180801 & 15.6 (P) 
22.3 (M1) 22.3 (M2) & 12/30/01 & 7.8 \\

SDSS~J133735.02$-$012815.7 & 2.41 & 0.3292 & 8.71 & \xmm-0502060101 & 2.4 
(M2) & 07/11/07 & & $f$\\

SDSS~J134733.36+121724.3 & 1.90 & 0.1204 & 8.65 & \ch-836 & 28.0 & 02/24/00 
& \\

SDSS~J141120.52+521210.0 & 1.33 & 0.4617 & 8.41 & \ch-2254 & 92.1 & 05/18/01 
& \\

SDSS~J143027.66$-$005614.9 & 3.35 & 0.3177 & 8.42 & \xmm-0502060301 & 1.4 
(P) 5.0 (M1) 5.0 (M2) & 08/03/07 && $f$\\

SDSS~J143156.38+325137.7 & 1.07 & 0.4198 & 9.52 & \ch-10457 & 34.6 & 10/30/08 & 6.0 \\

SDSS~J144642.29+011303.0 & 3.55 & 0.7259 & 9.54 & \ch-7755 & 10.2 & 03/22/07 
& & $c$ \\

SDSS~J144920.72+422101.3 & 1.53 & 0.1784 & 8.85 & \ch-5717 & 4.4 & 10/04/05 
& \\

SDSS~J150719.93+002905.1 & 4.48 & 0.1819 & 8.98 & \xmm-0305750801 & 10.5 (P) 
13.4 (M1) 13.4 (M2) & 07/20/05 & 1.1 \\

SDSS~J151711.47+033100.2 & 3.78 & 0.6128 & 9.10 & \ch-7756 & 10.0 & 03/28/07 
& & $c$ \\

SDSS~J160641.42+272556.9 & 3.89 & 0.5411 & 9.44 & \xmm-0304070701 & 2.2 (M1) 
1.9 (M2) & 07/29/05 & 9.2 \\

SDSS~J164131.73+385840.9 & 1.16 & 0.5957 & 10.04 & \xmm-0204340101 & 12.2 (P) 
16.8 (M1) 17.1 (M2) & 08/20/04 & & $b,e$ \\

SDSS~J171350.32+572954.9 & 2.48 & 0.1128 & 8.95 & \xmm-0305750401 & 6.2 (P) 
8.7 (M1) 8.7 (M2) & 06/23/05 & \\

SDSS~J235818.86$-$000919.4 & 3.25 & 0.4025 & 9.27 & \xmm-0303110301 & 1.9 
(P) 5.8 (M1) 5.7 (M2) & 12/04/05 & \\

& & & & \xmm-0303110801 & 6.9 (P) 9.5 (M1) 9.5 (M2) & 06/20/06 & & $b$ \\

SDSS~J235831.16$-$002226.5 & 3.29 & 0.6277 & 9.68 & \ch-5699 & 6.3 & 
08/08/05 & & $b$ \\

\enddata

\tablecomments{Column 1: J2000 coordinate; Column 2: Galactic column density 
calculated by HEAsoft $N_{\rm{H}}$ tool; Column 3: redshift; Column 4: 
$\oiii$$\lambda$5007\AA\ line luminosity in units of solar (from 
\cite{reyes08}); Column 5: \ch~ and \xmmn~observation ID; Column 6: exposure 
times after filtering in units of ks (for \xmmn\ observations, the exposure 
times are listed separately for PN (P) and MOS1,2 (M1,2) instruments); Column 
7: date of observation; Column 8: separation from the center of field of view 
in units of arcminute; Column 9: references that have the source included: 
$a$-\cite{vignali04} (V04); $b$-\cite{vignali06} (V06); $c$-\cite{vignali10} 
(V10); $d$-\cite{lamassa09} (LM09); $e$-\cite{ptak06} (P06); $f$-\cite{lamastra09}(L09).}
\end{deluxetable}
\clearpage
\end{landscape}

\clearpage
\begin{landscape}
\begin{deluxetable}{lccccccccccc}
\tablewidth{0pt}
\tabletypesize{\scriptsize}
\setlength{\tabcolsep}{1pt}
\tablecaption{X-ray spectral properties of SDSS type 2 AGN \label{t:fits}}
\tablehead{Source ID & Total counts and & $N_{\rm{H,1}}$ & $\Gamma$ & $N_{\rm{H,2}}$  & PL1/PL2
  & $\chi^2$/\rm{dof} & $L_{\rm X}$  &
  $L_{\rm X,in}$  & \lx/\loiii\ &
  $L_{\rm X,in}$/\loiii & Compton- \\
& estimated background counts & ($10^{22}$ cm$^{-2}$) & & ($10^{22}$ cm$^{-2}$)  & & or \rm{c-stat/dof} 
& ($10^{44} \ \rm erg\ s^{-1}$) & ($10^{44} \ \rm erg\ s^{-1}$) & & & thick\\
(1) & (2) & (3) & (4) & (5) & (6) & (7) & (8) & (9) & (10) & (11) & (12)}
\startdata
0011+0056 & 77(62.6)/.../57(31.3) & $<2.70$ & $0.60_{-1.15}^{+1.17}$  & & &
123.3/122 & 0.031 & 0.031 & 1.7 & 1.7 & $\surd$ \\

0050$-$0039 & 45(0.4) & $35.5_{-26.0}^{+34.7}$ & $1.73_{-1.66}^{+1.86}$ & & &
51.0/39 & 1.83 & 7.21 & 4.2 & 16.4 & $\surd$\\

0056+0032 & 25(18.4)/16(8.8)/18(10.5) & $<0.96$ & $1.84_{-1.41}^{+2.46}$ & & &
69.3/54 & 0.04 & 0.04 & 0.59 & 0.59 & $\surd$ \\ 


0123+0044 & 161(0.3) & $6.92_{-2.80}^{+3.28}$ & $0.69_{-0.61}^{+0.63}$ & & &
115.1/128 & 1.81 & 2.44 & 34.2 & 46.0 & \\ 


0157$-$0053 & 23(0.2) & $N_{\rm{H,G}}$ & $2.03_{-1.56}^{+1.57}$ & $48.5_{-28.0}^{+106.5}$ & 0.011 & 10.2/19
 & 0.30 & 1.63 & 5.0 & 27.4 \\

& 351(322.2)/72(47.6)/83(46.3) & $<0.11$ & $1.64_{-0.63}^{+0.81}$ & &  & 443.8/439 
& 0.13 & 0.13 & 2.2 & 2.2  &\\ 

0210$-$1001 & 189(31.2)/78(8.1)/77(8.5) & $3.03_{-1.42}^{+2.06}$ & $0.89_{-
0.35}^{+0.38}$ & & & 325.9/312 & 1.81 & 2.0 & 6.3 & 7.0 & \\

0304+0007 & .../29(18.2)/28(20.3) & $43.4_{-20.4}^{+73.2}$ & $2.10_{-3.39}^{+2.07}$  & &
& 58.1/51 & 0.31 & 1.63 & 4.4 & 23.0  &\\


0758+3923 & 90(43.7)/20(8.9)/20(9.3) & $<0.24$ & $1.38_{-0.70}^{+0.96}$ & & &
8.6/8 & 0.02 & 0.02 & 0.44 & 0.44  & $\surd$\\

& 85(69.4)/45(38.3)/46(38.3) & $0.26_{-0.21}^{+0.42}$ & $2.04_{-
1.15}^{+2.82}$ & &  &142.1/164 & 0.07 & 0.07 & 1.5 & 1.5 & \\

& & $<0.25$ & $1.68_{-0.71}^{+0.94}$ & & & 21.3/29 &  &  &  &  & \\ 


0801+4412 & 47(2.4) & $N_{\rm{H,G}}$ & $1.08_{-1.29}^{+1.28}$ & $40.8_{-24.9}^{+38.8}$ & 0.035 & 44.9/40 
 & 0.93 & 2.90 & 5.5 & 17.2  \\

0812+4018 & 201(0.8) & $0.93_{-0.42}^{+0.45}$ & $1.91_{-0.36}^{+0.37}$ && & 
104.9/125 & 1.56 & 1.70 & 16.4 & 18.0 & \\

0834+5534 & 174(57.9) & $0.054_{-0.043}^{+0.048}$ & $1.64_{-0.32}^{+0.36}$ & & &
101.9/113 & 0.17 & 0.17 & 9.0 & 9.0 & \\ 

& 2967 (3.0) & $0.11_{-0.03}^{+0.03}$ & $2.09_{-0.10}^{+0.10}$ &&  &107.9/100 & 
0.21 & 0.22 & 11.1 & 11.2 &\\

 & 2514(238.8)/1079(74.5)/1110(69.9) & $0.12_{-0.02}^{+0.02}$ & $2.24_{-
0.09}^{+0.10}$ &&  &236.2/200 & 2.67 & 2.71 & 142 & 144 & \\

& & $0.12_{-0.03}^{+0.02}$ & $2.12_{-0.10}^{+0.11}$ &&  &128.6/122 &  &  &  &  
&\\ 

0839+3843 & 363(137.6)/133(37.9)/111(41.5) & $2.01_{-1.05}^{+1.57}$ & 
$1.21_{-0.39}^{+0.45}$ & & & 54.6/55 & 1.36 & 1.56 & 89.0 & 102.0 & \\

0840+3838 & 91(64.7)/30(21.9)/29(20.9) & $<0.38$ & $2.08_{-1.17}^{+1.68}$ & & &
130.4/137 & 0.008 & 0.008 & 0.71 & 0.71  & $\surd$\\


0853+1753 & 134(28.3)/169(52.9)/124(15.7)  & $N_{\rm{H,G}}$ & $2.42_{-0.38}^{+0.44}$ & $55.7_{-11.7}^{+14.9}$ & 0.007
& 299.8/364 & 0.08 & 0.62 & 2.5 & 19.4 & $\surd$ \\

0855+3709 & 26(1.6) & $3.27_{-3.05}^{+4.66}$ & $1.14_{-1.29}^{+1.47}$ & &  &
26.6/23 & 0.23 & 0.28 & 8.6 & 11.3 & \\




0900+2053 &  2017(2.0) & $N_{\rm{H,G}}$ & $1.83_{-0.15}^{+0.25}$ & $37.4_{-7.8}^{+10.4}$ & 0.066 & 73.1/76 
 & 1.10 & 3.52 & 30.0 & 96.0 &  \\ 

& 336(0.3) & $N_{\rm{H,G}}$ & $1.54_{-0.46}^{+0.52}$ & $52.9_{-26.6}^{+50.1}$ & 0.110 & 11.5/12
& 1.21 & 4.42 & 33.0 & 120.5 \\

& 7871(23.6)/3705(7.4)/3098(9.3) & $0.12_{-0.02}^{+0.02}$ & $2.30_{-0.09}^{+0.09}$ & $80.0_{-27.5}^{+33.0}$ & 0.265 &
567.7/535 & 2.50 & 9.14 & 68.2 & 249.3 \\

& & $N_{\rm{H,G}}$ & $1.81_{-0.11}^{+0.15}$ & $37.3_{-5.8}^{+7.9}$ & 0.075 & 87.5/91\\ 




0913+4056 & 250(50.0) & $0.08_{-0.03}^{+0.04}$ & $2.24_{-0.53}^{+0.69}$ & $29.2_{-
13.3}^{+31.6}$ & 0.113 & 135.9/139 & 1.74 & 5.07 & 2.1 & 6.1 & $\surd$\\ 

& 2298 (2.3) & $N_{\rm{H,G}}$ & $1.93_{-0.17}^{+0.19}$ & $62.1_{-19.7}^{+28.2}$ & 0.142 &101.8/86 & 2.30 & 
9.28 & 2.8 & 11.2 \\

& 6259(275.4)/2470(86.5)/2574(75.6) & $0.09_{-0.03}^{+0.03}$ & $1.98_{-0.13}^{+0.07}$ & $78.0_{-51.4}^{+60.6}$ & 1.233 &
455.9/423 & 9.61 & 16.0 & 35.1 & 58.4 \\ 

& & & $1.89_{-0.12}^{+0.17}$ & $58.3_{-13.0}^{+22.9}$ & 0.158 & 134.6/108\\ 

0920+4531 & 17(2.6) & $<0.31$ & $1.38_{-0.93}^{+1.32}$ & &  &17.1/15 & 0.04 & 
0.04 & 0.72 & 0.72  & $\surd$ \\

0923+0101 & 171(120.2)/38(31.5)/24(25.4) & $<0.08$ & 1.7 & &  &188.1/205 & 0.026 
& 0.026 & 1.1 & 1.1 & $\surd$ \\


0924+3028 & 53(38.2)/24(6.2)/... & $N_{\rm{H,G}}$ & $1.50_{-2.20}^{+3.19}$ & $35.3_{-32.7}^{+53.2}$ & 0.006  & 88.9/67 
 & 0.28 & 0.93 & 11.6 & 38.5 \\


0939+3553 & 782(136.9)/536(94.3)/544(97.4) & $N_{\rm{H,G}}$ & $1.73_{-0.24}^{+0.26}$ & $11.4_{-3.0}^{+4.6}$ & 0.148 & 108.6/86
& 0.19 & 0.32 & 8.9 & 14.9  \\

0945+0355 & .../40(31.8)/34(25.5) & $<0.55$ & 1.7 & & & 62.8/65 & 0.015 & 0.015 & 
0.96 & 0.96 & \\

1003+5541 & 141(120.7)/103(91.7)/107(94.4) & $<1.55$ & $0.80_{-1.33}^{+2.02}$ 
& &  & 277.8/321 & 0.04 & 0.04 & 6.0 & 6.0 & \\ 

1022+1929 & 21(4.5) & $1.06_{-0.84}^{+2.18}$ & $1.50_{-1.38}^{+1.40}$ & & &
25.0/17 & 0.11 & 0.12 & 2.1 & 2.3 & \\


1034+6001\tablenotemark{a} & 560(49.8)/124(9.3)/123(12.4) & $0.06_{-0.06}^{+0.18}$ & $1.75_{-1.22}^{+1.81}$ 
& $26.3_{-26.3}^{+42.1}$ & 0.403 & 84.3/68 &  0.009 & 0.02 & 0.39 & 
0.87 & $\surd$ \\


1034+3939 & 859(280.9)/307(113.6)/299(120.8) & $N_{\rm{H,G}}$ & $2.89_{-0.23}^{+0.25}$ & $77.8_{-52.6}^{+82.2}$ & 0.010 & 145.1/133 
 & 0.02 & 0.21 & 0.5 & 5.0 & $\surd$\\

1039+6430 & 11(4.3) & $<0.32$ & 1.7 & & &12.2/10 & 0.02 & 0.02 & 0.19 & 0.19 & 
$\surd$\\


1044+0637 & 263(133.9)/100(42.2)/110(52.3) & $N_{\rm{H,G}}$ & $2.54_{-1.44}^{+1.72}$ & $87.1_{-33.9}^{+50.9}$ & 0.002 & 42.0/40 & 
 0.07 & 0.96 & 12.4 & 170.1  \\

1106+0357 & 26(3.6) & $<0.20$ & $0.81_{-0.53}^{+0.58}$ & & & 16.3/20  & 0.046 & 
0.046 & 1.2 & 1.2 & $\surd$ \\

1119+6004 & 1301(1010.9)/326(215.8)/266(167.0) & $<0.02$ & $1.99_{-
0.31}^{+0.34}$ & & & 129.9/90 & 0.10 & 0.10 & 13.3 & 13.3 & \\

1131+3106 & .../.../54(49.9) & $<1.44$ & $2.56_{-1.54}^{+4.88}$ &&  &
38.4/51 & 0.03 & 0.03 & 2.0 & 2.0 & $\surd$ \\

1145+0241 & 146(100.0)/.../... & $<0.05$ & $3.12_{-1.26}^{+1.30}$ & &  & 153.7/127 
& 0.004 & 0.004 & 0.71 & 0.71 & $\surd$ \\

1153+0326 & 91(2.8) & $<0.43$ & $0.73_{-0.33}^{+0.42}$ & &  &
87.5/74 & 1.30 & 1.30 & 7.7 & 7.7 & \\ 


1218+4706 & 90(38.8)/144(41.6)/170(50.5) & $N_{\rm{H,G}}$ & $2.55_{-0.30}^{+0.39}$ & $80.2_{-41.0}^{+55.8}$ & 0.011 & 21.8/31 
 & 0.006 & 0.02 & 0.4 & 1.7 & $\surd$\\

1226+0131 & 221(27.4)/186(32.6)/216(50.0) & $2.42_{-0.61}^{+0.70}$ & $1.69_{-
0.24}^{+0.30}$ & & & 96.9/93 & 3.24 & 3.93 & 13.4 & 16.2  &\\ 


1227+1248 & 221(141.9)/62(26.2)/50/(37.0) & $N_{\rm{H,G}}$ & $2.26_{-0.66}^{+0.84}$ & $76.7_{-41.4}^{+81.3}$ & 0.007 & 276.1/303
 & 0.04 & 0.41 & 3.2 & 34.2 & $\surd$ \\

& 66(0) & $20.6_{-8.3}^{+11.7}$ & $1.86_{-1.13}^{+1.02}$ && & 58.2/59 & 0.07 & 
0.18 & 5.8 & 15 &\\

& 27(2.0) & $26.6_{-19.1}^{+35.7}$ & $2.33_{-2.27}^{+2.34}$ && & 20.0/23 & 0.04 & 
0.13 & 3.3 & 10.8 & \\

& 22(0) & $6.66_{-3.85}^{+9.44}$ & 1.7 && & 16.4/20 & 0.03 & 0.04 & 2.5 & 3.3 
&\\

&& $19.9_{-8.6}^{+10.5}$ & $1.78_{-0.96}^{+0.96}$ & & & 98.0/103 & &&&& 
\\

1228+0050 & 54(3.3) & $13.2_{-8.9}^{+12.1}$ & $1.55_{-1.38}^{+0.67}$ && & 
51.3/45 & 1.17 & 2.21 & 15.8 & 30.6 & \\

1232+0206 & 12(2.8) & $7.45_{-5.52}^{+13.8}$  & $2.11_{-1.62}^{+2.01}$ && & 
17.8/13 & 0.09 & 0.33 & 0.14 & 0.87 & $\surd$ \\


1238+0927 & 1616(150.3)/540(57.2)/545(53.4) & $N_{\rm{H,G}}$ & $2.26_{-0.23}^{+0.29}$ & $45.3_{-4.7}^{+6.3}$ & 0.004
& 313.0/246 & 0.18 & 1.00 & 14.5 & 80.6  \\

1243$-$0232 & 11(0.6) & $<2.84$ & 1.7 &&  & 12.8/8 & 0.007 & 0.008 
& 0.16 & 1.17 & $\surd$ \\

1301$-$0058 & 50(4.0) & $11.1_{-5.9}^{+8.4}$ & $2.16_{-1.40}^{+1.59}$ &&  &
74.1/42 & 0.18 & 0.39 & 3.5 & 7.8 & $\surd$ \\

1311+2728 & 385(125.5)/102(33.3)/101(33.4) & $<0.11$ & $2.48_{-0.20}^{+0.58}$  &
& & 416.7/434 & 0.015 & 0.015 & 1.4 & 1.4 & $\surd$ \\

& 19(0) & $0.21_{-0.18}^{+0.27}$ & $2.55_{-1.24}^{+2.35}$ && & 5.6/13 & 0.01 & 
0.01 & 0.9 & 0.9 & \\

1324+0537 & 61(42.8)/20(15.3)/50(29.2) & $<0.12$ & $1.69_{-0.86}^{+1.68}$ & & & 
128.1/123 & 0.02 & 0.02 & 1.7 & 1.7 & $\surd$ \\

1329+1140 & 344(254.9)/131(111.6)/140(123.8) & $0.25_{-0.11}^{+0.17}$ & 
$2.73_{-0.94}^{+1.47}$ & & & 426.9/472 & 0.13 & 0.14 & 1.5 & 1.6 & \\

1337$-$0128 & .../.../12(5.0) & $<2.02$ & 1.7 && & 19.6/10 & 0.065 & 0.065 & 3.3 
& 3.3 & \\


1347+1217 & 1110(5.6) & $0.22_{-0.10}^{+0.11}$ & $1.59_{-0.32}^{+0.32}$ & $4.43_{-
0.85}^{+0.94}$ & 0.049 & 360.7/378 & 0.35 & 0.47 & 17.1 & 20.6  \\


1411+5212 & 6159(43.1) & $N_{\rm{H,G}}$ & $3.56_{-0.05}^{+0.11}$ & $19.52_{-
1.37}^{+1.59}$ & 0.058 & 416.5/238  & 2.35 & 10.22 & 238.0 & 1036.0 \\

1430$-$0056 & 15(9.5)/6(8.3)/10(6.1) & $<0.23$ & 1.7 & & & 38.5/28 & 0.023 & 
0.023 & 2.3 & 2.3  & $\surd$ \\

1431+3251 & 124(1.5) & $39.9_{-16.5}^{+30.4}$ & $1.85_{-1.02}^{+1.71}$ && & 
9.1/9 & 0.69 & 3.01 & 5.4 & 23.6 & $\surd$ \\


1449+4221 & 31(0.5) & $N_{\rm{H,G}}$ & 1.7 & $17.23_{-8.0}^{+15.9}$ & 0.040 & 43.2/33 & 0.17 & 
0.38 & 6.2 & 13.9 & \\


1507+0029 & 754(492.4)/162(90.7)/161(84.2) & $6.04_{-4.79}^{+9.56}$ & $2.51_{-1.23}^{+1.11}$ & $66.8_{-
27.9}^{+32.7}$ & 0.052 & 96.4/100 & 0.23 & 2.18 & 6.3 & 59.2\\

1641+3858 & 991(68.4)/438(25.0)/450(25.7) & $2.28_{-0.41}^{+0.48}$ & $1.34_{-
0.14}^{+0.14}$ & & & 210.9/174 & 5.31 & 6.20 & 12.6 & 14.7 & \\

1713+5729 & 314(241.2)/71(45.2)/82(46.9) & $<0.03$ & $2.53_{-0.43}^{+0.42}$ && & 
75.1/43 & 0.008 & 0.008 & 0.26 & 0.26 & $\surd$ \\

2358$-$0009 & 39(34.6)/22(14.9)/14(13.9) & $<1.30$ & $2.27_{-
0.23}^{+0.48}$ && & 58.9/72 & 0.033 & 0.033 & 0.45 & 0.45 & $\surd$ \\

& 42(27.9)/12(7.4)/15(10.5) & $<0.27$ & $3.68_{-1.98}^{+5.60}$ && & 55.9/63 & 
0.015 & 0.015 & 0.06 & 0.06 & \\

&& $<0.37$ & $2.24_{-1.17}^{+2.32}$ & & & 114.8/136 & &&&& \\

\enddata
\tablenotetext{a}{SDSS~J1034+6001: The photon index of the two power-law components 
are not tied together in the spectral fits. The other photon indice is 
$3.01_{-0.58}^{+1.51}$.}
\tablecomments{Column 1: Source ID in \textit{hhmm+ddmm} notation; Column 2: total 
and background photon counts for each detector; Column 3: column 
density of the first absorber; Column 4: photon index of power-law; 
Column 5: column density of the second absorber;
Column 6: the ratio of power-law norms;
Column 7: $\chi^2$ or $C$-statistic and degrees of freedom; 
Column 8: observed hard X-ray (2-10 keV in rest frame) luminosity 
derived from spectral fit; 
Column 9: intrinsic hard X-ray luminosity after correction for absorption; 
Column 10: observed X-ray to $\oiii$ luminosity 
ratio; Column 11: intrinsic X-ray to $\oiii$ luminosity ratio; 
Column 12: Compton-thick or not (see Section \ref{ct}).}
\end{deluxetable}
\clearpage
\end{landscape}

\clearpage
\begin{deluxetable}{lcccccc}
\tablewidth{0pt}
\tabletypesize{\scriptsize}
\tablecaption{X-ray counts, count rates, 3-$\sigma$ upper limits of 
marginally detected AGNs. \label{t:upper}}
\tablehead{Source ID & observed counts & $S_{\rm max}$ & count rates & $f_{\rm 2-
10keV}$ & $L_{\rm 2-10keV}$ 
& Compton-thick \\
(1) & (2) & (3) & (4) & (5) & (6) & (7)}
\startdata

0028$-$0014\tablenotemark{a} & 12 (15.2) (M2) & 12.3 & 0.0081 & $5.3\times 
10^{-13}$ & $1.2\times 10^{44}$
\\


0120$-$0055 & 2 (0.3) & 9.7 & 0.0010 & $4.1\times 10^{-14}$ & $3.9\times 
10^{43}$ & $\surd$
\\
0134+0014 & 3 (1.3) & 10.4 & 0.0010 & $2.3\times 10^{-14}$ & $1.2\times 
10^{43}$  & $\surd$
\\
0149$-$0048 & 1 (1.2) & 7.3 & 0.0007 & $1.6\times 10^{-14}$ & $1.3\times 
10^{43}$ 
&$\surd$ \\
0319$-$0058 & 9 (2.9) & 18.0 & 0.0016 & $3.5\times 10^{-14}$ & $3.6\times 
10^{43}$ 
& $\surd$ \\

0737+4021 & 3 (0.2) & 11.5 & 0.0012 & $2.6\times 10^{-14}$ & $2.6\times 
10^{43}$ 
& $\surd$ \\

0815+4304 & 2 (0.3) & 9.7 & 0.0012 & $2.7\times 10^{-14}$ & $1.8\times 
10^{43}$ 
& $\surd$ \\

0842+3625 & 8 (2.2) & 17.3 & 0.0009 & $4.4\times 10^{-14}$ & $3.6\times 
10^{43}$ 
& $\surd$ \\

0921+5153 & 1 (0.7) & 7.5 & 0.0007 & $1.6\times 10^{-14}$ & $1.4\times 
10^{43}$ 
& $\surd$ \\

1027+0032 & 6 (2.0) & 14.4 & 0.0015 & $4.3\times 10^{-14}$ & $4.3\times 
10^{43}$ & $\surd$
\\
1151+0049 & 5 (2.4) & 12.5 & 0.0027 & $8.0\times 10^{-14}$ & $7.1\times 
10^{42}$ 
\\
1157+6003 & 4 (3.3) & 10.4 & 0.0015 & $3.5\times 10^{-14}$ & $2.1\times 
10^{43}$ 
& $\surd$ \\

1243+1220 & 6 (1.9) & 14.5 & 0.0014 & $3.6\times 10^{-14}$ & $2.2\times 
10^{43}$ 
& $\surd$ \\


1446+0113 & 10 (3.7) & 18.6 & 0.0019 & $3.7\times 10^{-14}$ & $5.2\times 
10^{43}$ 
\\

1517+0331 & 8 (4.4) & 15.1 & 0.0015 & $3.2\times 10^{-14}$ & $3.1\times 
10^{43}$ 
\\
1606+2725\tablenotemark{a} & 15 (15.2) (M1) & 15.1 & 0.0068 & $3.6\times 
10^{-13}$ & $2.7\times 10^{44}$ & $\surd$
\\
2358$-$0022 & 5 (2.2) & 12.7 & 0.0020 & $4.6\times 10^{-14}$ & $4.8\times 
10^{43}$ 
& $\surd$ \\

\enddata
\tablenotetext{a}{Photons are obtained by three detectors on \xmmn~for 0028$-
$0014 and 1606+2725. We choose the lowest flux upper limit among PN/MOS1/MOS2 
as the flux limit.}
\tablecomments{Column 1: Source ID in \textit{hhmm+ddmm} notation; Column 2: 
observed total counts and the estimated mean background counts (in bracket); 
Column 3: upper limit of source counts at 3-$\sigma$ level; Column 4: count 
rates; Column 5: flux in 2-10 keV range; Column 6: observed hard X-ray (2-10 
keV in rest frame) luminosity; Column 7: Compton-thick or not (see Section \ref{ct}). 
Values reported in column 4, 5 and 6 are upper limits.}
\end{deluxetable}
\clearpage

\clearpage
\begin{table}[ht]
\centering
\caption{Fe K$\alpha$ features of the AGNs with visually-detected iron 
emission line. \label{t:fe}}
\begin{tabular}{c c c c c c}
\hline\hline
Source ID & $E_{\rm{line}}$\tablenotemark{a} & EW\tablenotemark{a}  (eV) & $L_{\rm{Fe}}$ ($10^{42} 
\rm{ergs}~\rm{s}^{-1}$) & $\chi^2$/dof & $\Delta \chi^2$\\
\hline
0834+5534 & $6.75_{-0.11}^{+0.14}$ & $598_{-308}^{+425}$ & $1.64_{-
0.84}^{+1.17}$ & 107.9/100 & 18.3 \\

0900+2053 & $6.34_{-0.07}^{+0.08}$ & $183_{-78.5}^{+81.1}$ & $4.36_{-
1.87}^{+1.93}$ & 73.1/76 & 15.6 \\

0913+4056 & $6.44_{-0.10}^{+0.10}$ & $457_{-289}^{+473}$ & $17.6_{-
11.1}^{+18.2}$ & 135.9/139 & 10.4 \\ 

0939+3553 & $6.47_{-0.09}^{+0.08}$ & $513_{-160}^{+163}$ & $1.56_{-
0.49}^{+0.50}$ & 108.6/88 & 30.8 \\ 

1034+6001  & $6.42_{-0.06}^{+0.18}$ & $1585_{-817}^{+897}$ & $0.20_{-
0.10}^{+0.11}$ & 84.3/68 & 18.2 \\

1034+3939 & $6.25_{-0.18}^{+0.14}$ & $452_{-294}^{+274}$ & $0.16_{-
0.10}^{+0.10}$ & 145.1/133 & 7.5 \\

1044+0637 & $6.30_{-0.11}^{+0.13}$ & $419_{-248}^{+254}$ & $0.75_{-
0.44}^{+0.45}$ & 42.0/40 & 9.2 \\ 

1218+4706 & $6.38_{-0.22}^{+0.19}$ & $1656_{-1435}^{+2428}$ & $0.15_{-
0.13}^{+0.22}$ & 21.8/31 & 8.1 \\

1238+0927 & $6.41_{-0.07}^{+0.07}$ & $111_{-51}^{+51}$ & $0.47_{-
0.22}^{+0.22}$ & 313.0/246 & 13.4 \\

1311+2728 & $6.45_{-0.12}^{+0.13}$ & $527_{-363}^{+363}$ & $0.36_{-
0.25}^{+0.25}$ & 416.7/434 & 26.5 \\

1347+1217 & $6.42_{-0.08}^{+0.07}$ & $195_{-122}^{+148}$ & $0.88_{-
0.55}^{+0.67}$ & 360.7/378 & 4.0 \\

\hline 
\tablenotetext{a}{In rest frame.}
\end{tabular}
\end{table}

\clearpage
\begin{deluxetable}{l c c l c c}
\tablewidth{0pt}
\tabletypesize{\scriptsize}
\setlength{\tabcolsep}{1pt}
\tablecaption{$N_{\rm{H}}$ from simulation and spectral fitting using the  
\textit{plcabs} model (cm$^{-2}$, in logarithmic scale) \label{t:sim}}
\tablehead{Source ID  & ~~~$N_{\rm{H,sim}}~(deviation)$~~~ & ~~$N_{\rm{H,plcabs}}$~~ & ID  & 
~~~$N_{\rm{H,sim}}~(deviation)$~~~ & ~~$N_{\rm{H,plcabs}}$~~ \\}
\startdata
0011+0056	&	24.22 (0.37)	&	22.07	&	1039+6430	&	24.41 (0.51)	&	20.00	\\
0028$-$0014	&	23.31 (0.60)	&		&	1044+0637	&	23.38 (0.36)	&	23.95  \\	
0050$-$0039	&	24.02 (0.40)	&	23.61	&	1106+0357	&	24.13 (0.62)     &	21.43	\\	
0056+0032	&	24.27 (0.37)	&	23.80	&	1119+6004	&	22.49 (0.60)	&	20.00 \\	
0120$-$0055	&	23.93 (0.44)	&		&	1131+3106	&	24.01 (0.55)	&	23.00	\\
0123+0044	&	23.10 (0.71)	&	22.90	&	1145+0241	&	23.93 (0.57)	&	23.41\\	
0134+0014	&	24.71 (0.32)	&		&	1151+0049	&	23.85 (0.28)	&	 \\	
0149$-$0048	&	$>23.79$	&		&	1153+0326	&	23.54 (0.34)	&	21.93 \\	
0157$-$0053	&	23.82 (0.32)	&	21.82	&	1157+6003	&	24.54 (0.40)	& 		\\
0210$-$1001	&	23.10 (0.80)	&	22.17	&	1218+4706	&	24.84 (0.24)	&	20.00	\\
0304+0007	&	23.88 (0.29)	&	23.61	&	1226+0131	&	23.44 (0.42)	&	22.49	\\
0319$-$0058	&	24.27 (0.35)	&		&	1227+1248	&	23.97 (0.46)	&	23.94	\\
0737+4021	&	24.40 (0.41)	&		&	1228+0050	&	23.45 (0.54)	&	23.13	\\
0758+3923	&	23.89 (0.50)	&	22.37	&	1232+0206	&	24.37 (0.48)	&	22.92	\\
0801+4412	&	23.89 (0.30)	&	23.25	&	1238+0927	&	23.54 (0.52)	&	23.66 \\	
0812+4019	&	22.98 (0.84)	&	22.07	&	1243+1220	&	24.19 (0.56)	&	$<22.52$	\\
0815$-$4304	&	$>22.95$	&		&	1243$-$0232	&	24.11 (0.62)	&	23.21	\\
0834+5534	&	22.90 (0.75)	&	21.04	&	1301$-$0058	&	23.92 (0.59)	&	23.07	\\
0839+3843	&	21.23 (1.03)	&	22.37	&	1311+2728	&	24.00 (0.54)	&	20.00	\\
0840+3838	&	23.94 (0.39)	&	20.48	&	1324+0537	&	24.12 (0.45)	&	21.81	\\
0842+3625	&	24.73 (0.34)	&		&	1329+1140	&	23.73 (0.34)	&	21.11	\\
0853+1753	&	24.04 (0.55)	&	23.01	&	1337$-$0128	&	22.12 (1.51)	&	21.72	\\
0855+3709	&	23.61 (0.45)	&	22.70	&	1347+1217	&	23.09 (0.38)	&	22.50	\\
0900+2053	&	21.69 (0.83)	&	21.11	&	1411+5212	&	19.48 (1.49) 	&	22.94	\\
0913+4056	&	23.81 (0.33)	&	23.56	&	1430$-$0056	&	23.98 (0.58)	&	22.39	\\
0920+4531	&	23.97 (0.42)	&	21.15	&	1431+3251	&	24.23 (0.53)	&		\\
0921+5153	&	$>23.41$	&		&	1446+0113	&	23.67 (0.30)	&		\\
0923+0101	&	24.00 (0.48)	&	22.89	&	1449+4221	&	23.59 (0.35)	&	23.26	\\
0924+3028	&	23.78 (0.31)	&	22.52	&	1507+0029	&	23.26 (0.64)	&	23.01	\\
0939+3553	&	22.68 (0.52)	&	22.55	&	1517+0331	&	20.90 (1.40)	&		\\
0945+0355	&	23.36 (0.41)	&	22.57	&	1606+2725	&	24.18 (0.41)	&		\\
1003+5541	&	22.49 (0.61)	&	21.58	&	1641+3858	&	23.16 (0.56)	&	22.29	\\
1022+1929	&	23.85 (0.32)	&	22.12	&	1713+5729	&	24.43 (0.51)	&	21.52	\\
1027+0032	&	24.04 (0.49)	&		&	2358$-$0009	&	24.14 (0.45)	&	22.48	\\
1034+6001	&	24.70 (0.36)	&	24.78	&	2358$-$0022	&	24.46 (0.43)	&		\\
1034+3939	&	24.25 (0.58)  	&	24.03	\\
\enddata
\tablecomments{We did not fit the sources reported in Table \ref{t:upper} using $plcabs$ model due to limited photon counts.}
\end{deluxetable}

\clearpage
\begin{deluxetable}{clcccc}
\tablewidth{0pt}
\tabletypesize{\scriptsize}
\setlength{\tabcolsep}{1pt}
\tablecaption{Properties of stacked Fe K$\alpha$ emission lines. \label{t:stack}}
\tablehead{& Source ID & \lx/\loiii\  & net counts & $E_{\rm{line}}$ (eV) & EW (eV) \\}
\startdata
\multirow{6}{*}{$-0.5<\log$ \lx/\loiii$<0$} & 0056+0032 & 0.59 & \multirow{6}{*}{84.4}&
\multirow{6}{*}{$6.43_{-0.04}^{+0.04}$} & \multirow{6}{*}{$1180_{-
638}^{+964}$} \\
& 0758+3923 & 0.44 & & \\
& 0840+3838 & 0.71 & & \\
& 0945+0355 & 0.96 & & \\
& 1145+0241 & 0.71 & & \\
& 2358$-$0009 & 0.45 & & \\
\hline 
\multirow{7}{*}{$0<\log$ \lx/\loiii$<0.5$} & 0011+0056 & 1.7 &  \multirow{7}{*}{255.2}&
\multirow{7}{*}{$6.45_{-0.33}^{+0.30}$} & \multirow{7}{*}{$<992$} \\
& 0157$-$0053 & 2.2 & & \\
& 0853+1753 & 2.5 & & \\
& 0923+0101 & 1.1 & & \\
& 1022+1929 & 2.1 & & \\
& 1324+0537 & 1.7 & & \\
& 1329+1140 & 1.5 & & \\
\hline
\multirow{8}{*}{$0.5<\log$ \lx/\loiii$<1.0$} & 0050$-$0039 & 4.2 &  \multirow{8}{*}{586.1}&
\multirow{8}{*}{$6.38_{-0.06}^{+0.06}$} & \multirow{8}{*}{$360_{-
166}^{+203}$} \\
& 0210$-1001$ & 6.3 & & \\
& 0801+4412 & 5.5 & & \\
& 0855+3709 & 8.6 & & \\
& 1003+5541 & 6.0 & & \\
& 1153+0326 & 7.7 & & \\
& 1301$-$0058 & 3.5 & & \\
& 1507+0029 & 6.3 & & \\
\hline
\multirow{6}{*}{$1.0<\log$ \lx/\loiii$<1.5$} & 0812+4018 & 16.4 &  \multirow{6}{*}{1740.4}&
\multirow{6}{*}{$6.40_{-0.06}^{+0.05}$} & \multirow{6}{*}{$148_{-73}^{+104}$} 
\\
& 0924+3028 & 11.6 & & \\
& 1119+6004 & 13.3 & & \\
& 1226+0131 & 13.4 & & \\
& 1347+1217 & 17.1 & & \\
& 1641+3858 & 12.6 & & \\
\enddata
\tablecomments{Net counts of the stacked spectra are in 3-8 keV band; $E_{\rm{line}}$ and EW are in rest frame.}
\end{deluxetable}

\clearpage
\begin{table}[ht]
\centering
\caption{The means of \oiii\ and X-ray luminosities and their ratios in 
\loiii\ bins. \label{t:loiii}}
\begin{tabular}{c c c c c}
\hline\hline
$\log L_{\rm OIII}$ range & $<\log L_{\rm OIII}>$ & $<L_{\rm X}>$ & $<L_{\rm 
X}/L_{\rm OIII}>$ & $<L_{\rm Fe}>$\\
($L_{\odot}$) & ($L_{\odot}$) & ($10^{44} {\rm ergs~s^{-1}}$) & & ($10^{42} 
{\rm ergs~s^{-1}}$)\\
\hline
8.0--8.5 & $8.35\pm 0.14$ & $0.04\pm 0.01$ & $6.01\pm 2.80$ & $0.23\pm 0.06$ 
\\
8.5--9.0 & $8.75\pm 0.15$ & $0.30\pm 0.21$ & $13.5\pm 8.17$ & $0.88\pm 0.42$ 
\\
9.0--9.5 & $9.21\pm 0.13$ & $0.38\pm 0.21$ & $5.73\pm 3.21$ & $1.26\pm 0.40$ 
\\
$>9.5$   & $9.88\pm 0.25$ & $2.04\pm 0.66$ & $6.51\pm 0.67$ & $3.85\pm 1.55$ 
\\
\hline 
\end{tabular}
\end{table}

\clearpage

\begin{figure*}
\centering
\begin{tabular}{ccc}
\epsfig{file=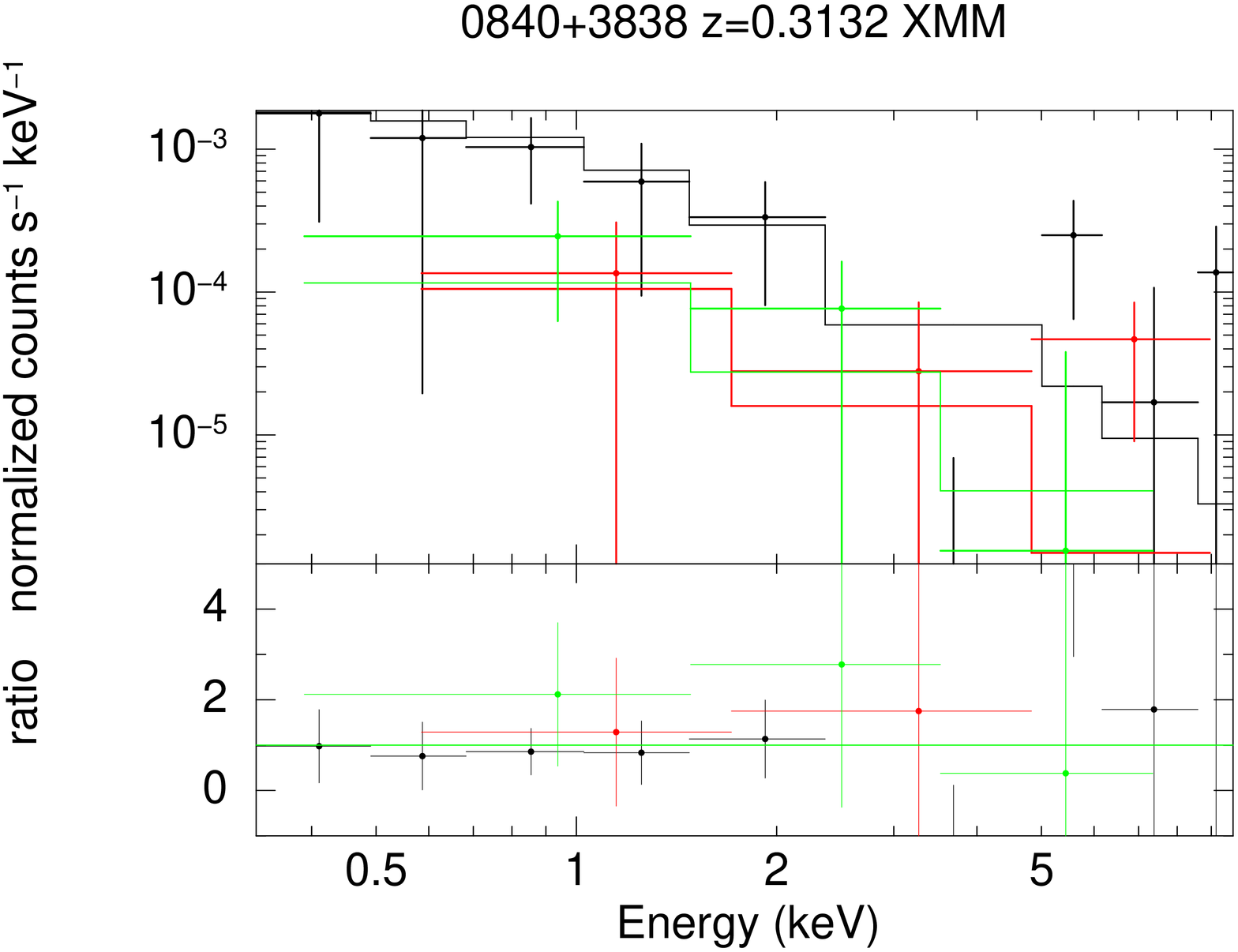,width=0.25\linewidth,angle=-90,clip=} & 
\epsfig{file=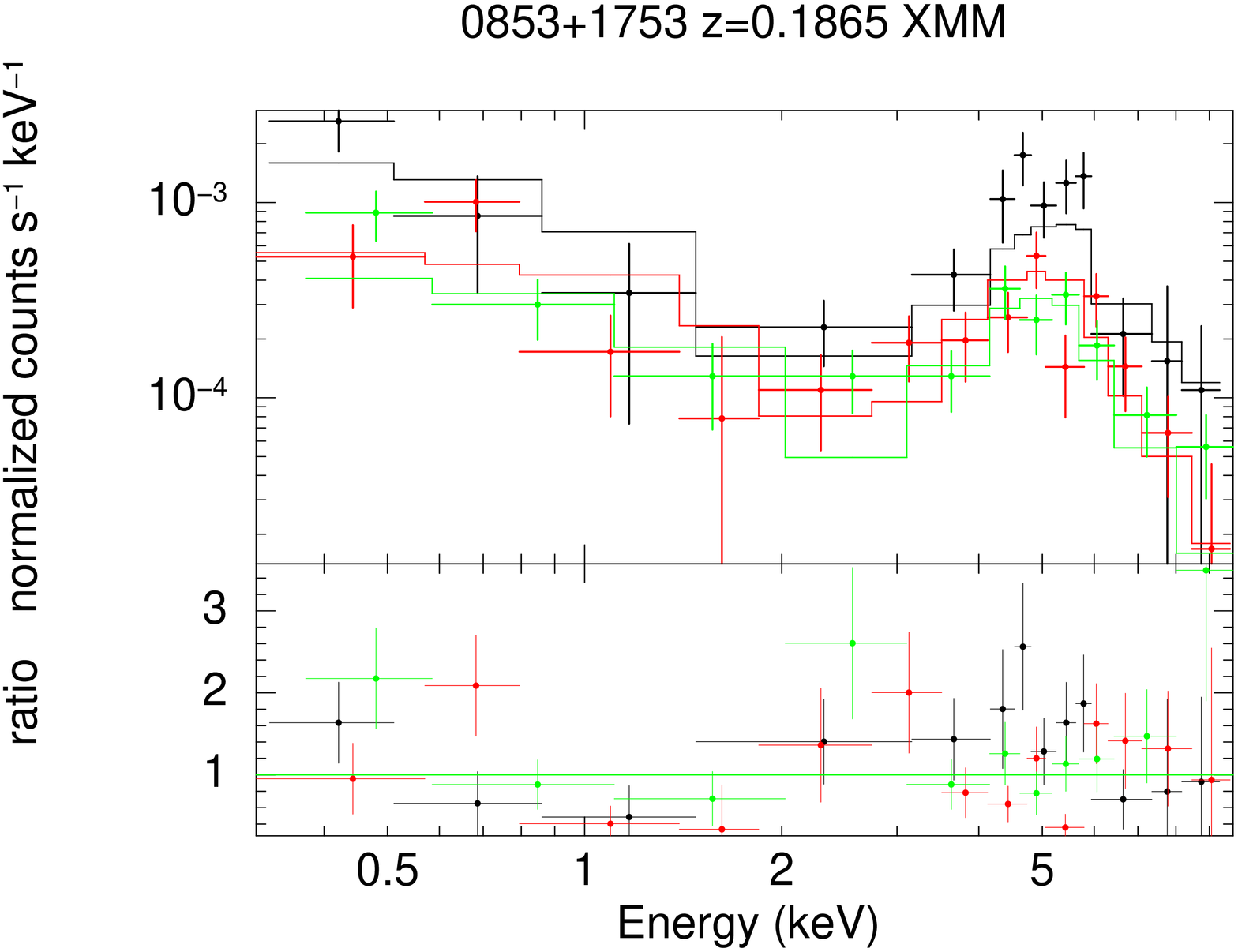,width=0.25\linewidth,angle=-90,clip=} &
\epsfig{file=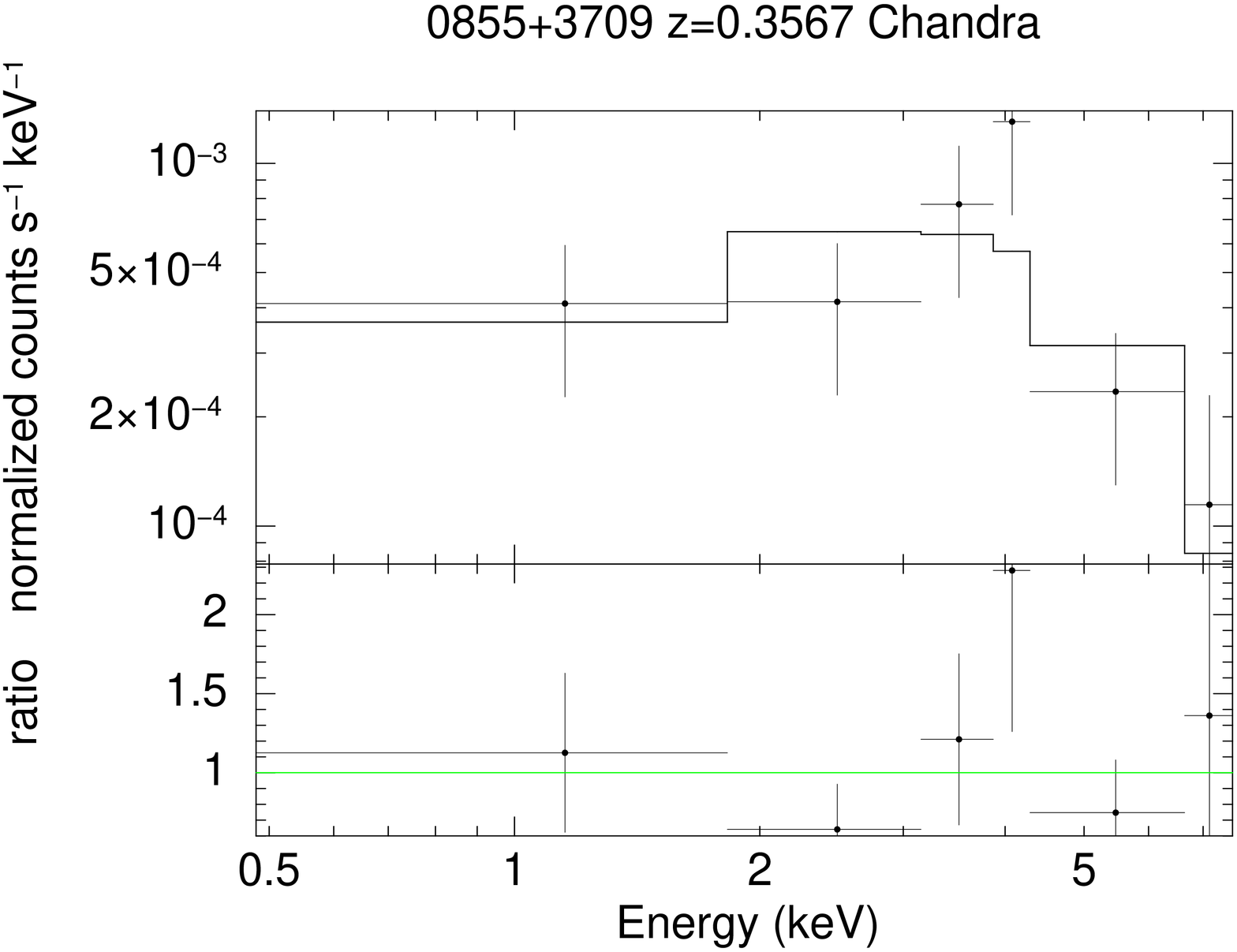,width=0.25\linewidth,angle=-90,clip=} \\
\epsfig{file=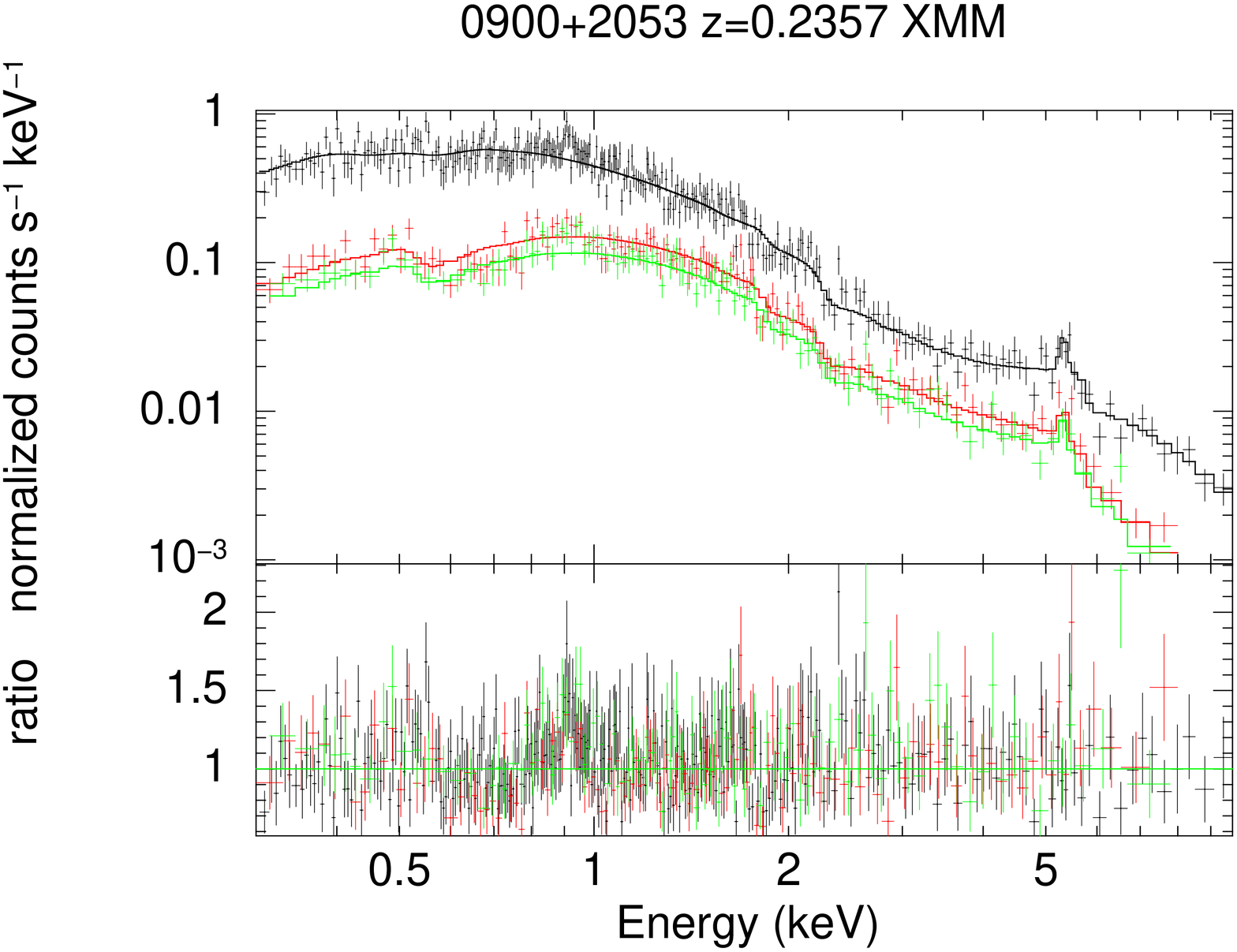,width=0.25\linewidth,angle=-90,clip=} &
\epsfig{file=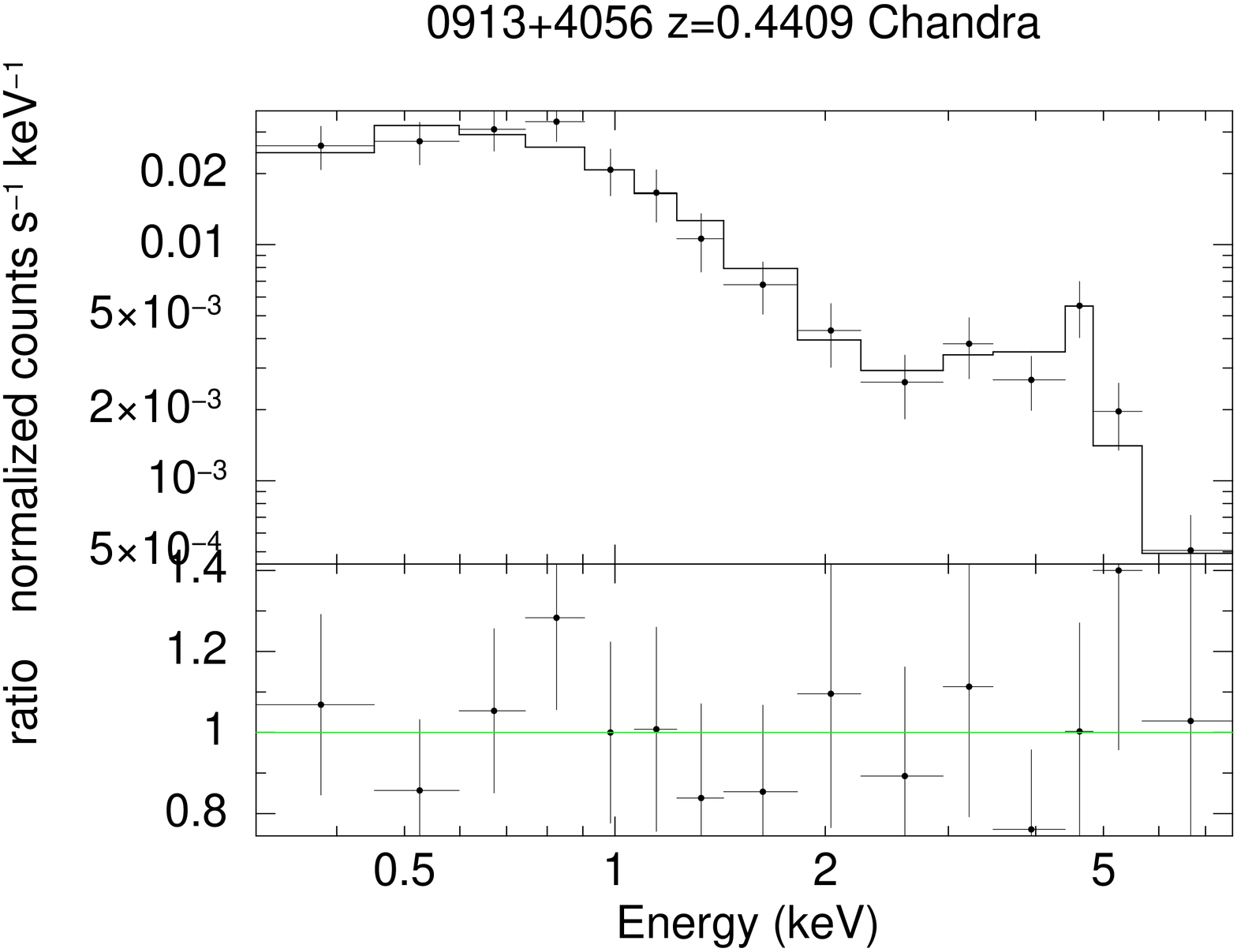,width=0.25\linewidth,angle=-90,clip=} &
\epsfig{file=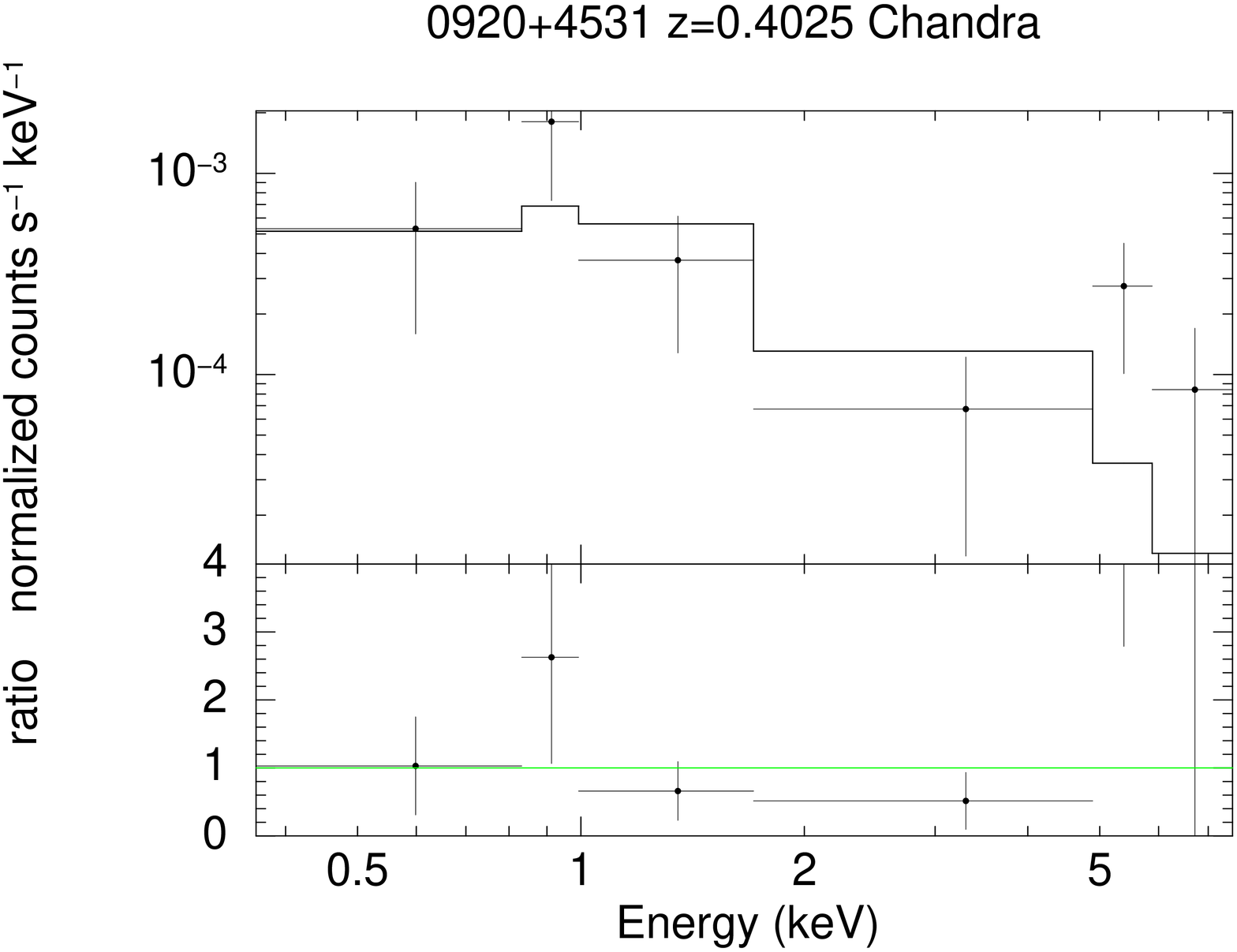,width=0.25\linewidth,angle=-90,clip=} \\
\epsfig{file=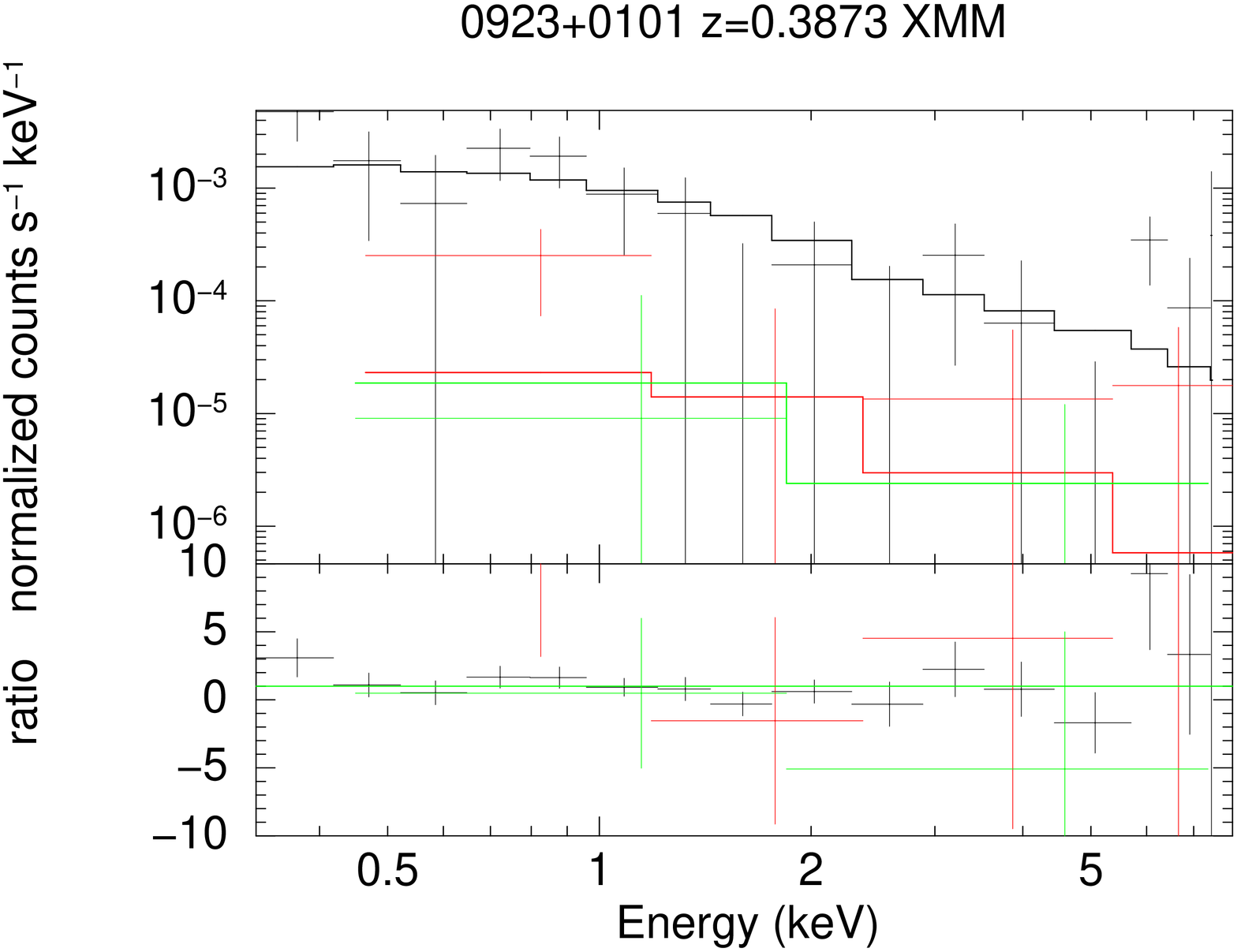,width=0.25\linewidth,angle=-90,clip=} & 
\epsfig{file=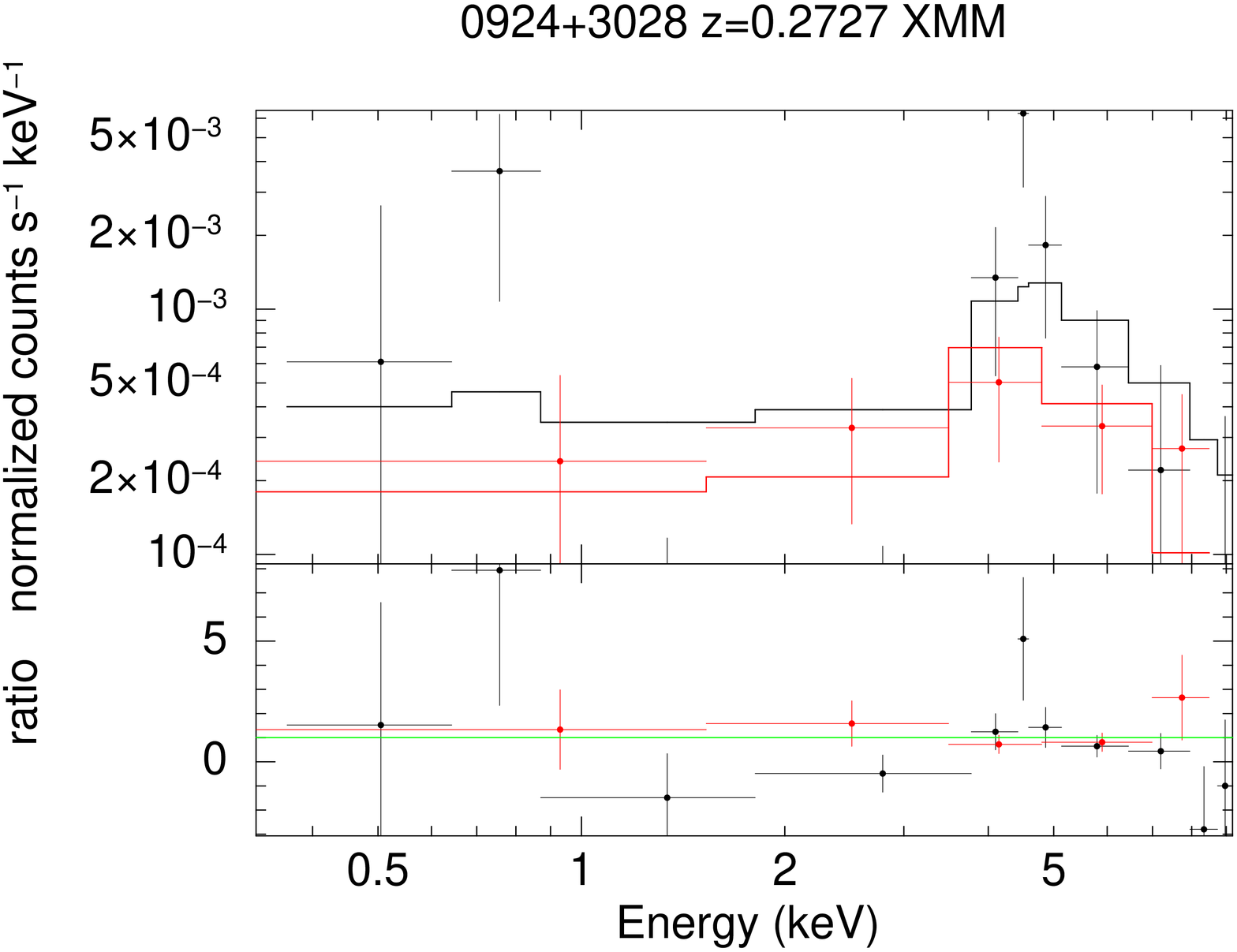,width=0.25\linewidth,angle=-90,clip=} &
\epsfig{file=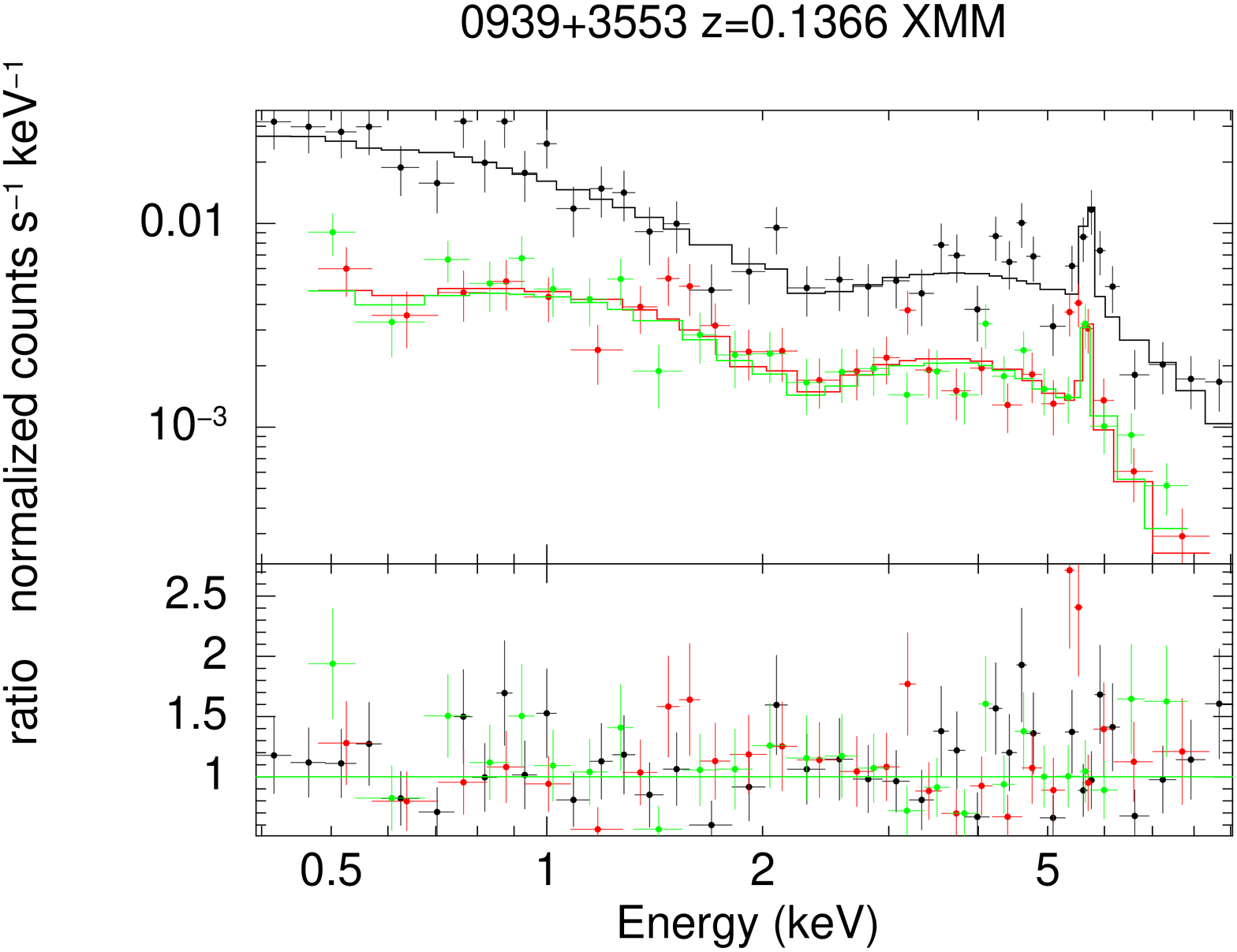,width=0.25\linewidth,angle=-90,clip=} \\
\epsfig{file=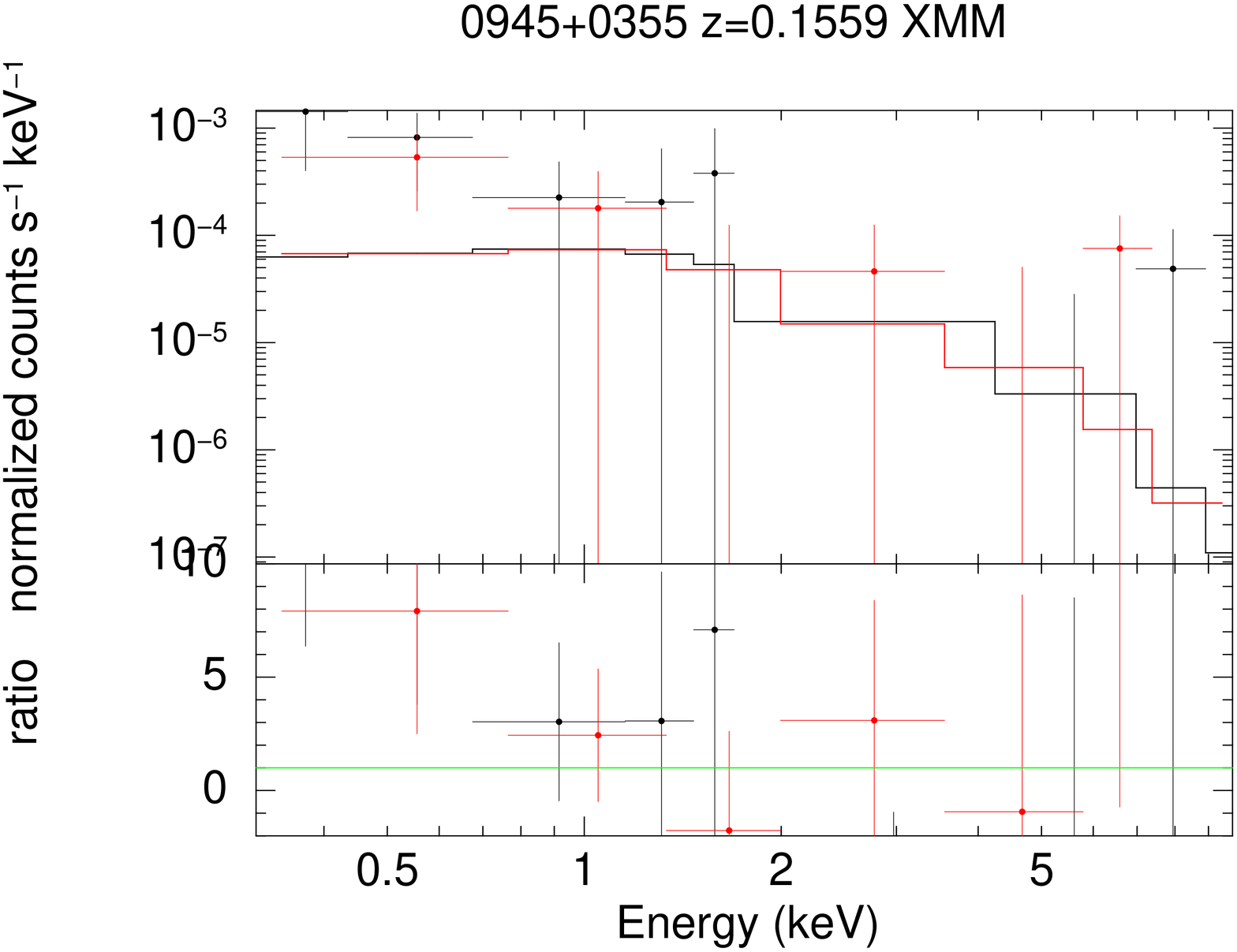,width=0.25\linewidth,angle=-90,clip=} &
\epsfig{file=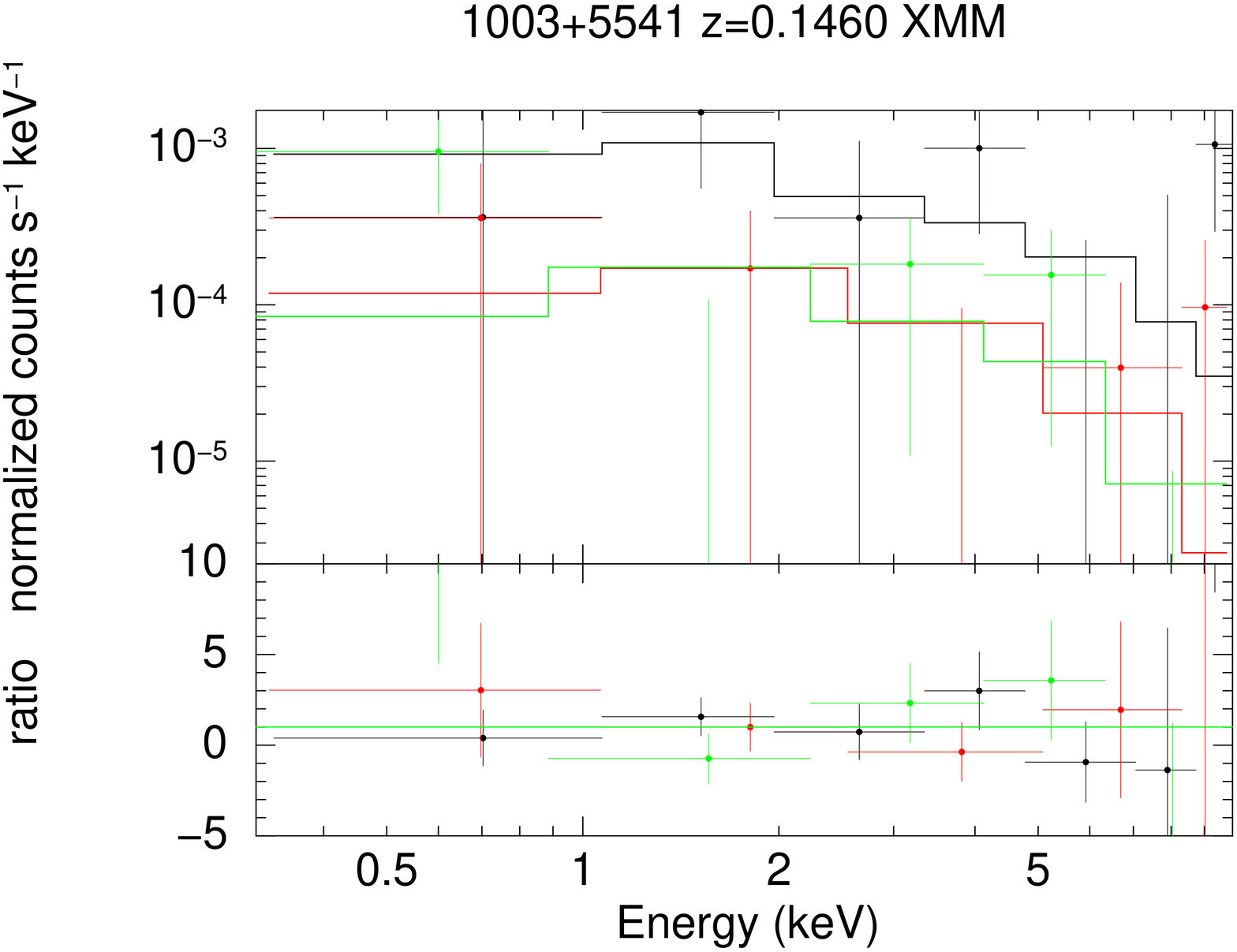,width=0.25\linewidth,angle=-90,clip=} &
\epsfig{file=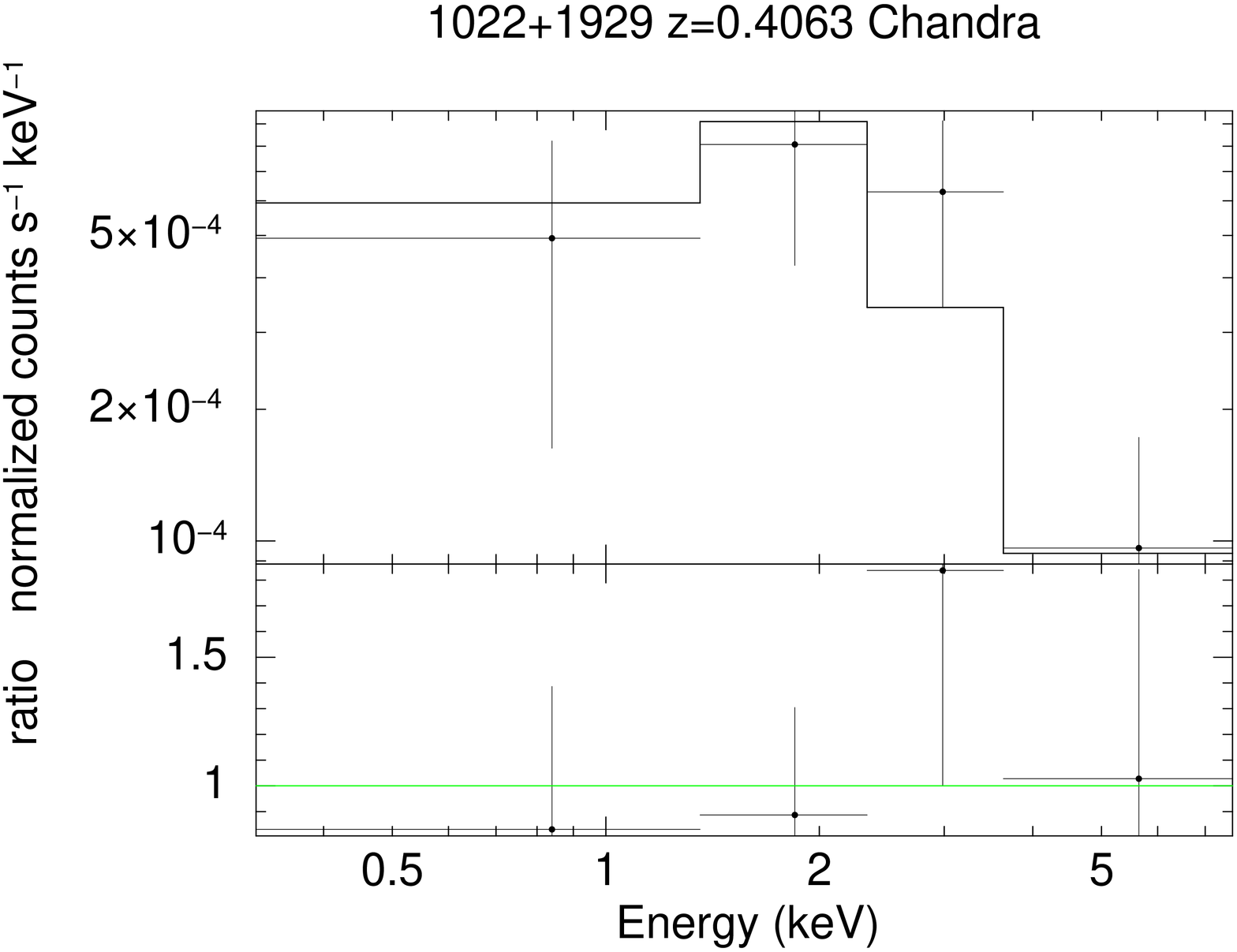,width=0.25\linewidth,angle=-90,clip=} \\
\end{tabular}
\centerline{Figure \ref{f:all}. --- {\it Continued}}
\end{figure*}

\begin{figure*}
\centering
\begin{tabular}{ccc}
\epsfig{file=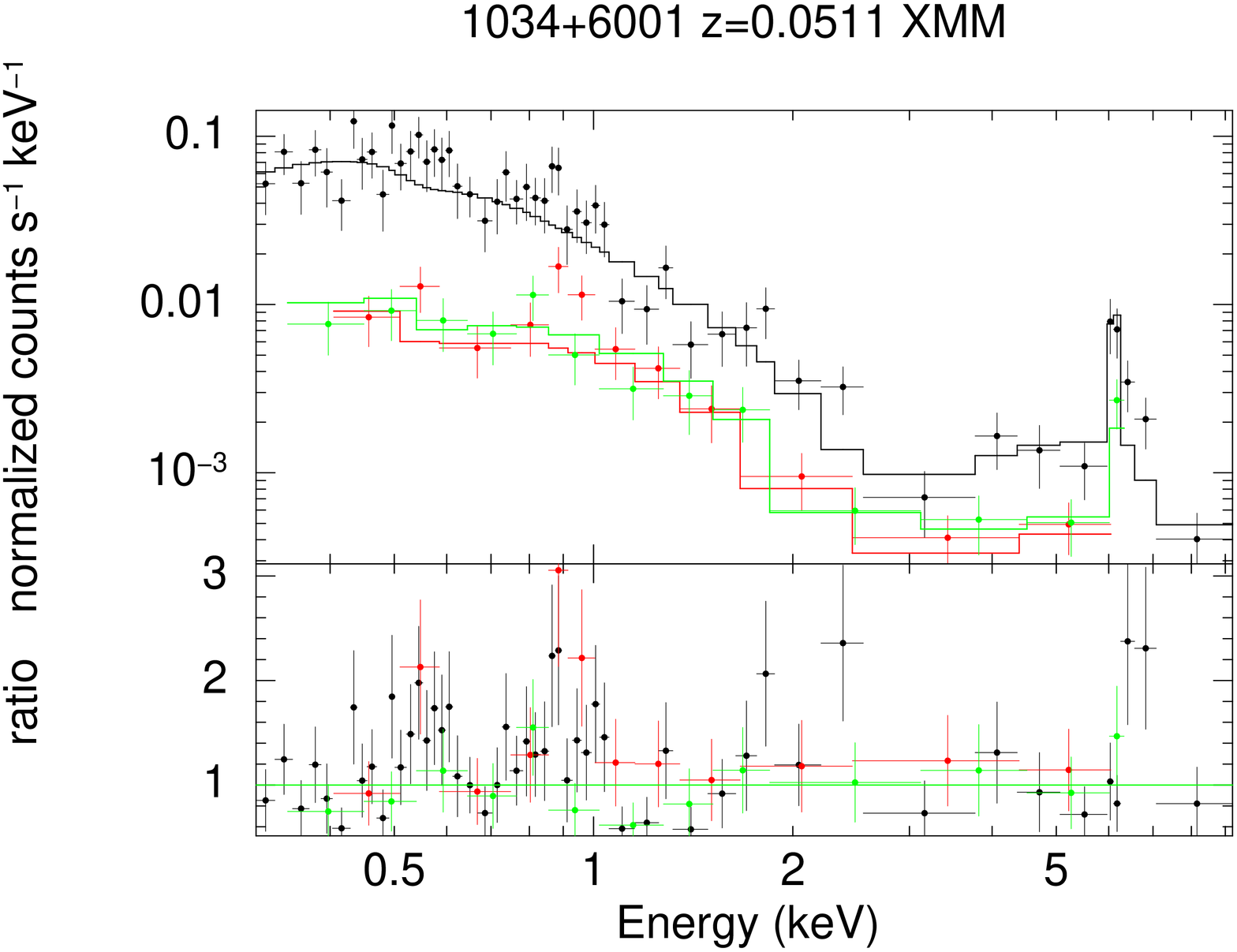,width=0.25\linewidth,angle=-90,clip=} & 
\epsfig{file=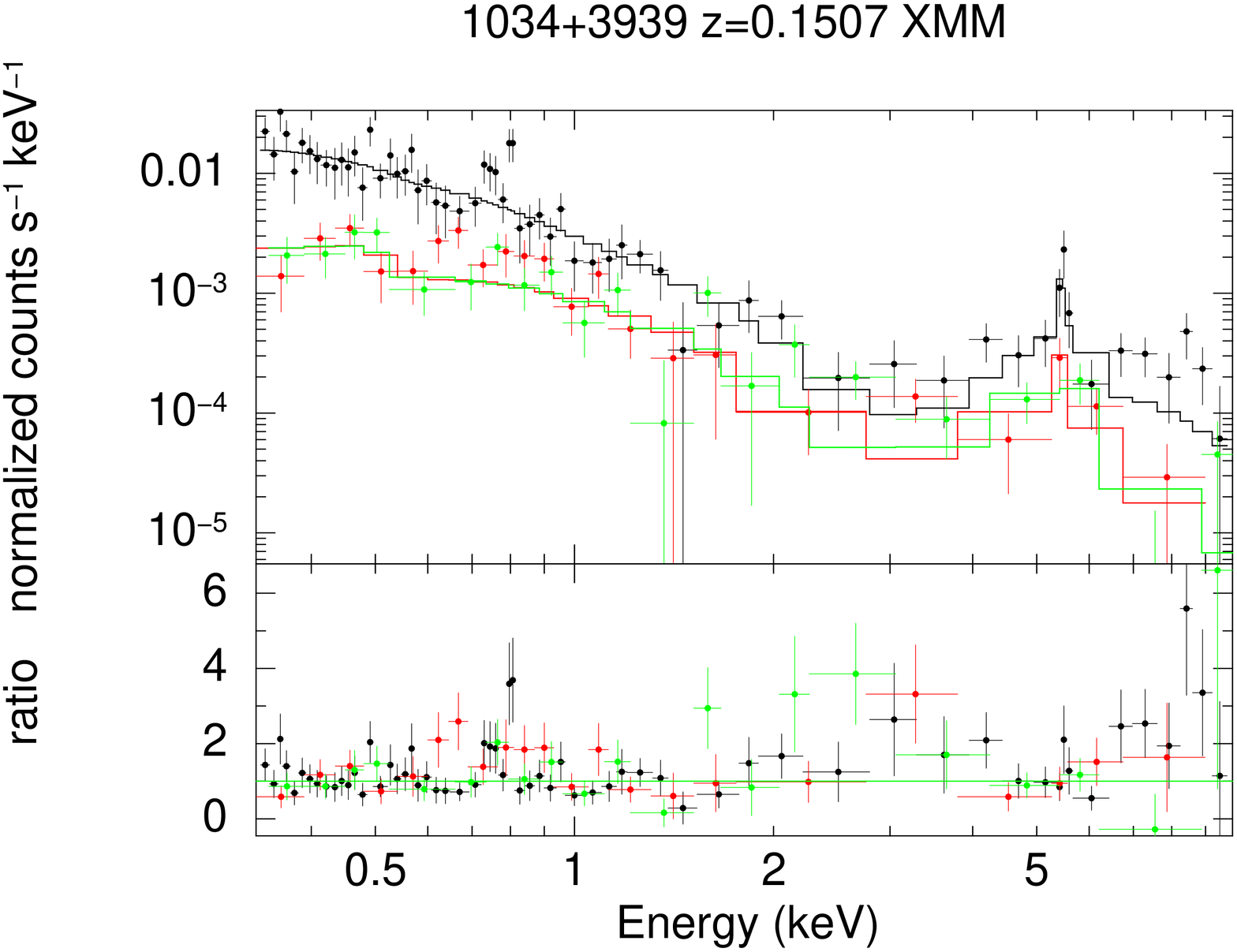,width=0.25\linewidth,angle=-90,clip=} &
\epsfig{file=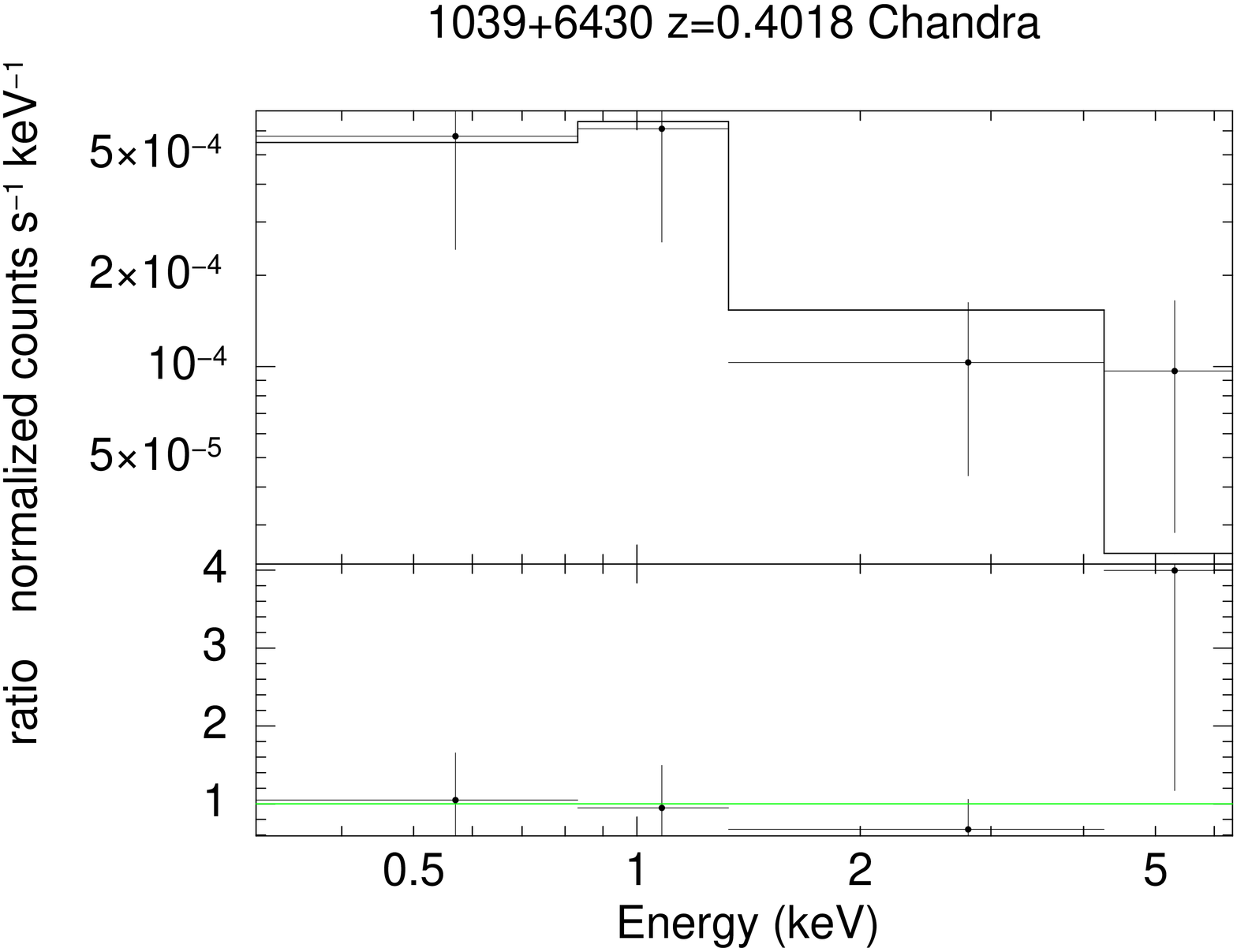,width=0.25\linewidth,angle=-90,clip=} \\
\epsfig{file=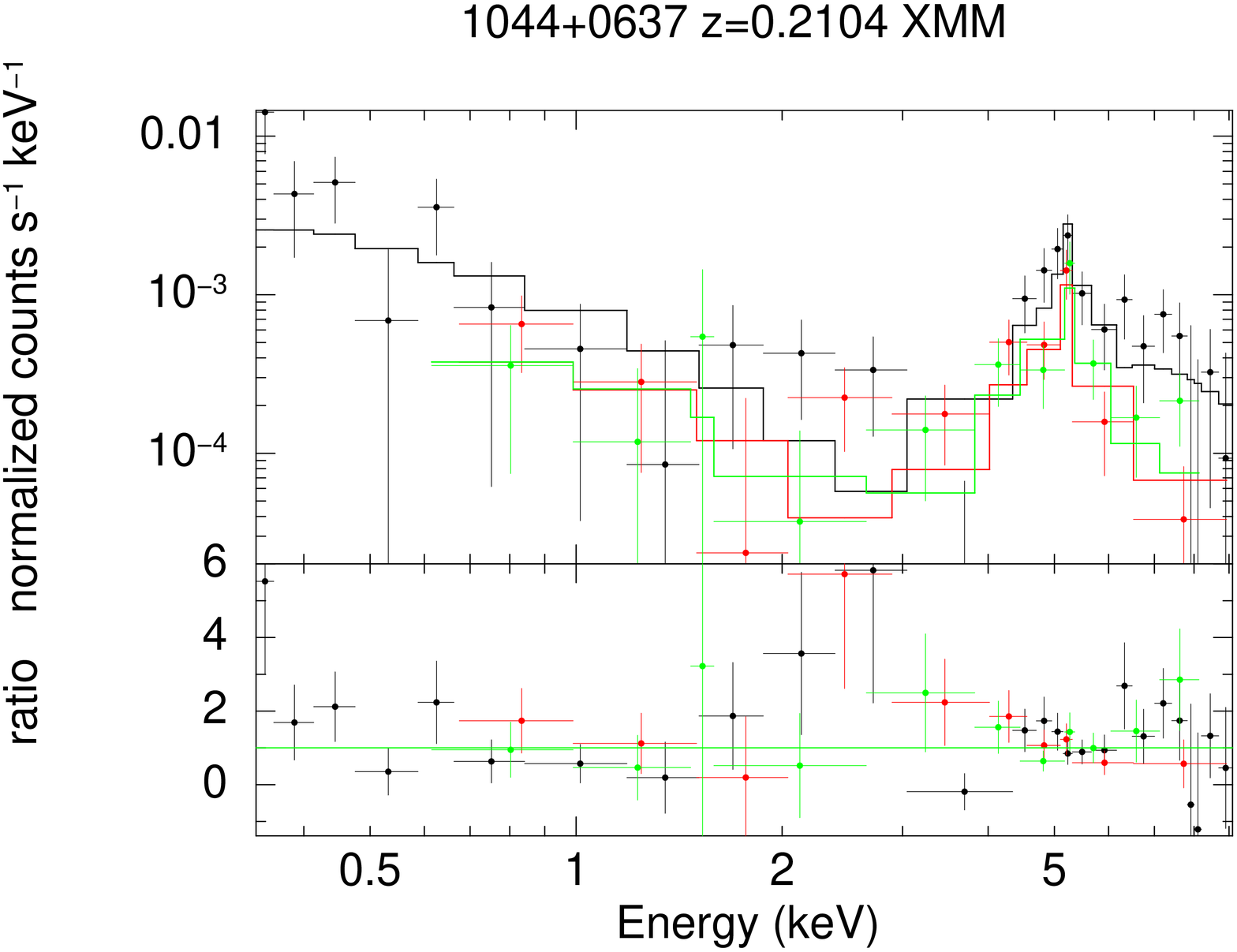,width=0.25\linewidth,angle=-90,clip=} &
\epsfig{file=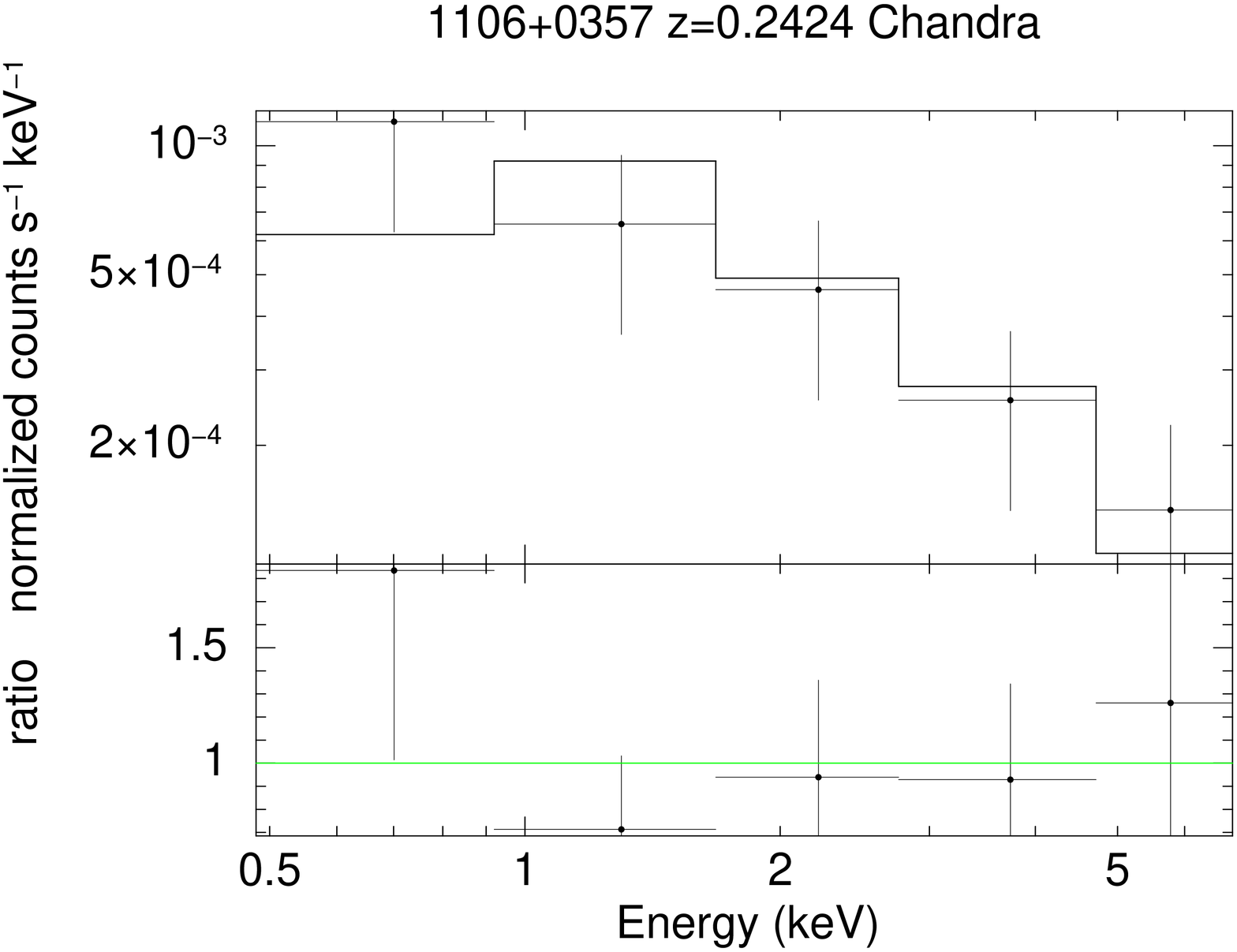,width=0.25\linewidth,angle=-90,clip=} &
\epsfig{file=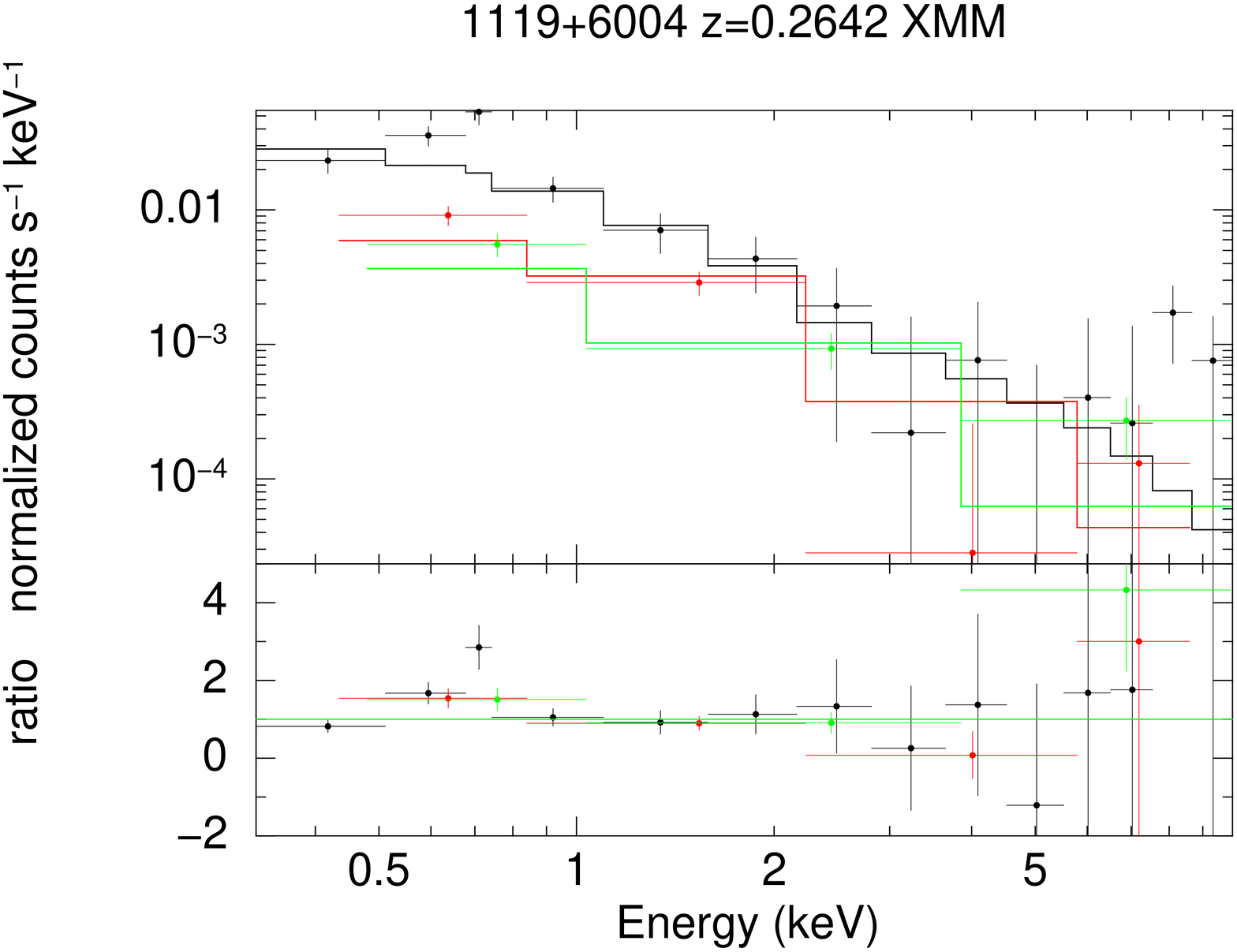,width=0.25\linewidth,angle=-90,clip=} \\
\epsfig{file=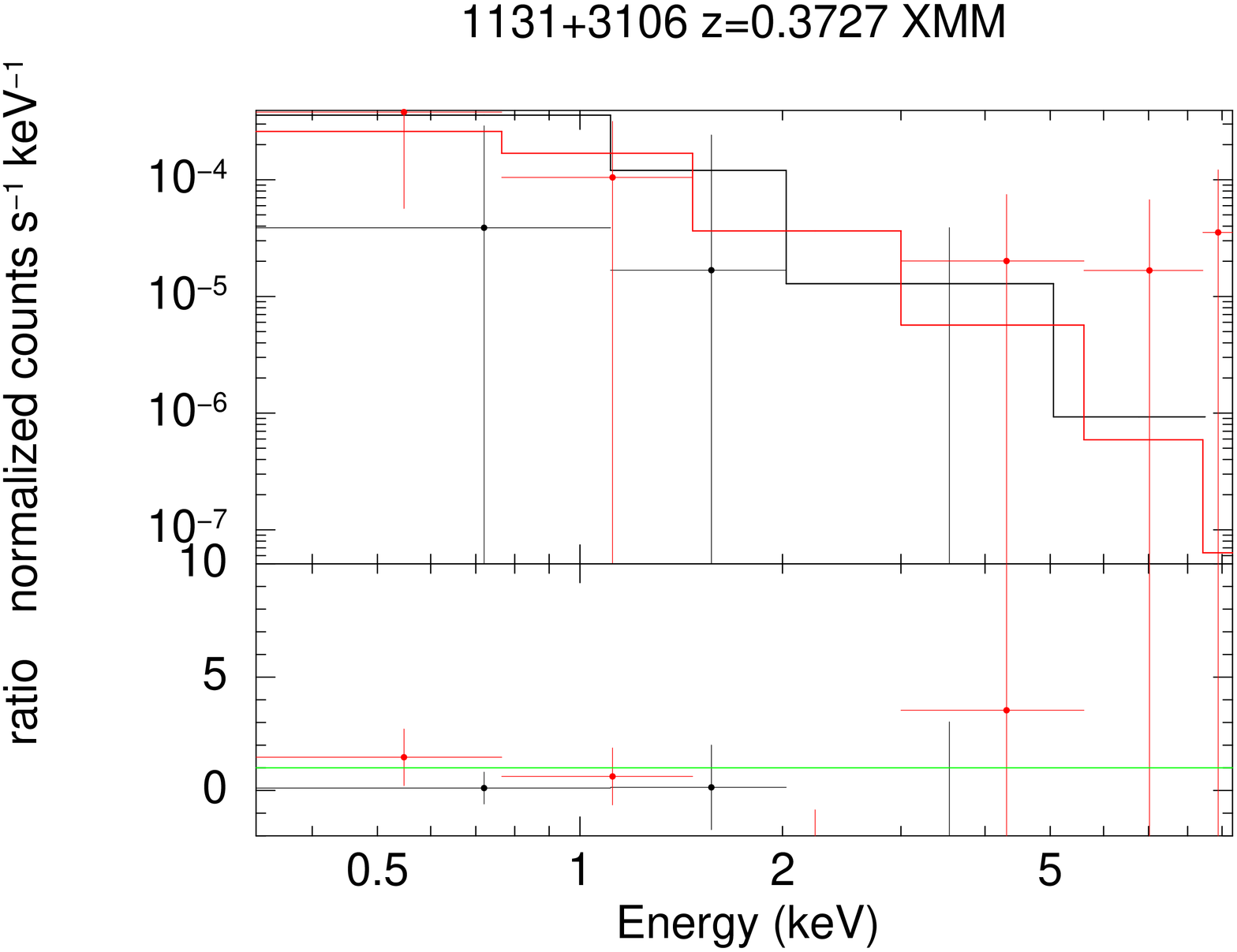,width=0.25\linewidth,angle=-90,clip=} & 
\epsfig{file=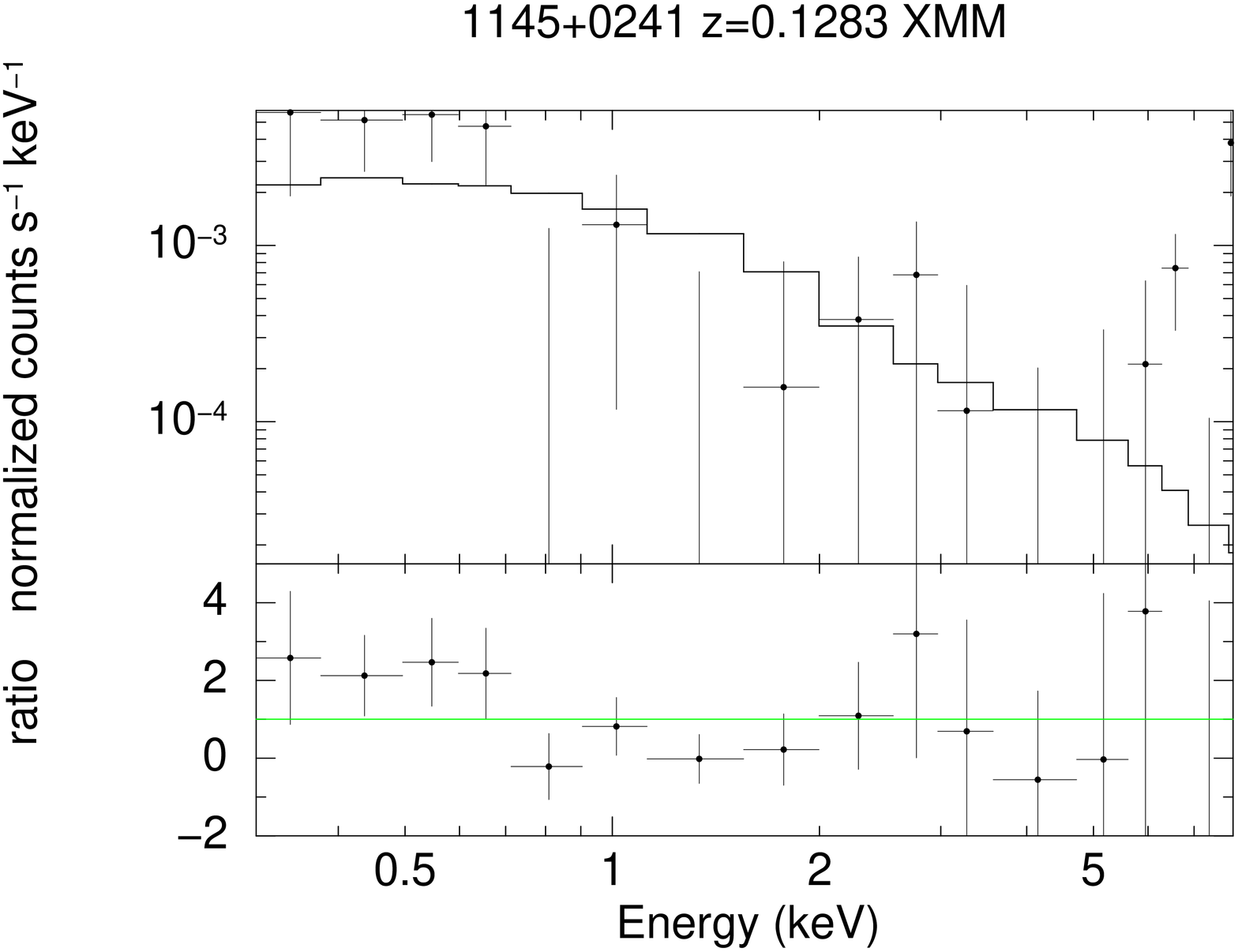,width=0.25\linewidth,angle=-90,clip=} &
\epsfig{file=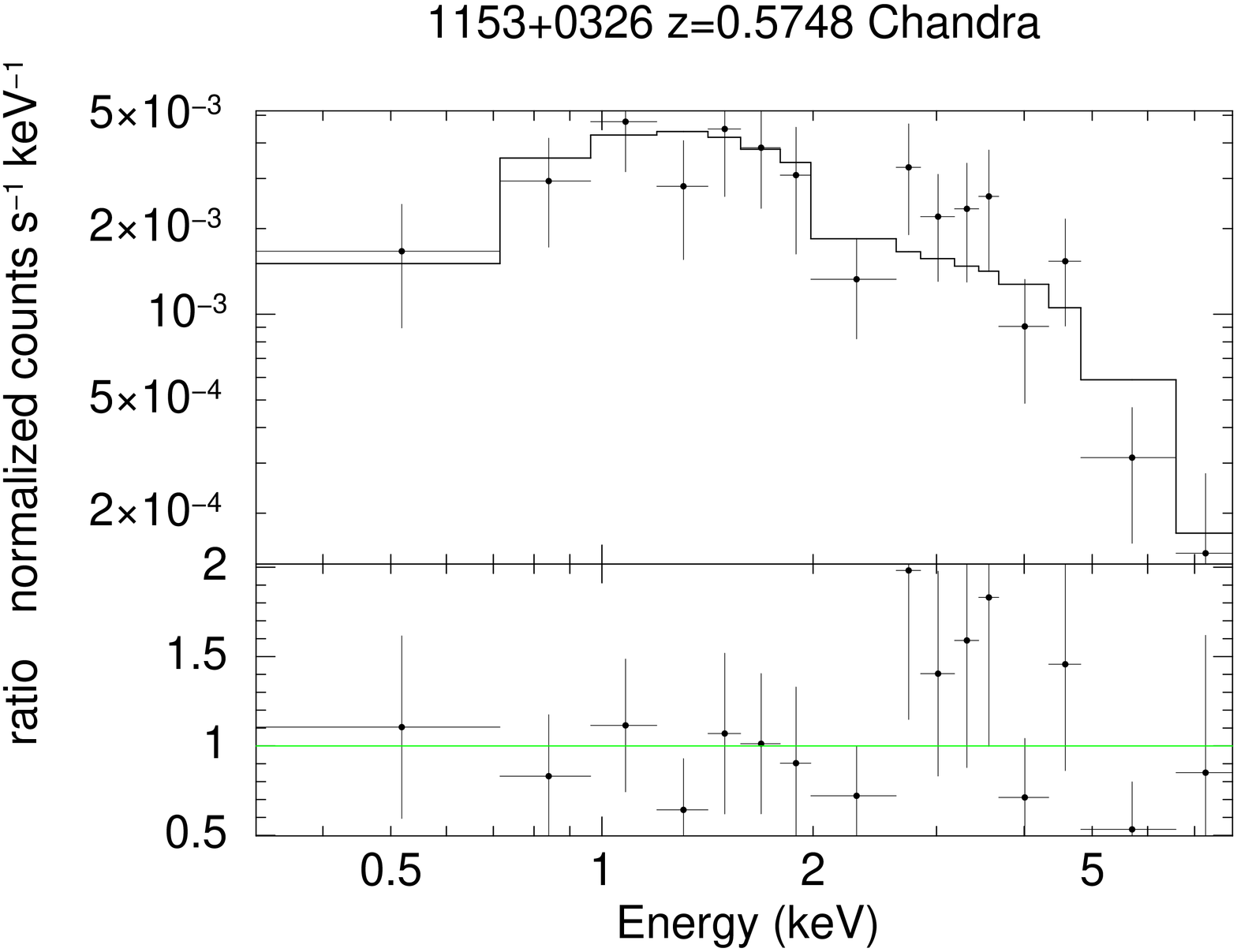,width=0.25\linewidth,angle=-90,clip=} \\
\epsfig{file=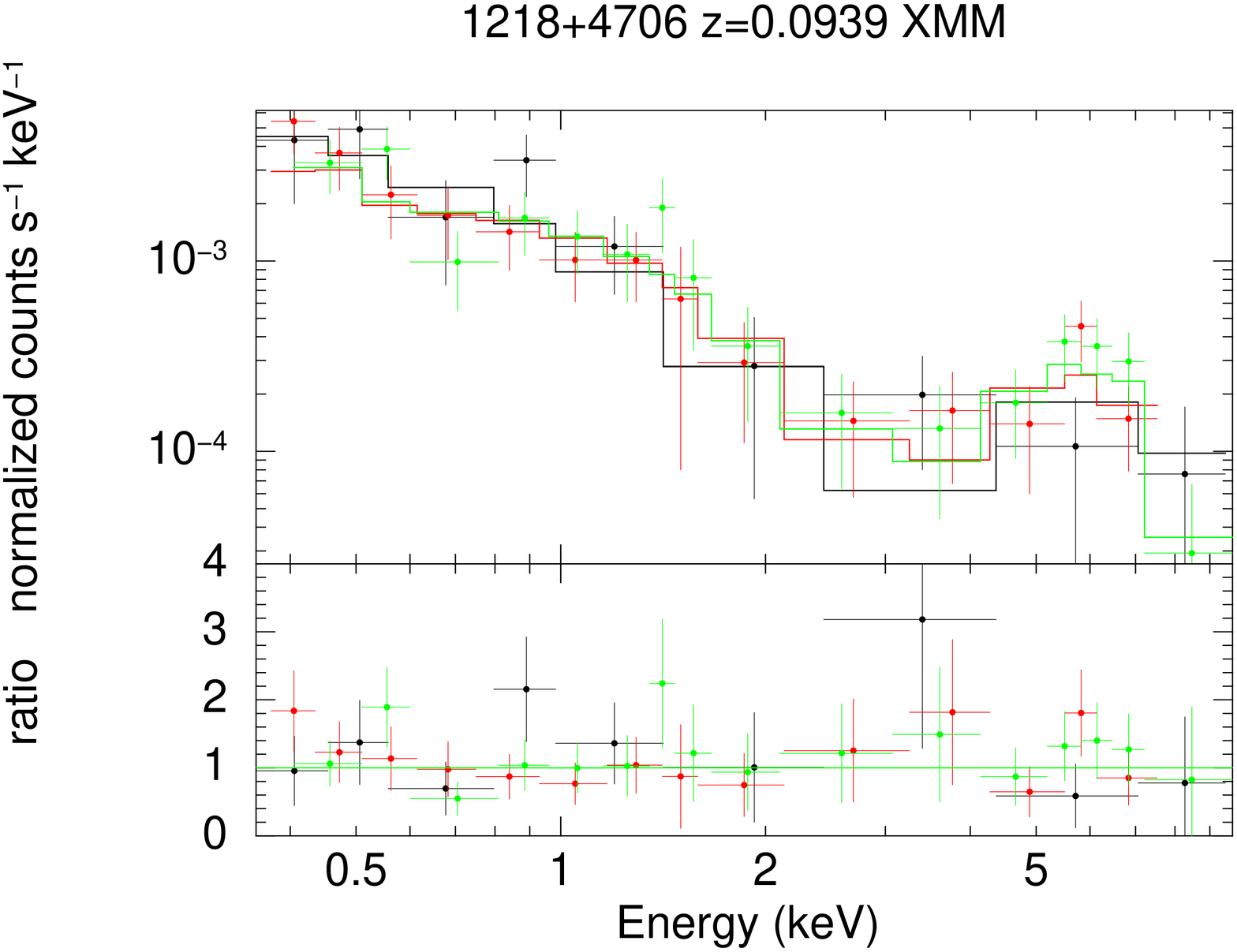,width=0.25\linewidth,angle=-90,clip=} &
\epsfig{file=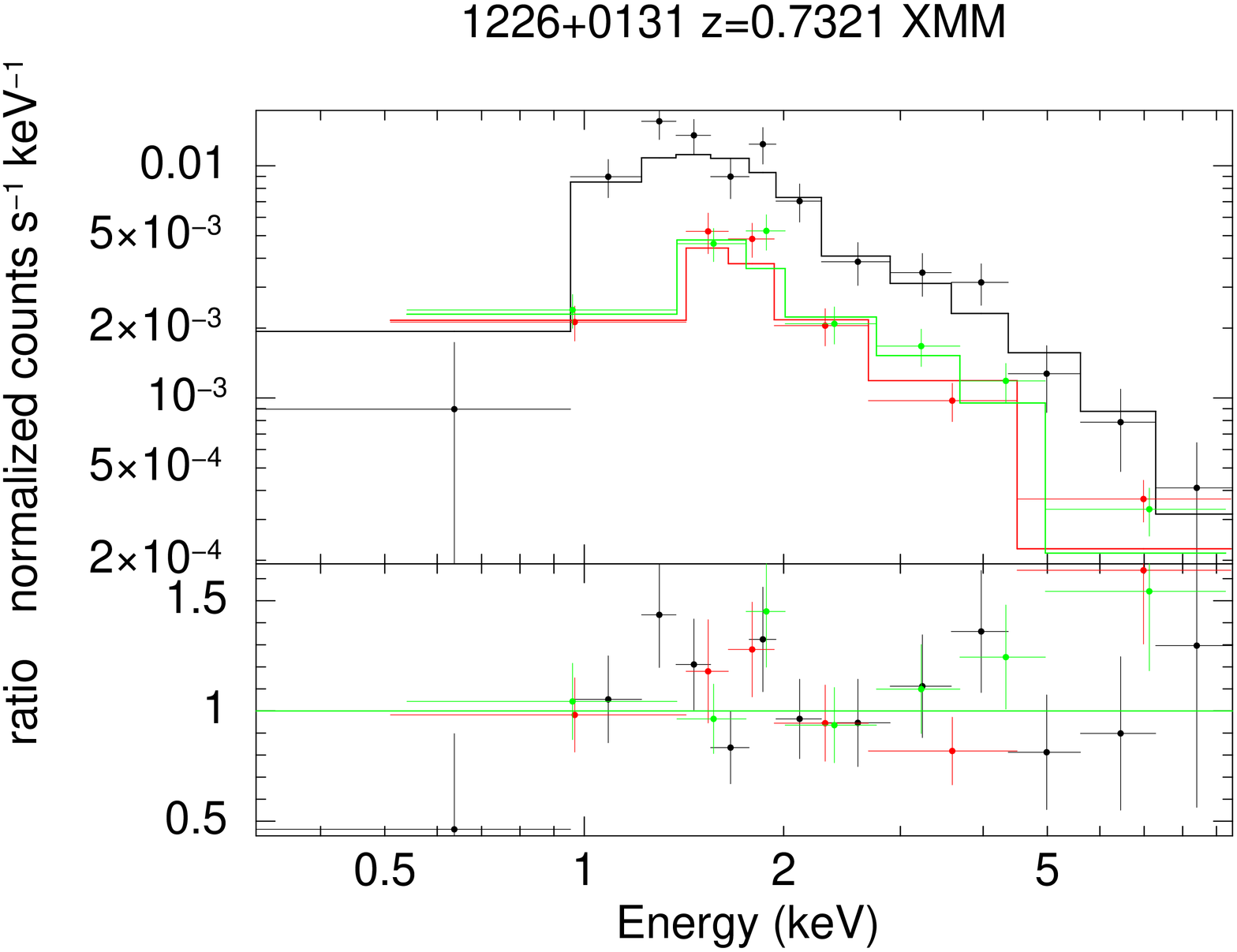,width=0.25\linewidth,angle=-90,clip=} &
\epsfig{file=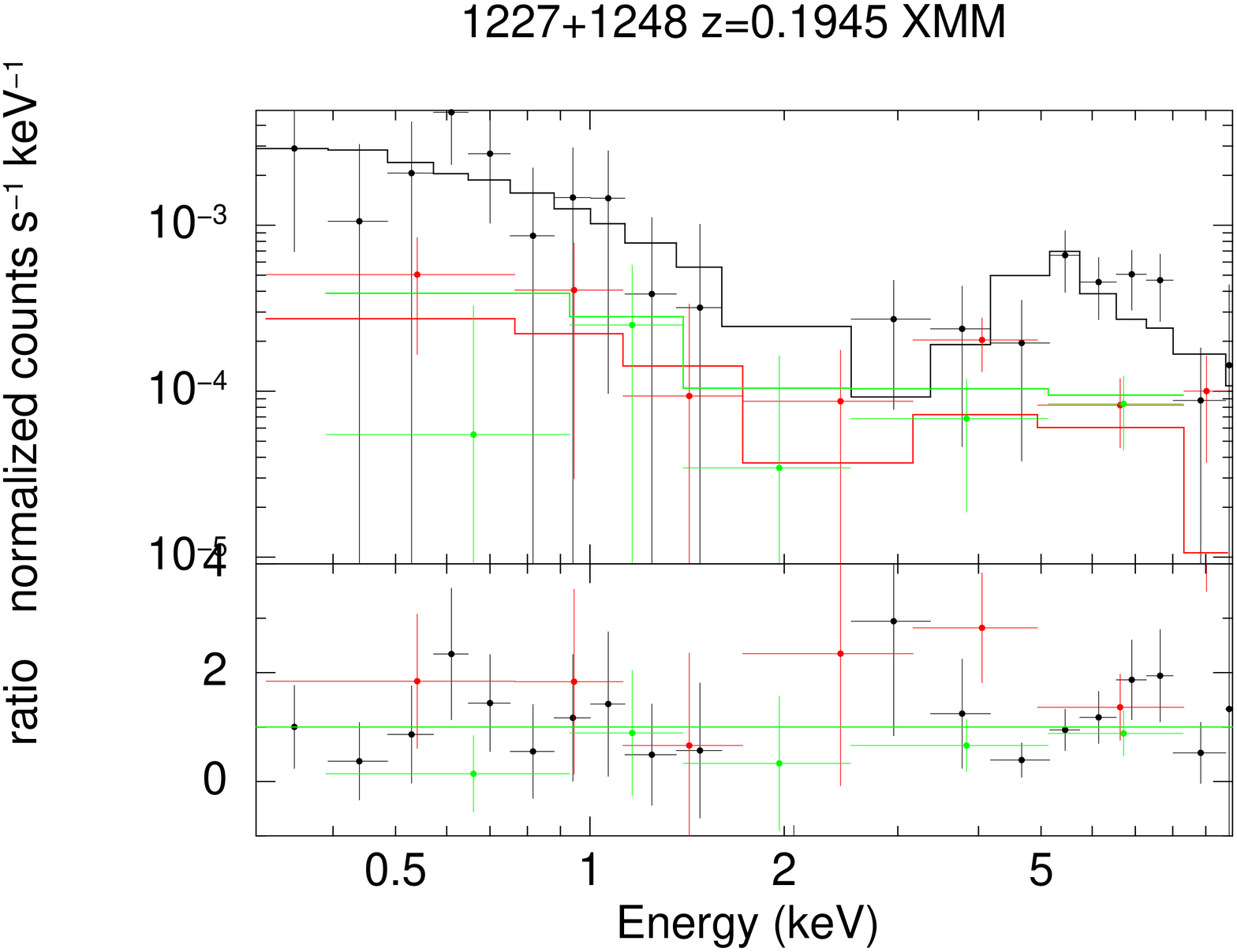,width=0.25\linewidth,angle=-90,clip=} \\
\end{tabular}
\centerline{Figure \ref{f:all}. --- {\it Continued}}
\end{figure*}

\begin{figure*}
\centering
\begin{tabular}{ccc}
\epsfig{file=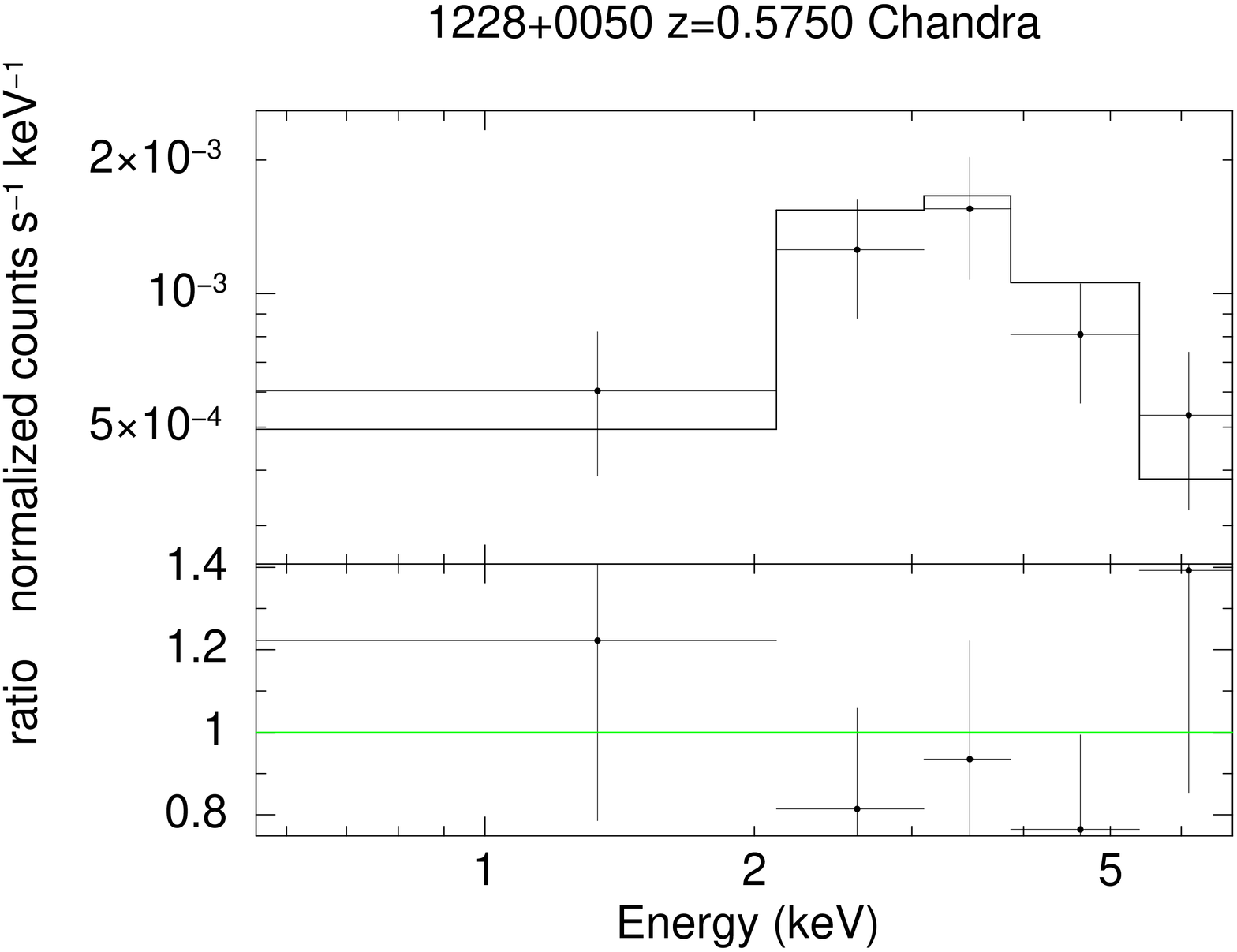,width=0.25\linewidth,angle=-90,clip=} & 
\epsfig{file=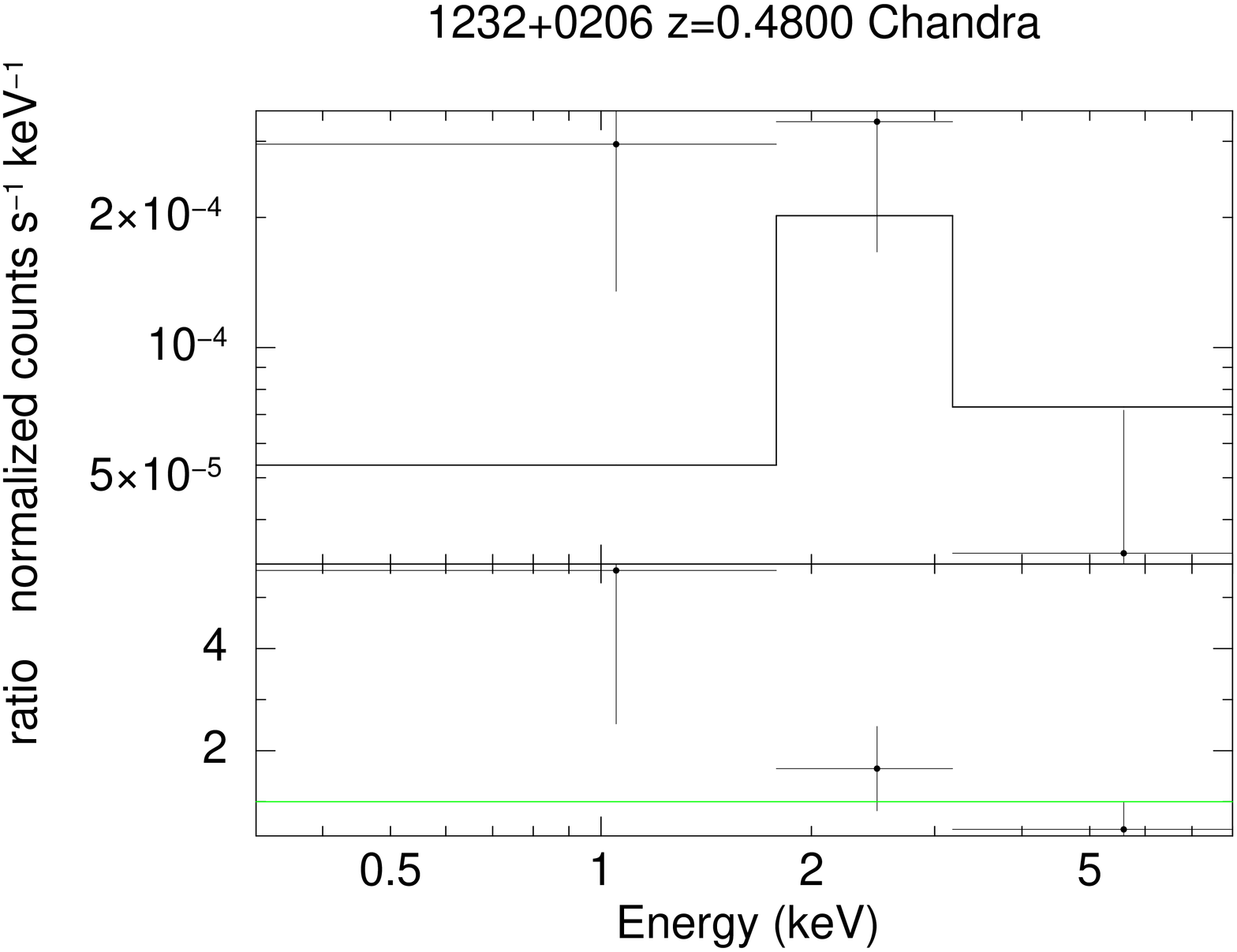,width=0.25\linewidth,angle=-90,clip=} &
\epsfig{file=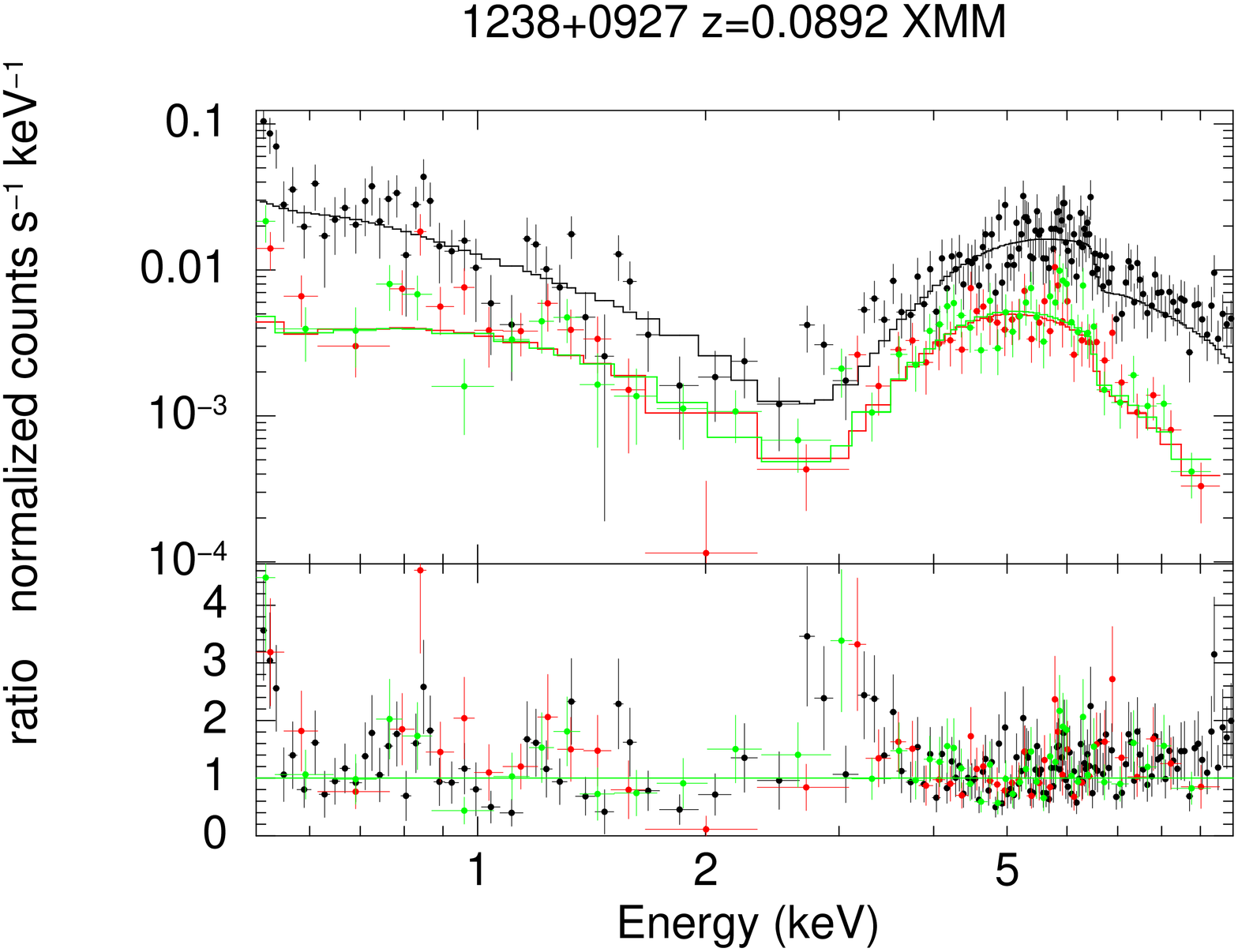,width=0.25\linewidth,angle=-90,clip=} \\
\epsfig{file=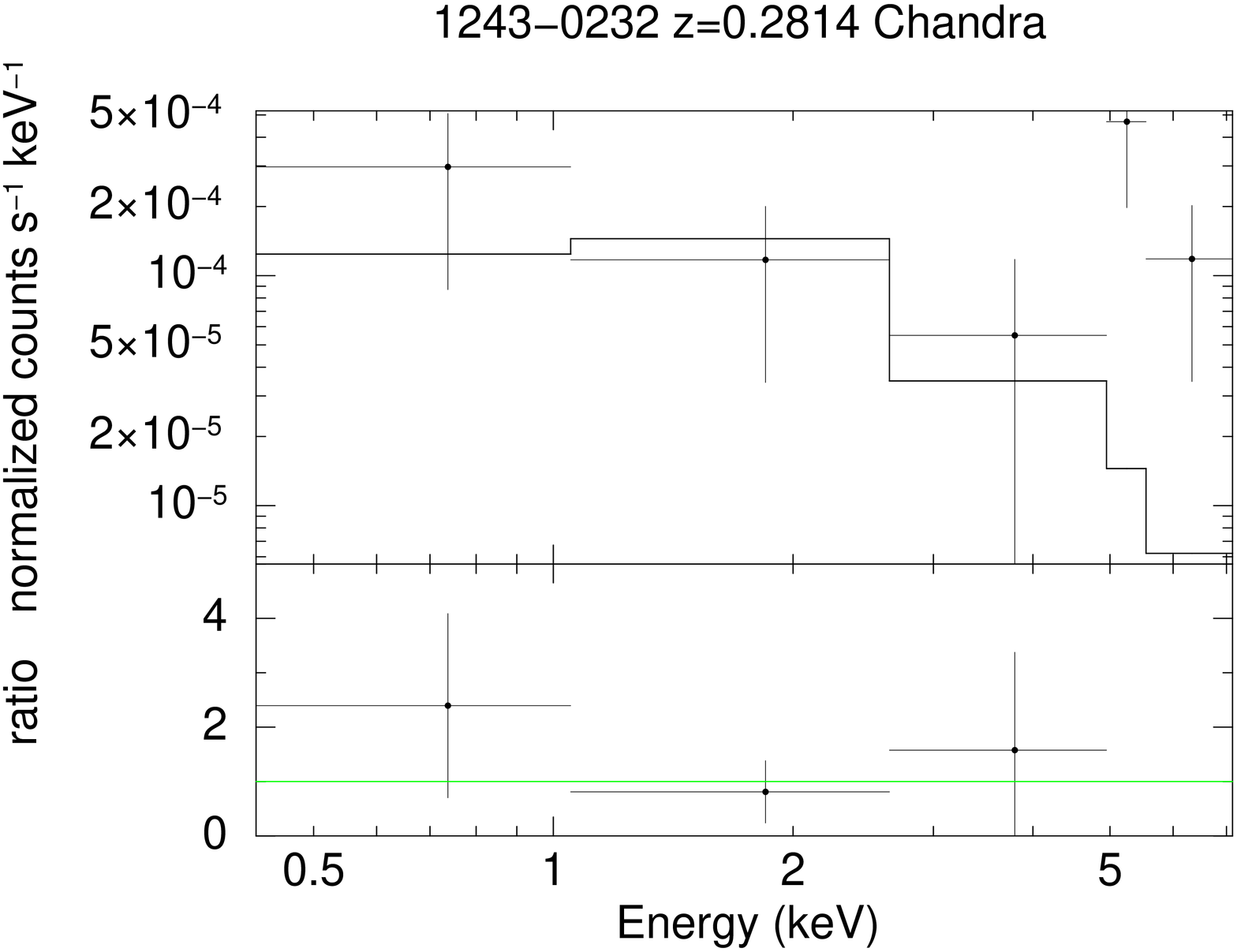,width=0.25\linewidth,angle=-90,clip=} &
\epsfig{file=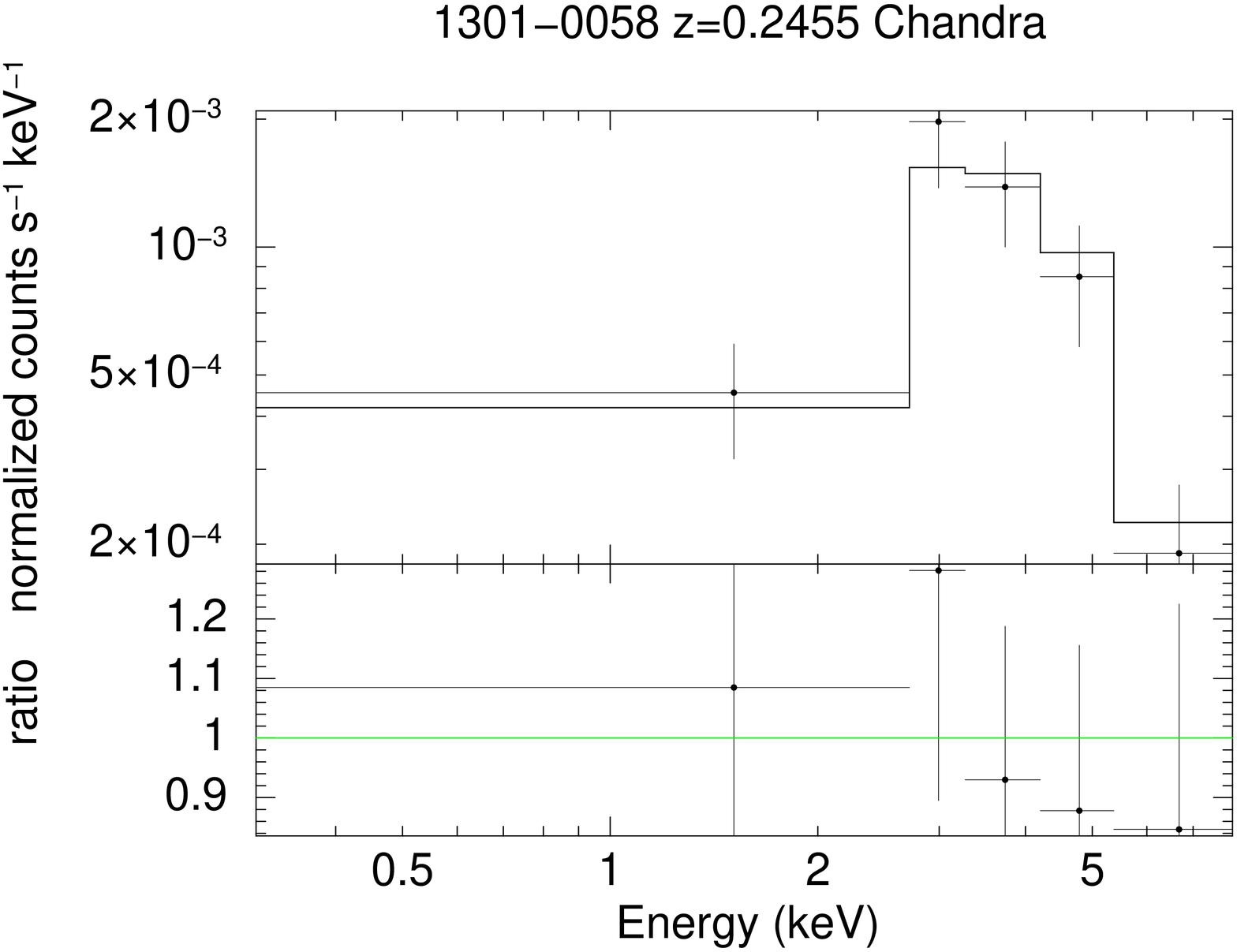,width=0.25\linewidth,angle=-90,clip=} &
\epsfig{file=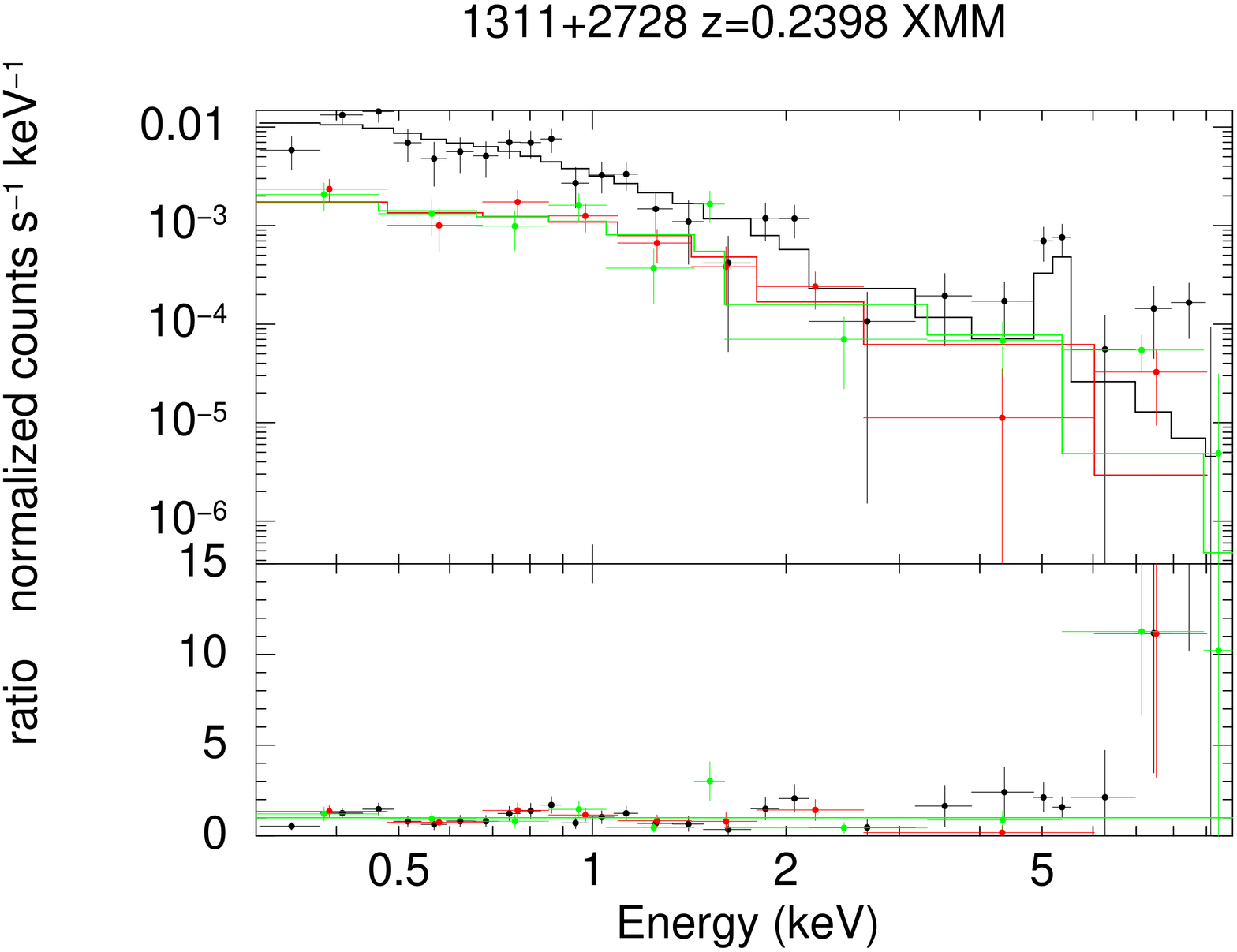,width=0.25\linewidth,angle=-90,clip=} \\
\epsfig{file=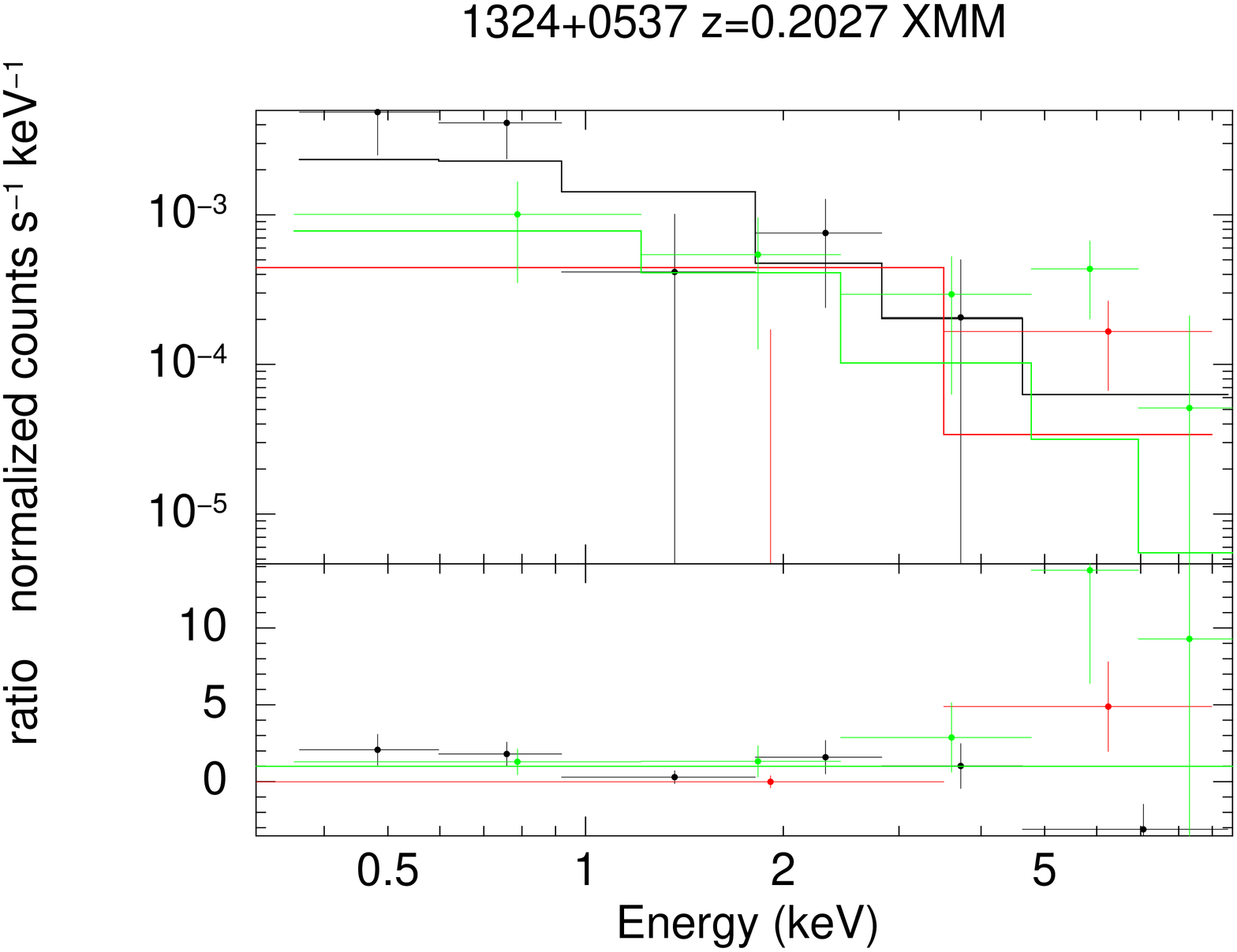,width=0.25\linewidth,angle=-90,clip=} &
\epsfig{file=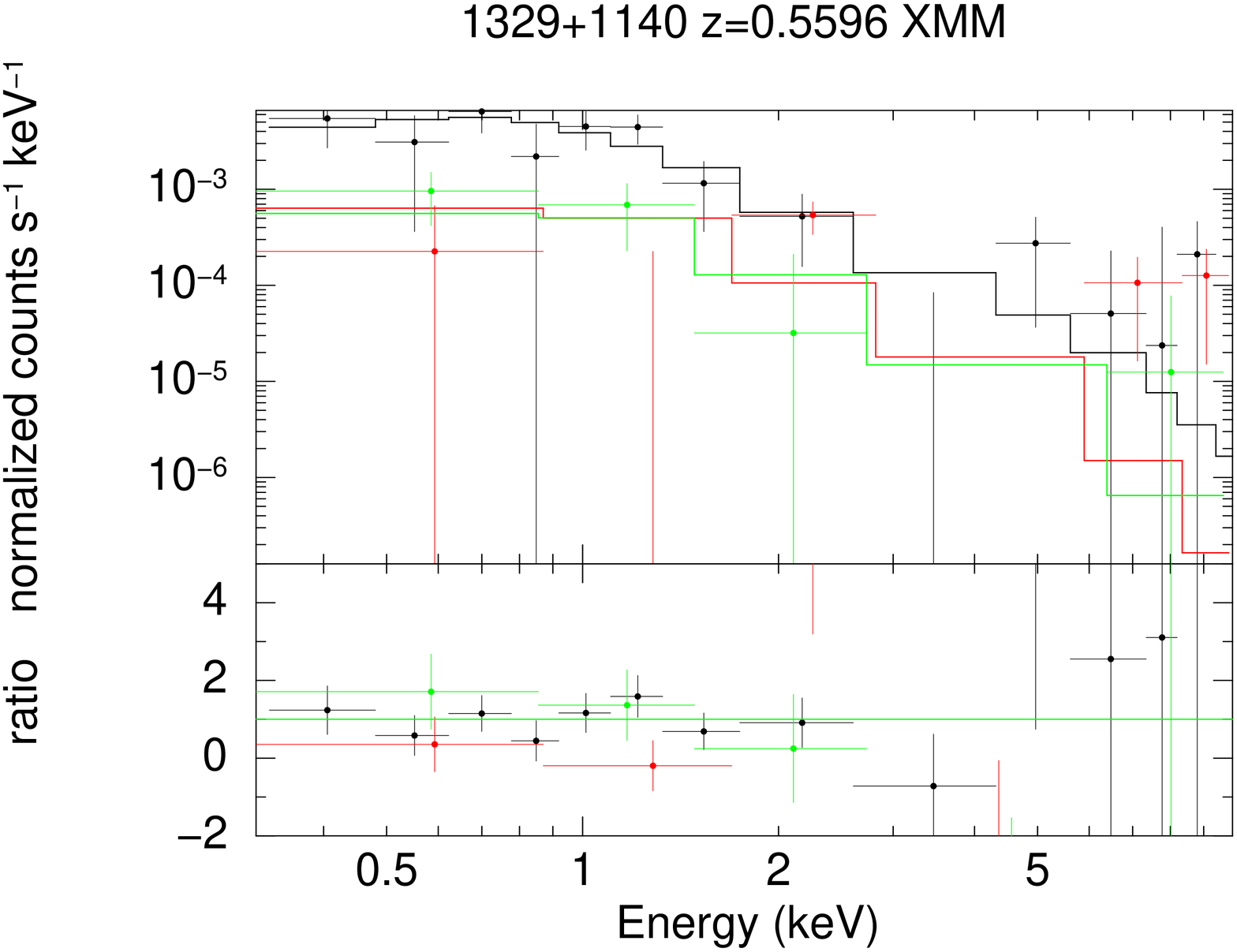,width=0.25\linewidth,angle=-90,clip=} &
\epsfig{file=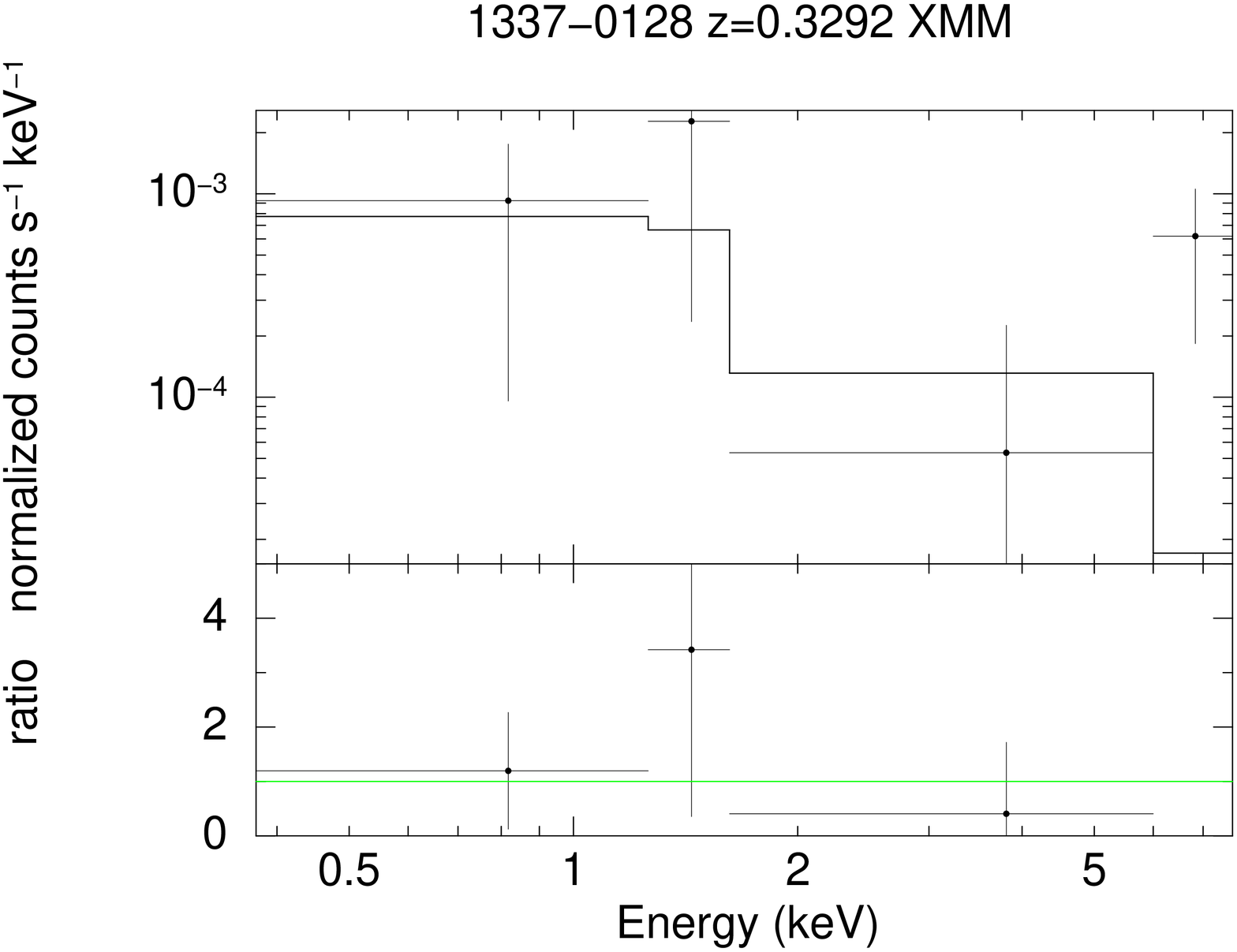,width=0.25\linewidth,angle=-90,clip=}\\
\epsfig{file=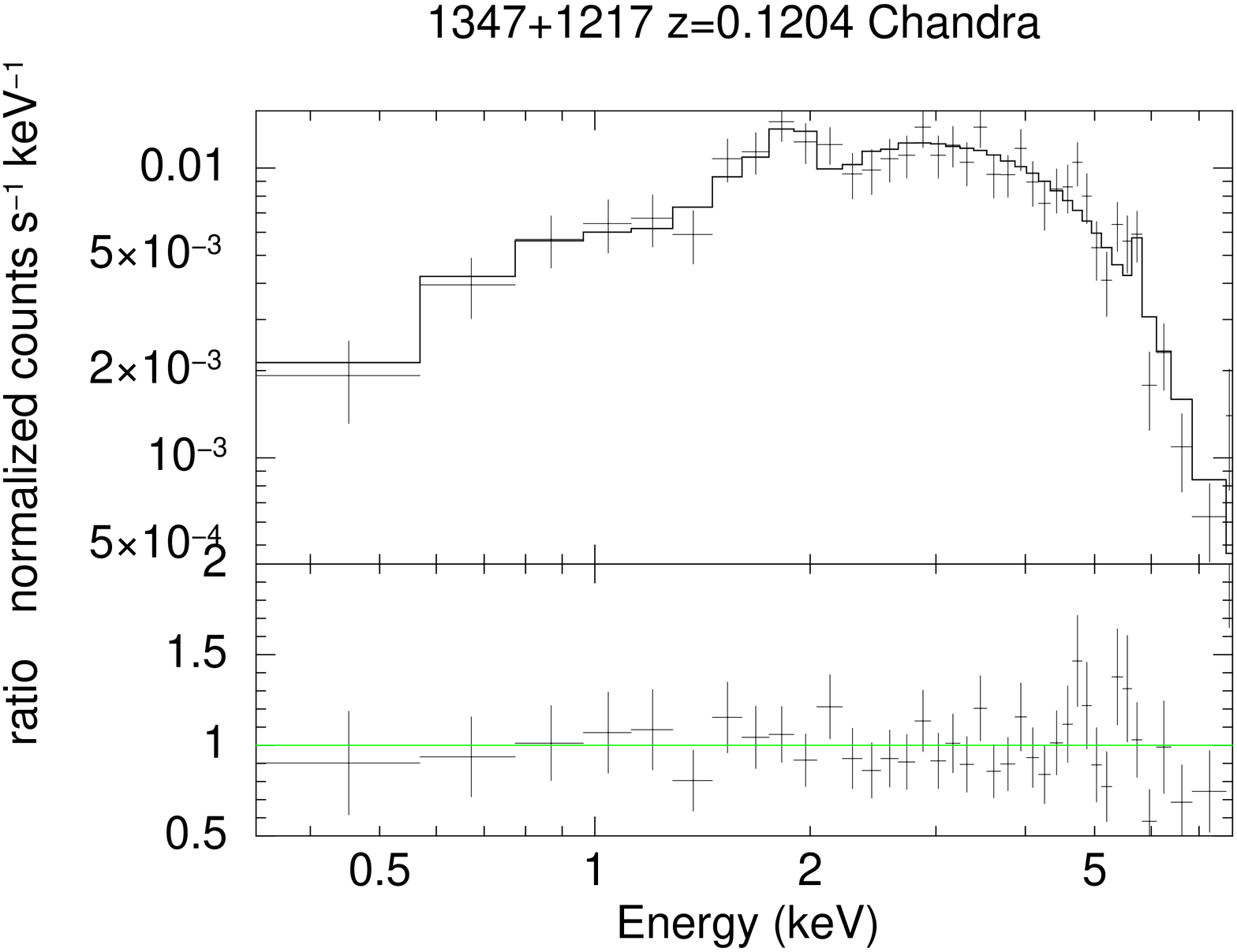,width=0.25\linewidth,angle=-90,clip=} &
\epsfig{file=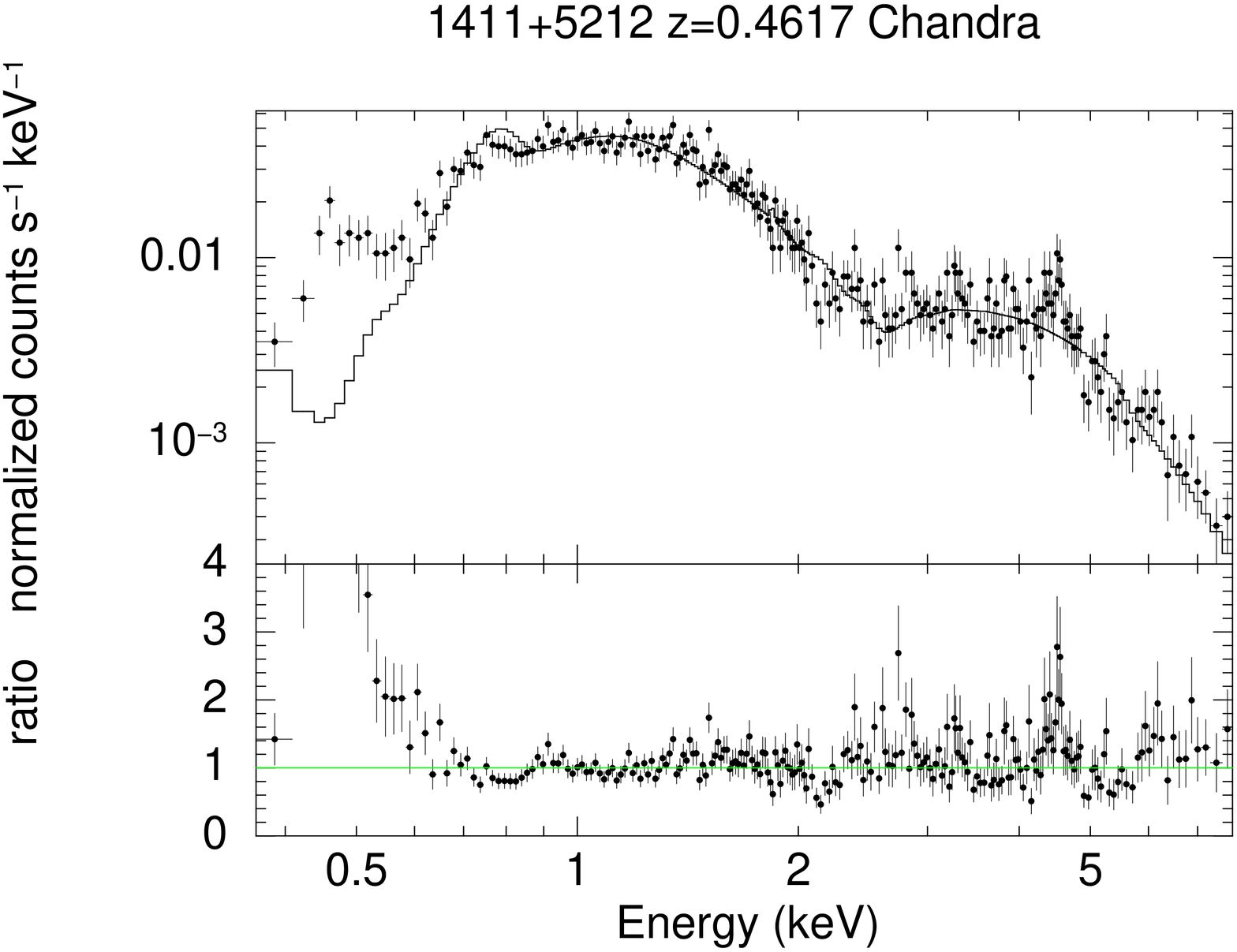,width=0.25\linewidth,angle=-90,clip=} & 
\epsfig{file=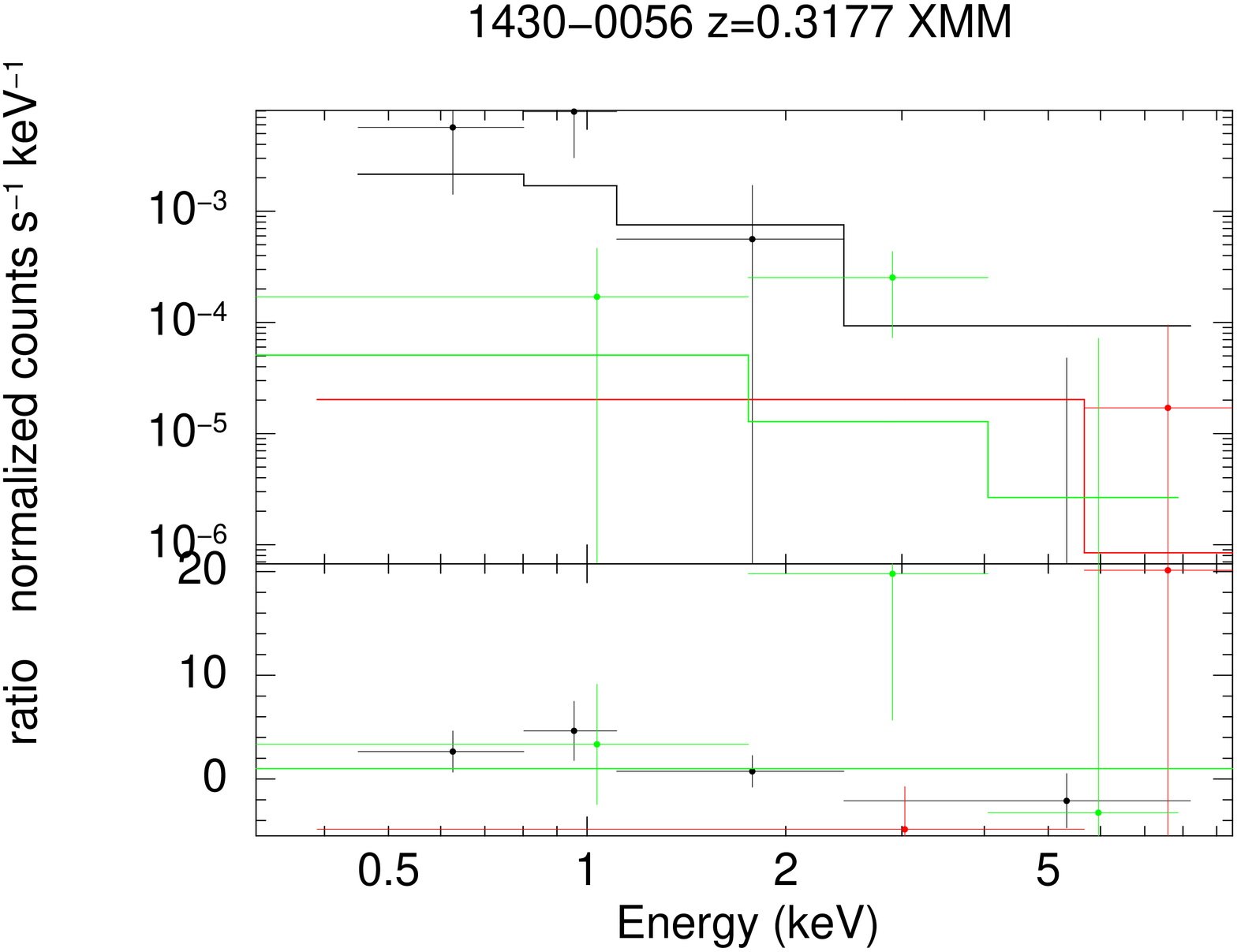,width=0.25\linewidth,angle=-90,clip=} \\
\end{tabular}
\centerline{Figure \ref{f:all}. --- {\it Continued}}
\end{figure*}

\begin{figure*}
\centering
\begin{tabular}{ccc}
\epsfig{file=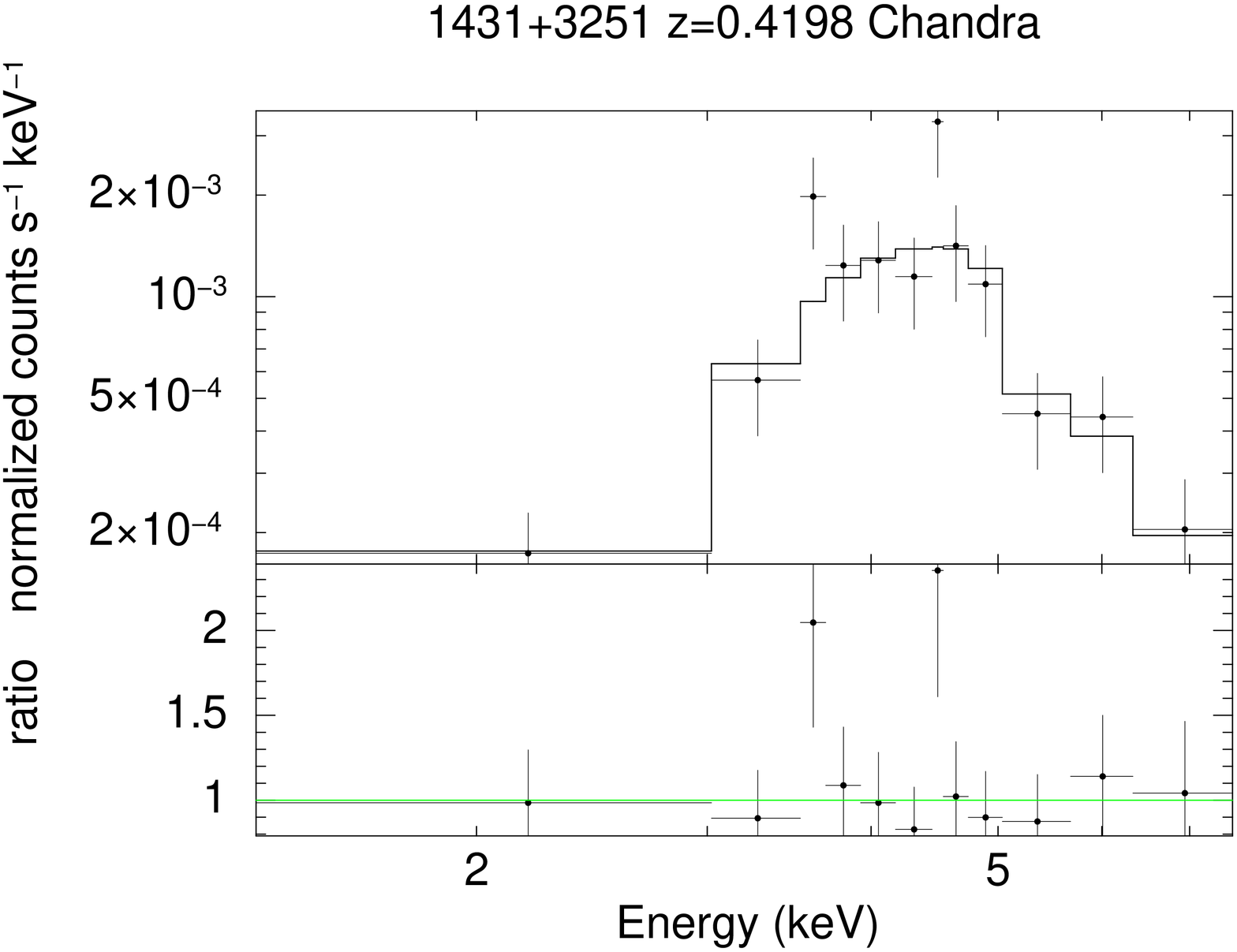,width=0.25\linewidth,angle=-90,clip=} &
\epsfig{file=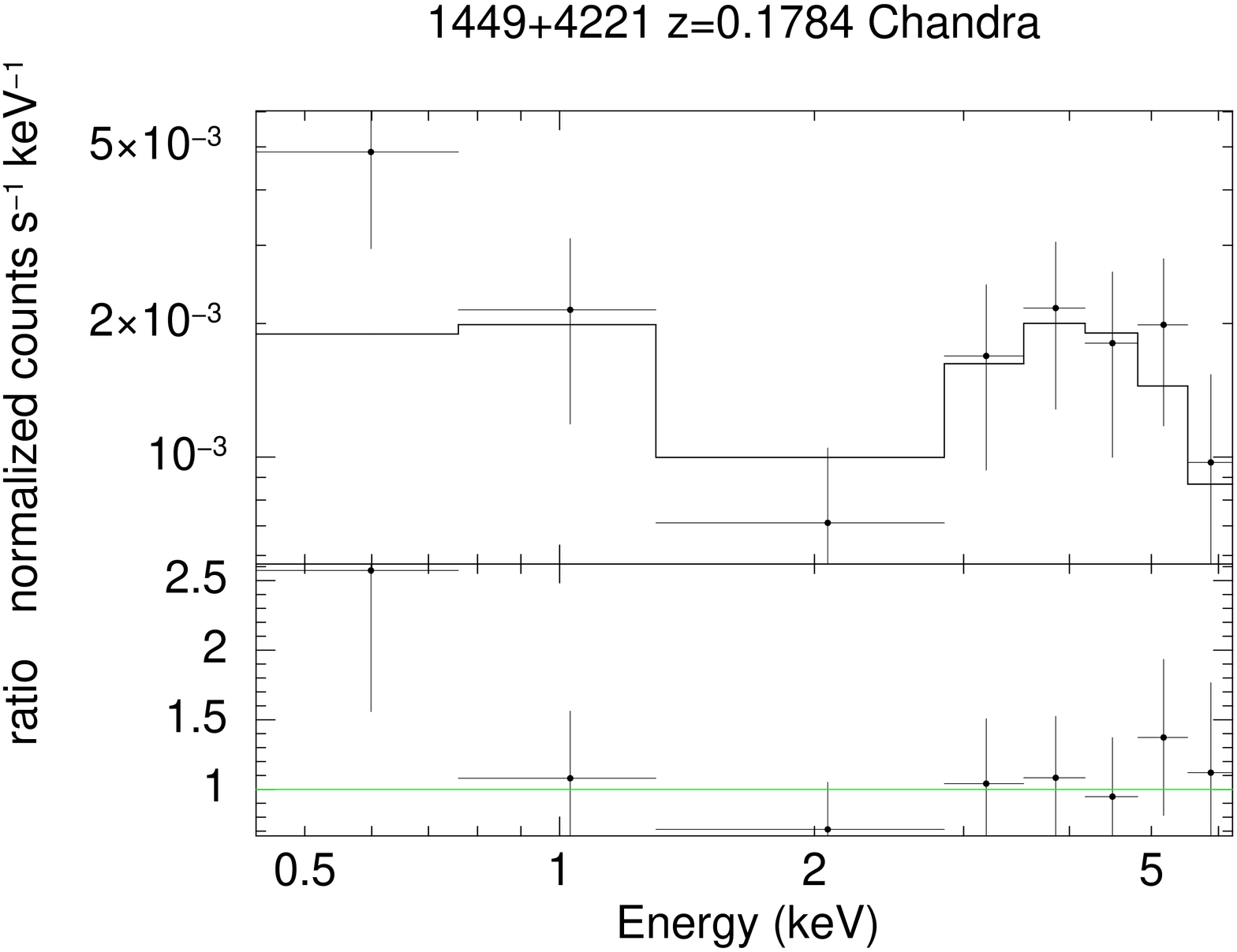,width=0.25\linewidth,angle=-90,clip=} &
\epsfig{file=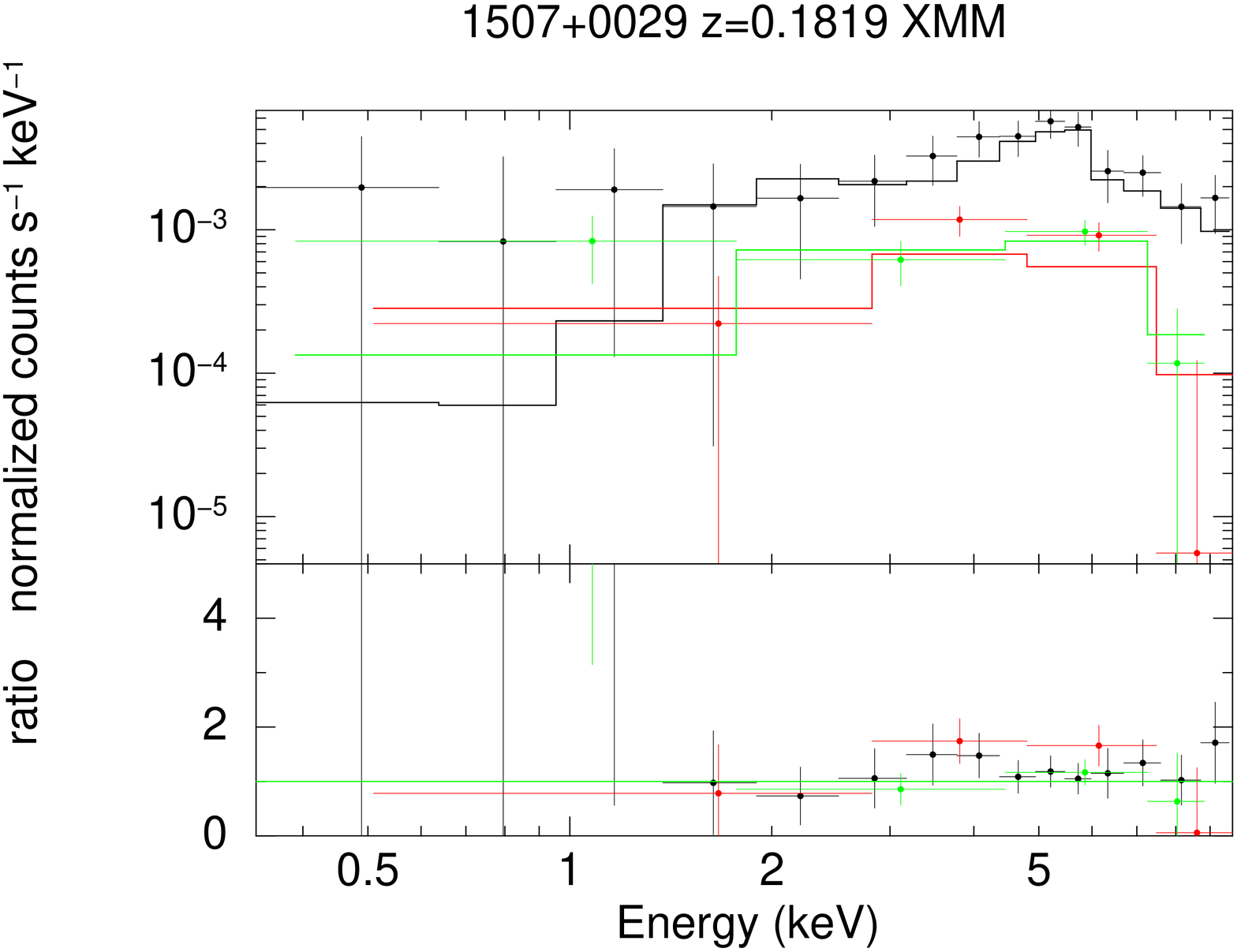,width=0.25\linewidth,angle=-90,clip=} \\
\epsfig{file=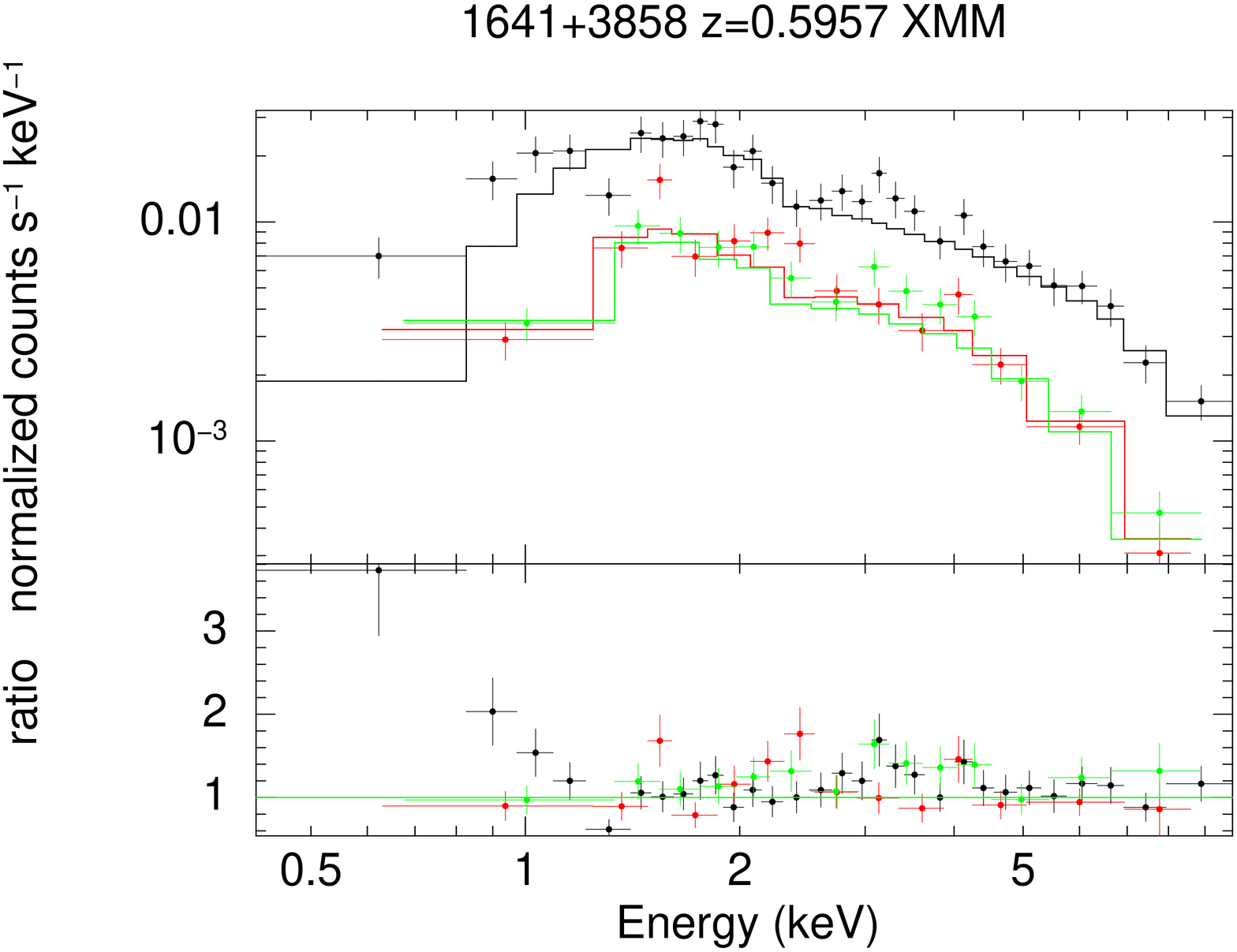,width=0.25\linewidth,angle=-90,clip=} &
\epsfig{file=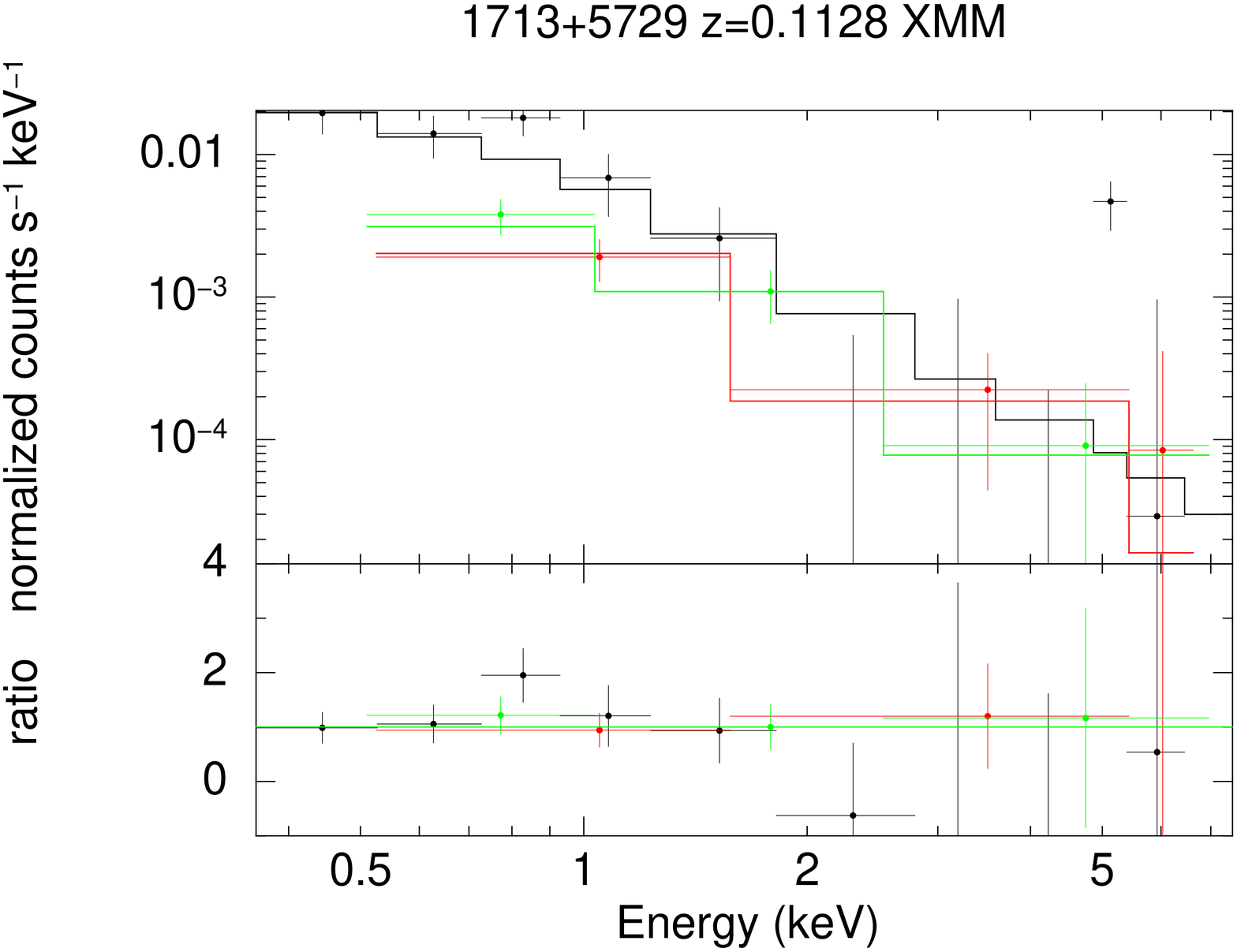,width=0.25\linewidth,angle=-90,clip=} &
\epsfig{file=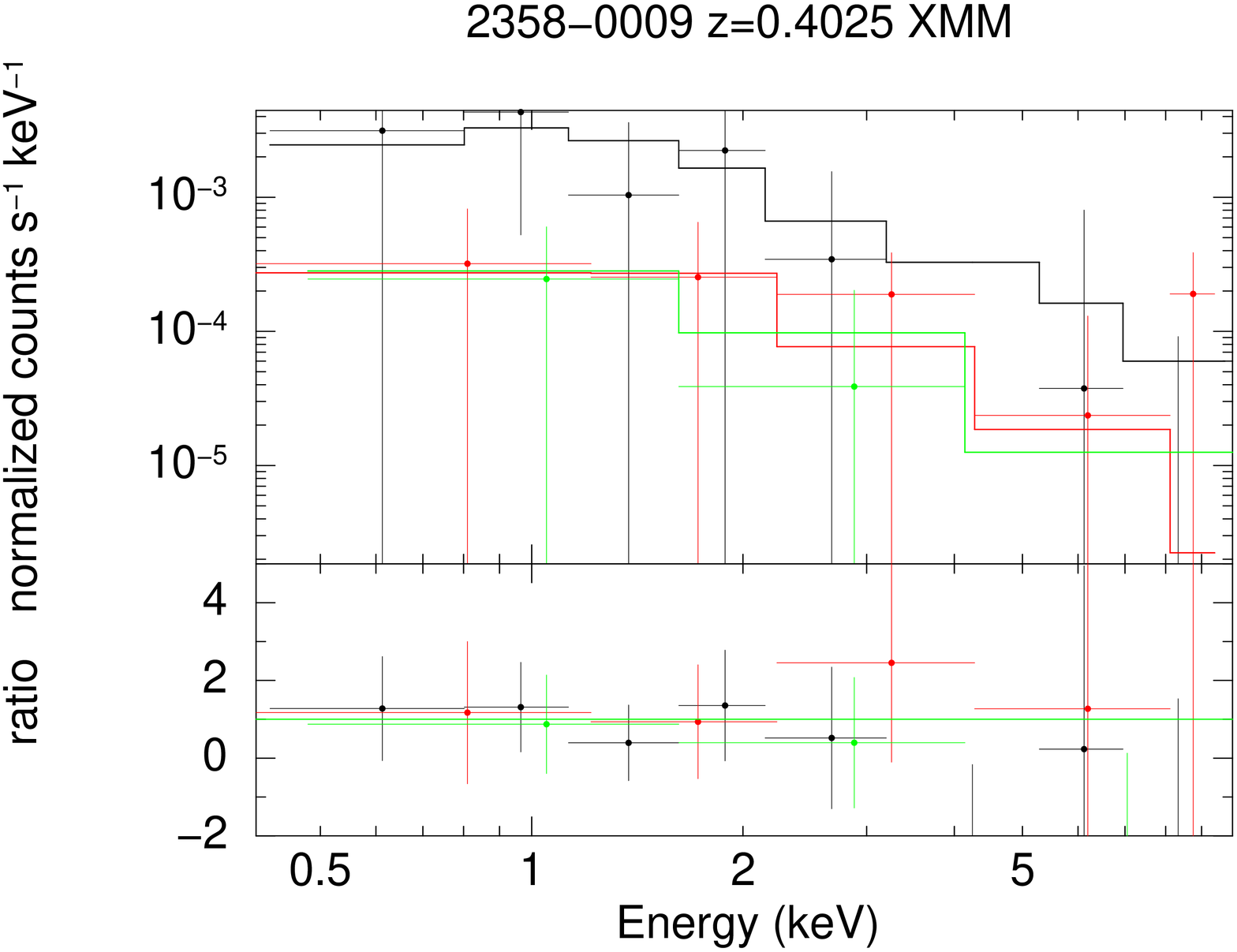,width=0.25\linewidth,angle=-90,clip=} \\
\end{tabular}
\centerline{Figure \ref{f:all}. --- {\it Continued}}
\end{figure*}

\begin{figure}
\centering
\epsfig{file=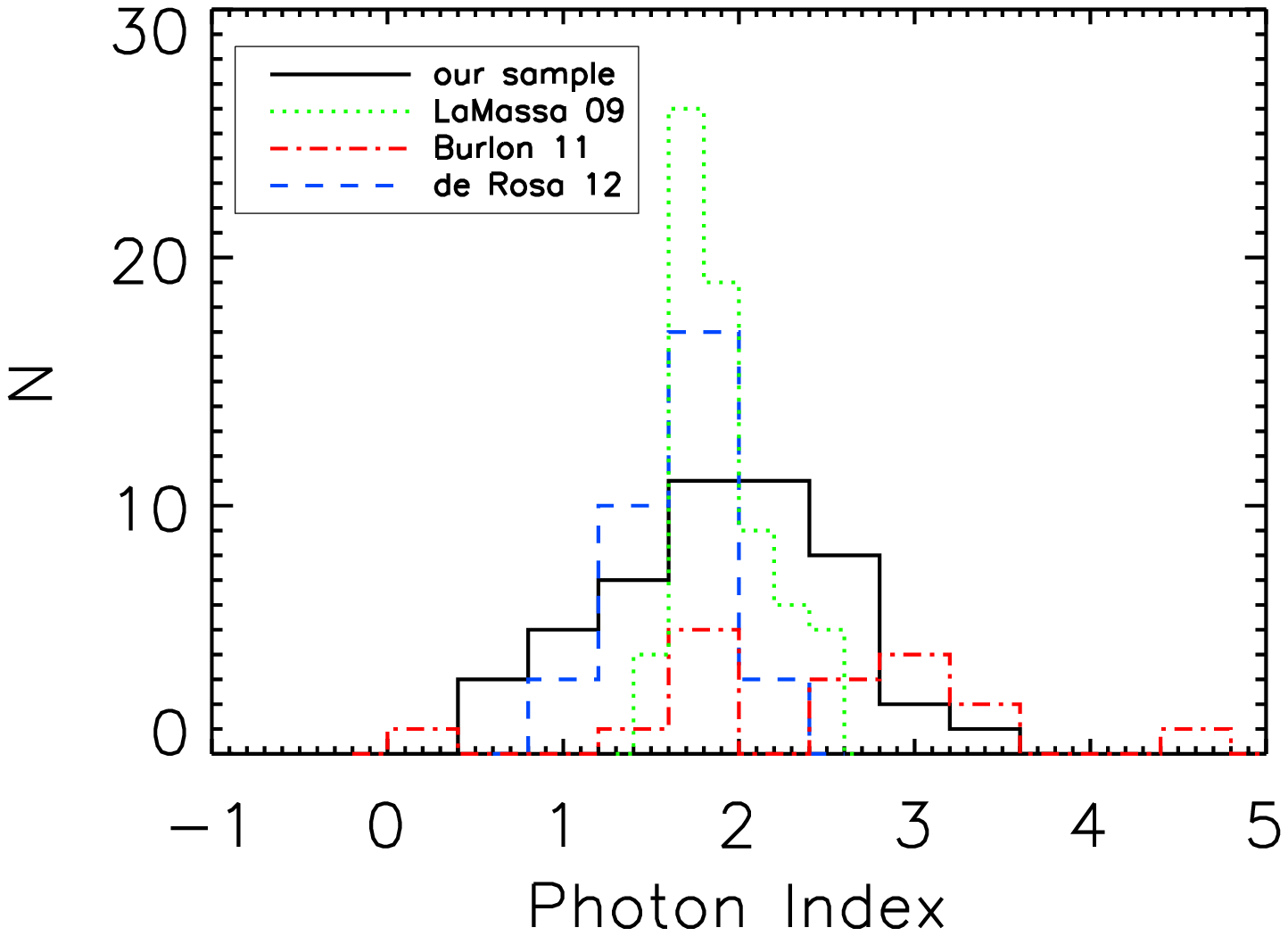,width=1.0\linewidth,clip=} 
\caption{Histograms of photon indices of the absorbed power-law spectral fits 
of our sample (solid black line).  We also show the sample of type 2 AGNs from 
{\it SWIFT}-BAT survey \citep[green dotted line]{2011ApJ...728...58B}, 
the sample of hard X-ray selected 
obscured AGNs from {\it INTEGRAL}
\citep[dashed blue line]{2012MNRAS.420.2087D} and the sample of 
optically selected local Seyfert 2s \citep[dot-dashed red line]{lamassa09} 
for comparison. \label{f:gamma}}
\end{figure}

\begin{figure}
\centering
\epsfig{file=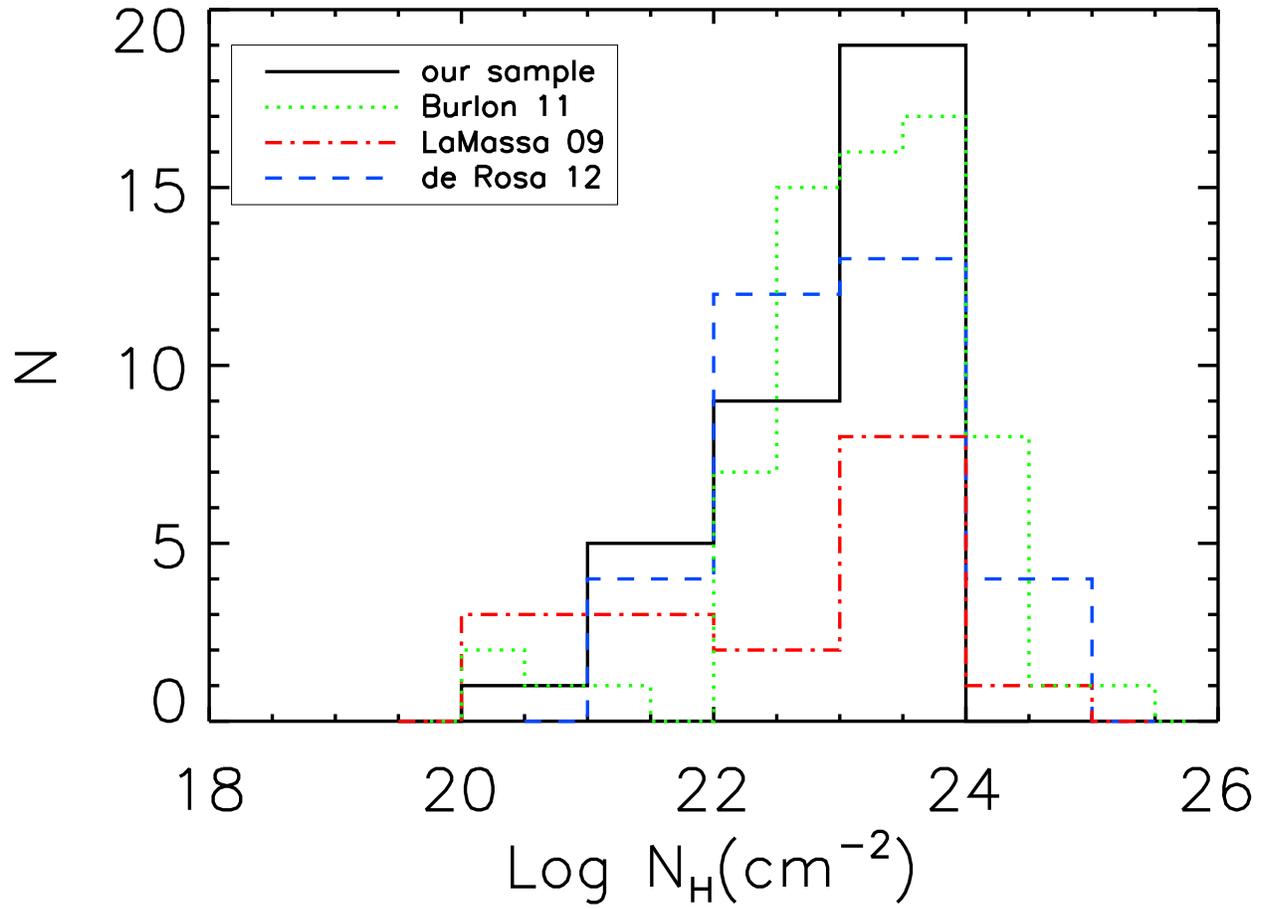,width=1.0\linewidth,clip=} 
\caption{Histograms of column densities of the absorbed power-law spectral 
fits. The samples and line styles are the same as indicated in Figure 
\ref{f:gamma}. \label{f:nh}}
\end{figure}

\begin{figure}
\centering
\epsfig{file=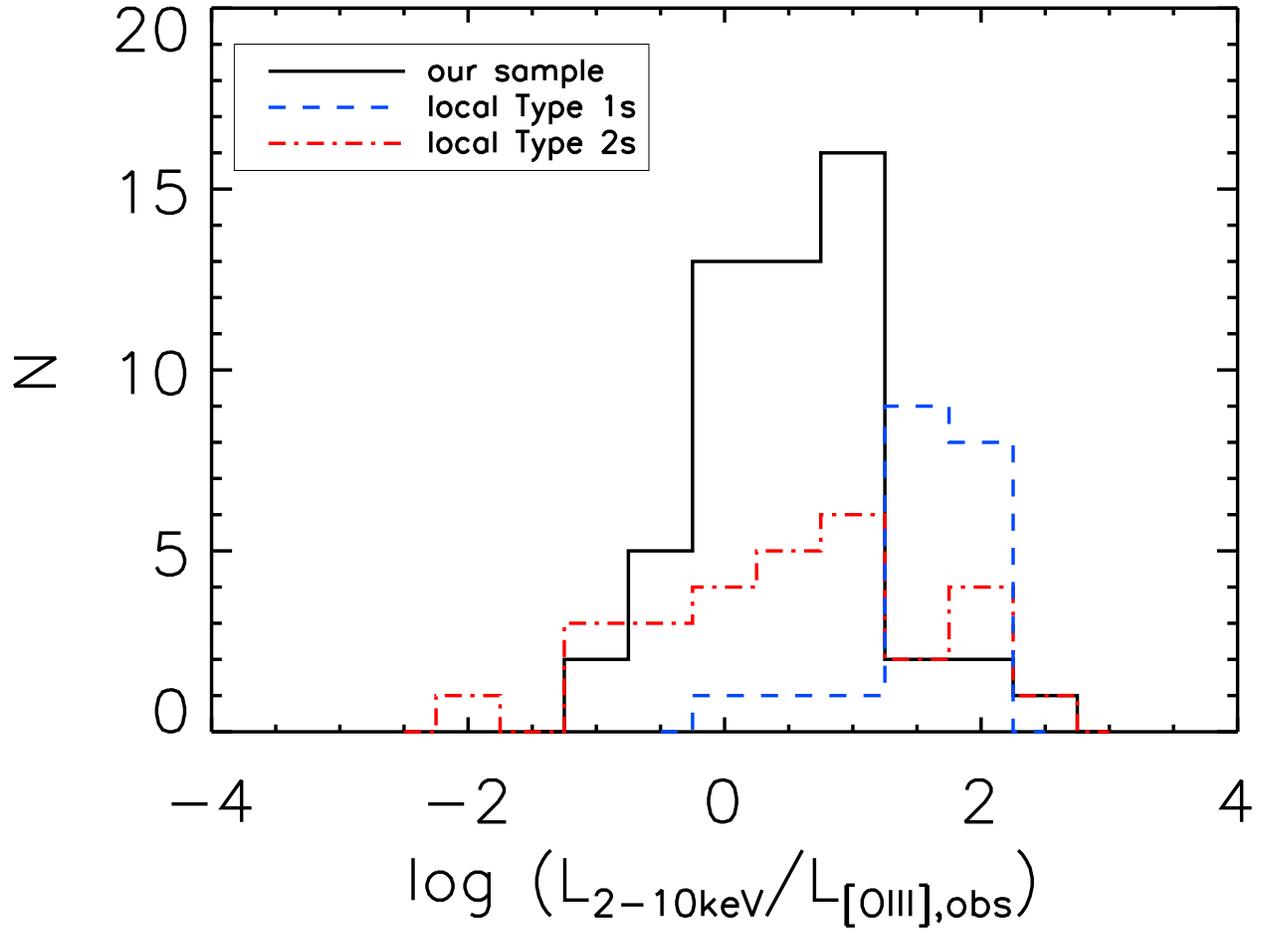,width=1.0\linewidth}
\caption{Histograms of the ratio of the hard X-ray and observed 
$\oiii$$\lambda 5007$ emission-line luminosity for local Type 1 (dashed blue 
line) and Type 2 (dash-dotted red line) of the samples in \cite{heckman05} 
and our type 2 quasar sample (solid black line). \label{f:lxlo}}
\end{figure}

\begin{figure}
\centering
\epsfig{file=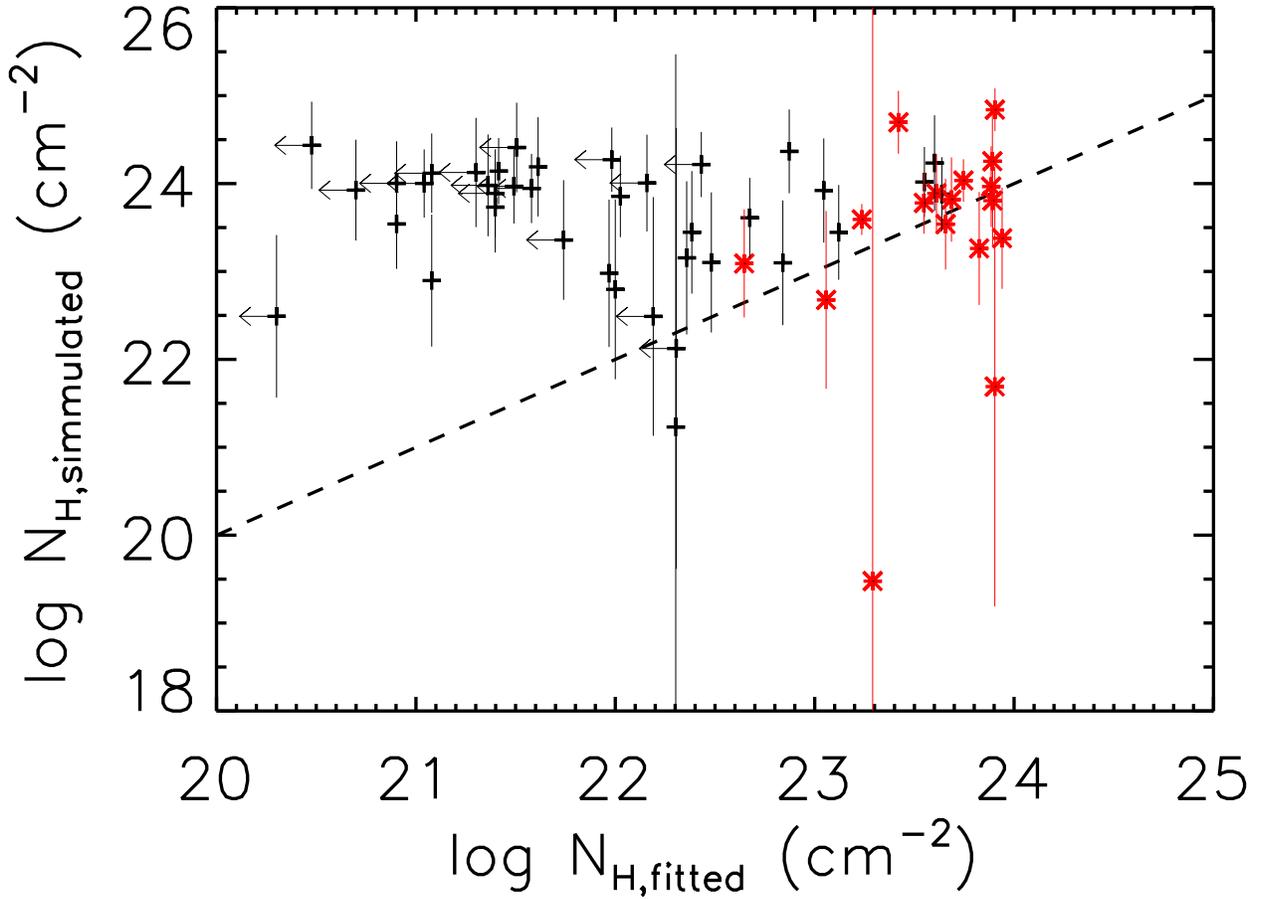,width=1.0\linewidth}
\caption{Simulated column densities vs. the values from the best-fitting spectral fits. 
The dashed line indicates where the two values are equal. Black and red symbols 
represent the single- and double-absorber model results, respectively.\label{f:nhsim}}
\end{figure}

\begin{figure}
\centering
\epsfig{file=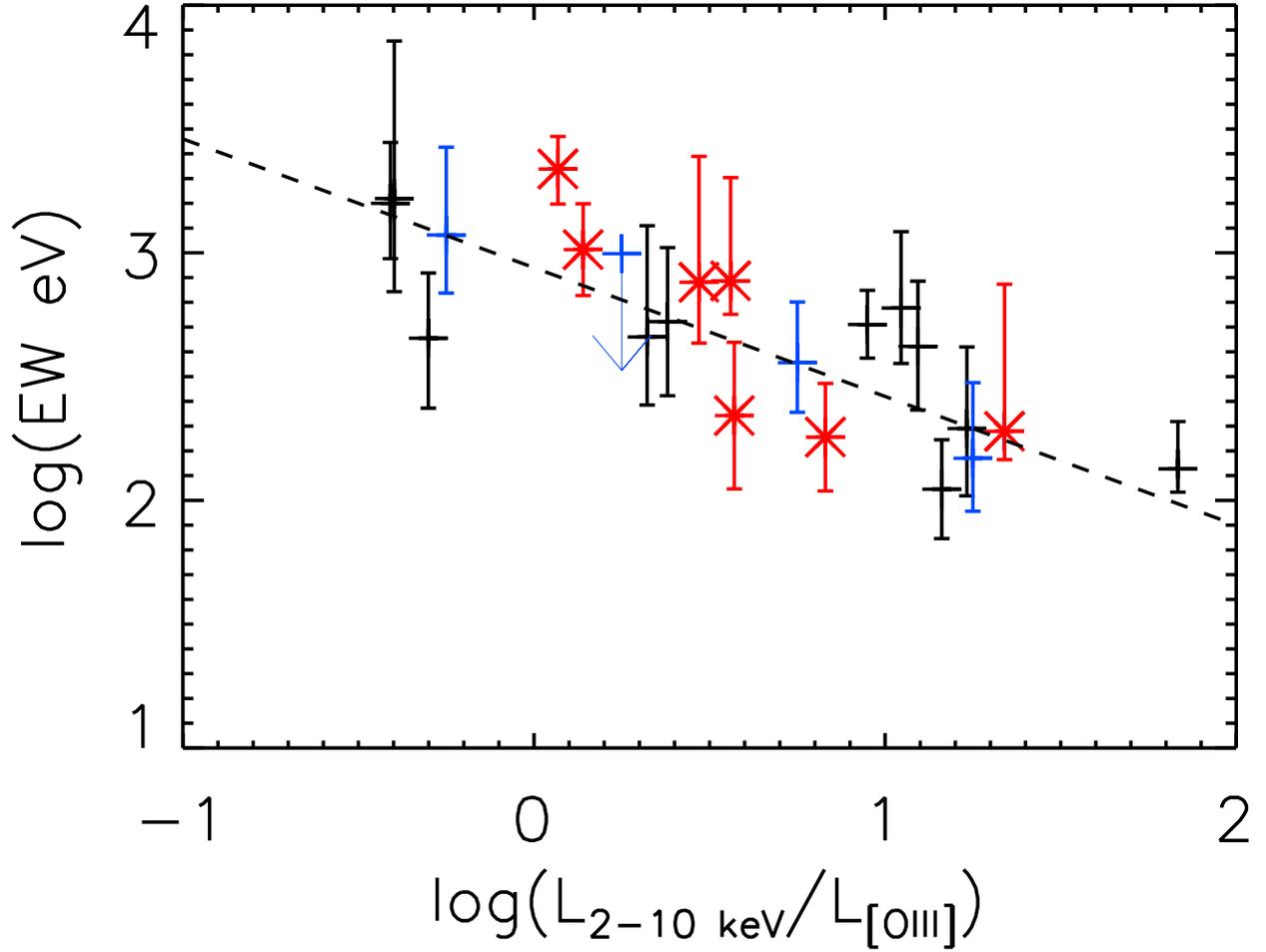,width=1.0\linewidth}
\caption{Equivalent width of Fe K$\alpha$ emission line vs. $L_{\rm 2-10keV}$/\loiii. 
The data in black and blue are from Table \ref{t:fe} and \ref{t:stack} in our 
sample, and those in red are from \cite{lamassa09}.\label{f:felx}}
\end{figure}

\begin{figure}
\centering
\epsfig{file=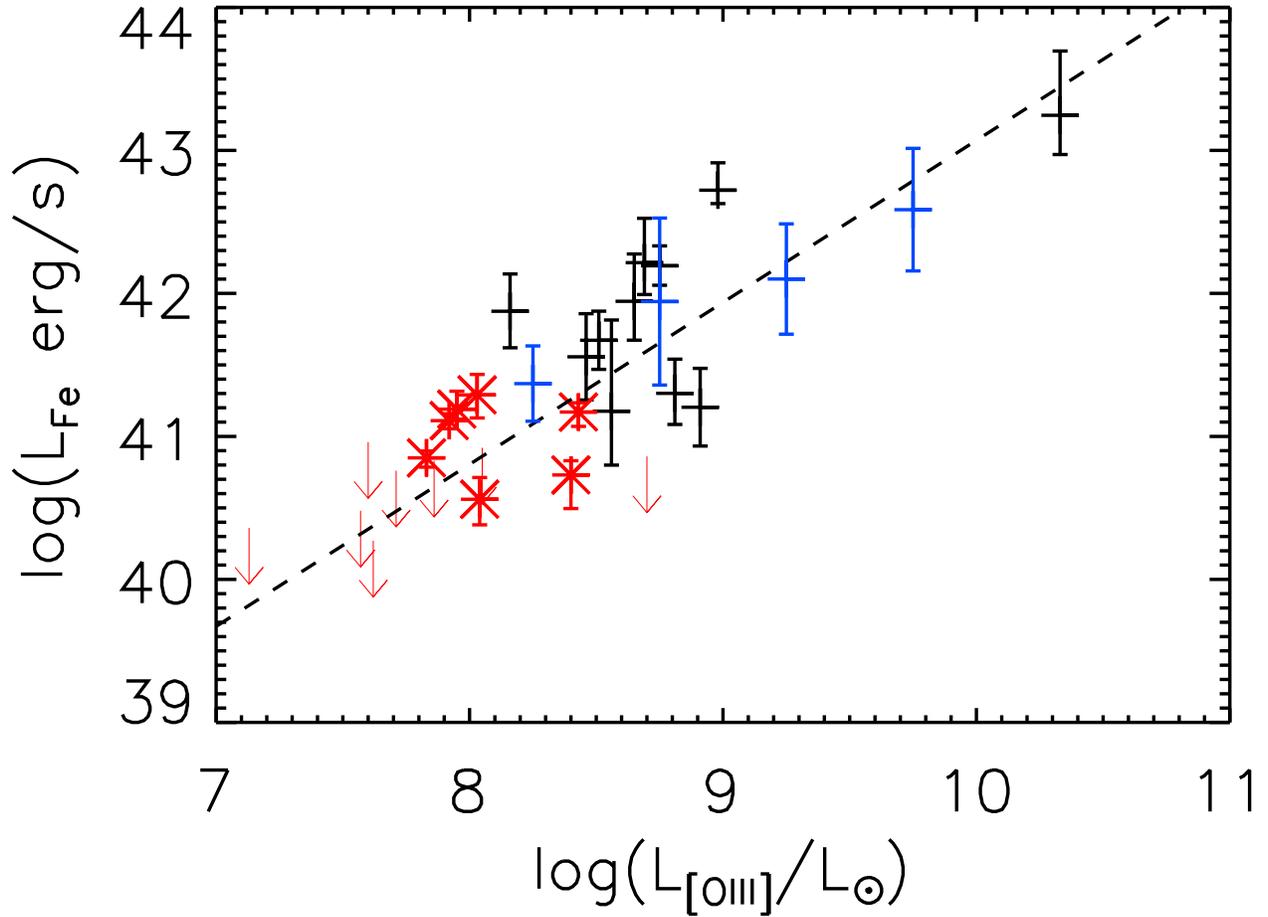,width=1.0\linewidth}
\caption{Fe K${\alpha}$ luminosity vs. $\oiii$ luminosity. The data in red 
are the sample of type 2 Seyfert galaxies from \cite{lamassa09}. The black 
symbols indicate the quasars having iron line detections listed in Table 
\ref{t:fe}, and the blue symbols indicate those from stacking. 
\label{f:feloiii}}
\end{figure}

\begin{figure}
\centering
\epsfig{file=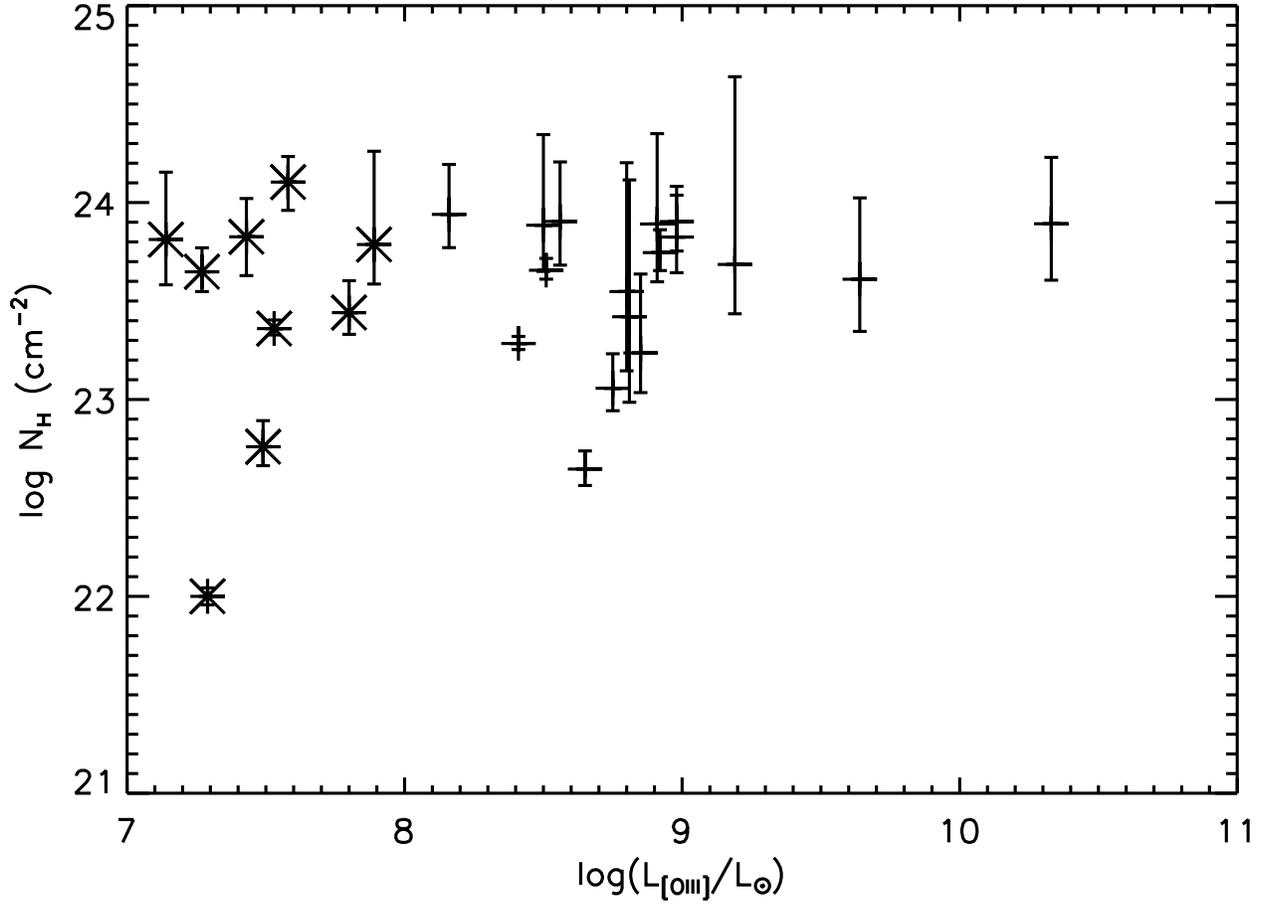,width=1.0\linewidth}
\caption{Column density of the second absorber ($N_{\rm H,2}$) in Table 
\ref{t:fits} vs. \oiii\ luminosity. The crosses are our type 2 quasar sample, 
while the asterisks are the type 2 Seyferts from \cite{lamassa09}. There 
is no correlation between column density and luminosity. \label{f:nh2loiii}}
\end{figure}

\begin{figure}
\centering
\epsfig{file=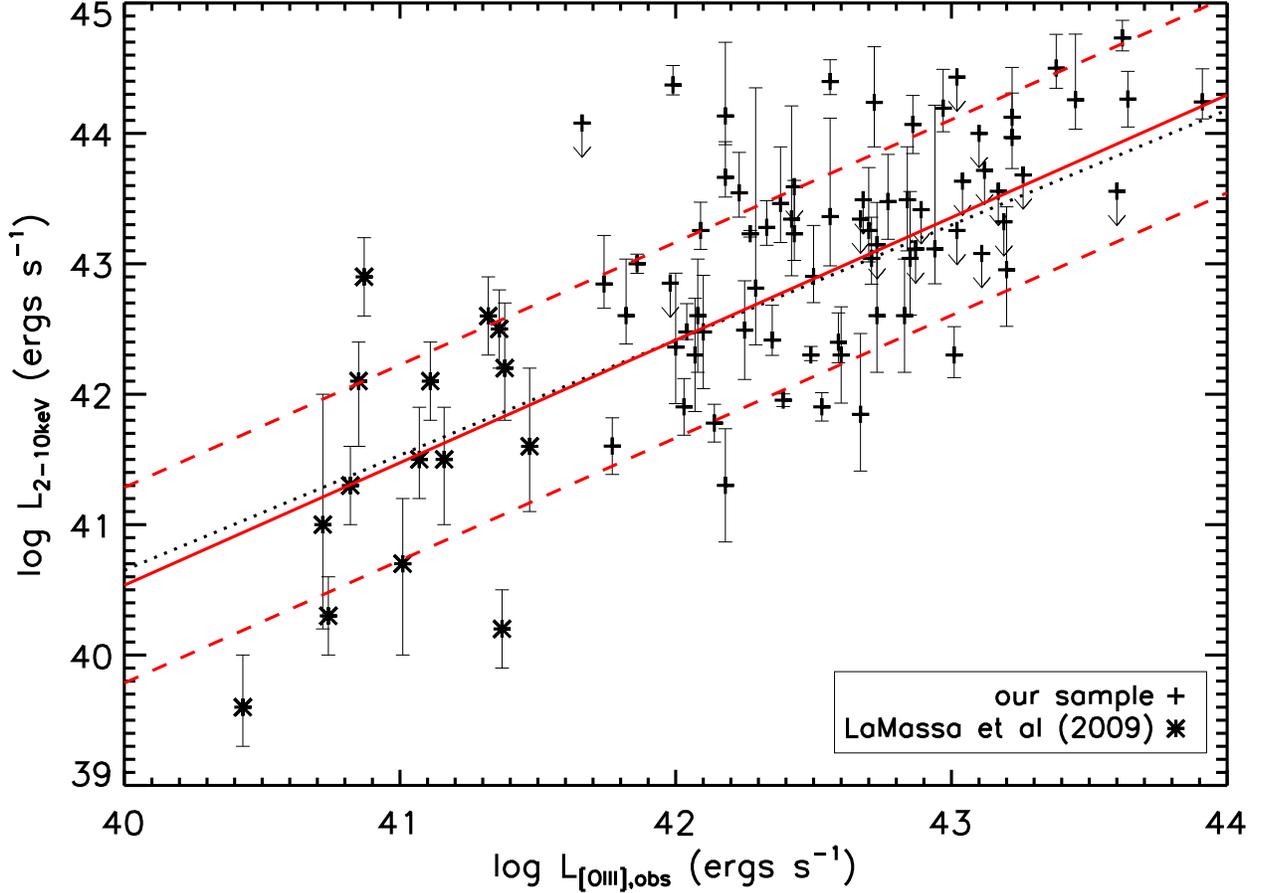,width=1.0\linewidth}
\caption{The log of the 2-10 keV X-ray luminosity plotted versus the
log of the \oiii\ luminosity. The pluses show our type 2 quasar sample, 
while the asterisks are the type 2 Seyfert galaxies in \cite{lamassa09}. 
The best fit (dotted line) slope (which includes the non-detections 
in X-rays) is $0.88\pm0.11$, and is not significantly different 
from unity. Thus the degree of X-ray obscuration does not depend on 
AGN luminosity. The solid-red line indicates the best fit slope of the
sample of type 1 AGNs given by \cite{jin12} with a shift of
1.26 dex downward to line up with the sample in our paper. The dashed-red 
lines indicate the $\pm 1\sigma$ deviation for the data points in this plot.
\label{f:lxlo2}}
\end{figure}

\begin{figure}
\centering
\begin{tabular}{c}
\epsfig{file=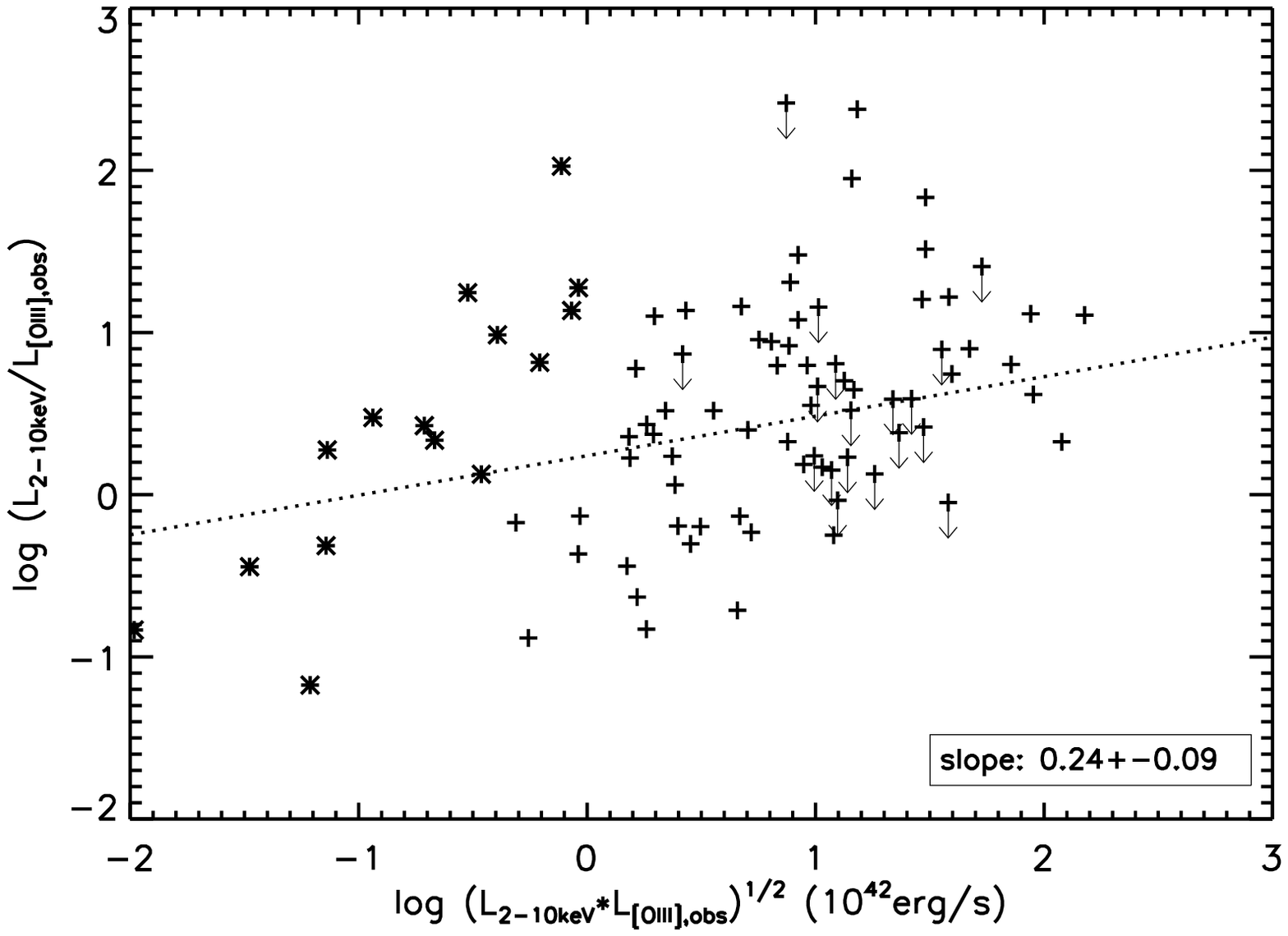,width=0.8\linewidth,angle=0,clip=} \\
\epsfig{file=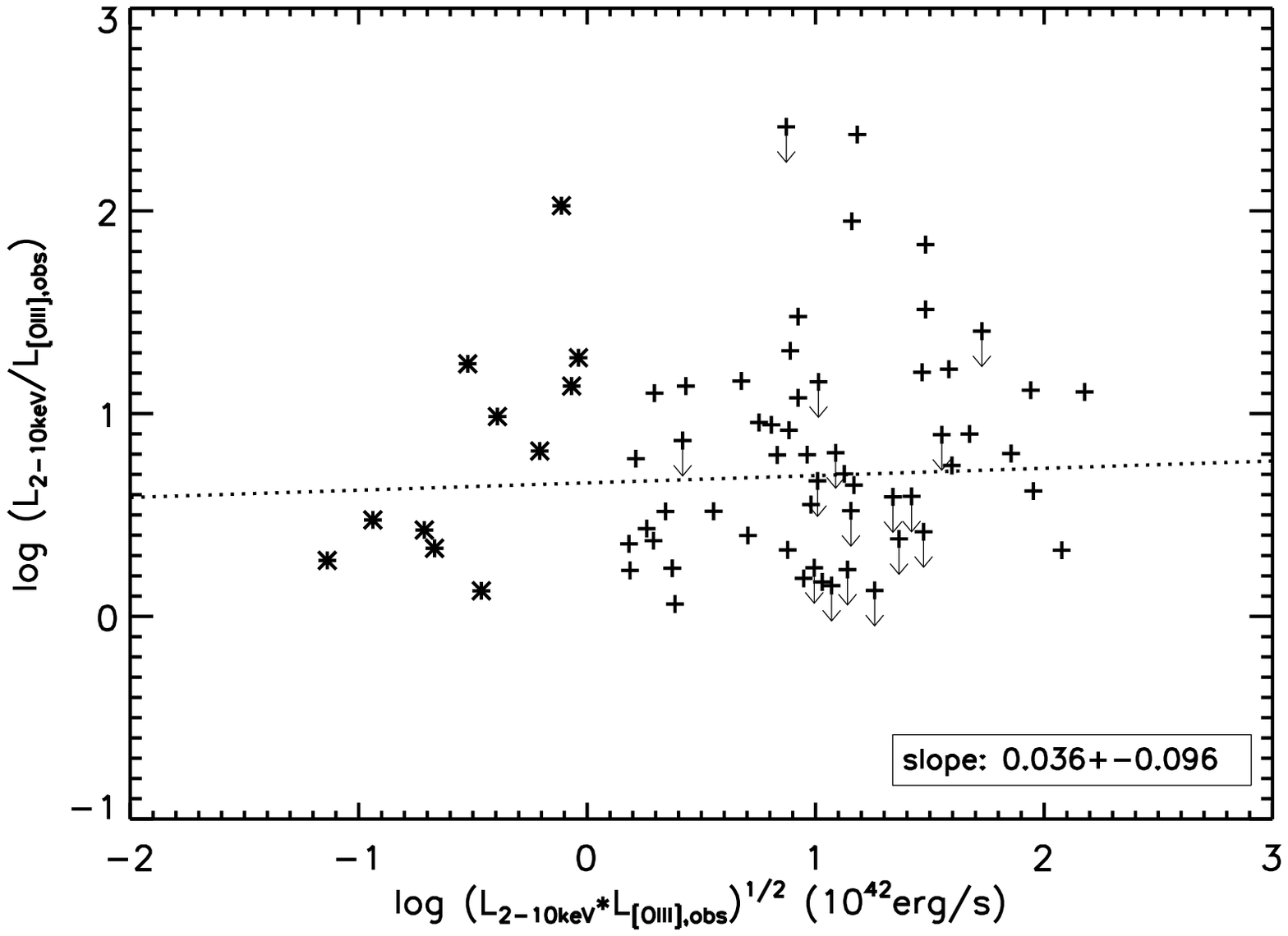,width=0.8\linewidth,angle=0,clip=} \\
\end{tabular}
\caption{\lx/\loiii\ vs. $(L_{\rm X}\cdot L_{\rm OIII})^{1/2}$. The upper panel 
includes all objects from our sample (plus symbols) and 
\citet[asterisk symbols]{lamassa09}. The lower panel excludes those
with $L_{\rm X}/L_{\rm OIII}<1$. \label{f:lxlo_geo}}
\end{figure}

\begin{figure}
\centering
\epsfig{file=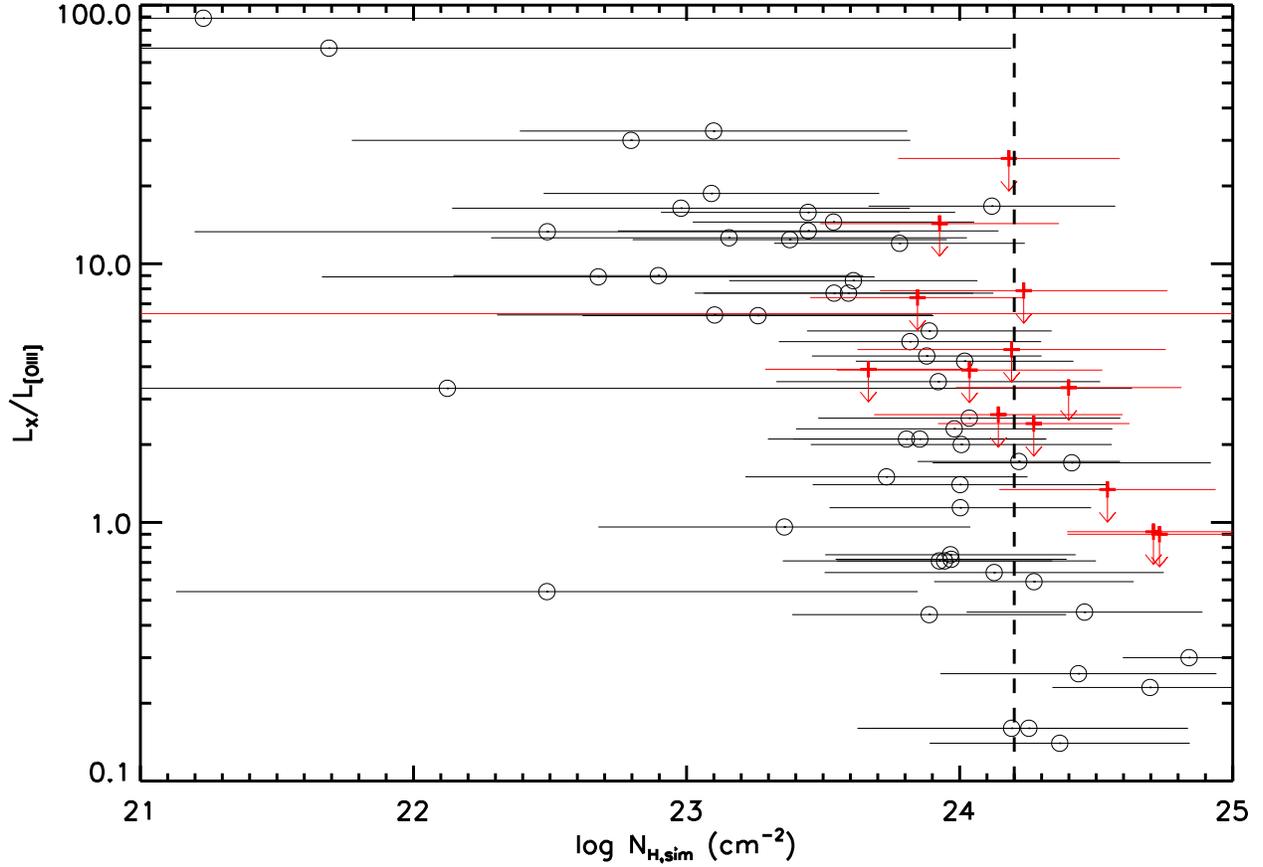,width=1.0\linewidth}
\caption{Simulated column density vs. observed hard X-ray to $\oiii$ 
luminosity ratio. The open circles represent the AGNs whose hard X-ray 
luminosities were derived from their spectral fits listed in Table 
\ref{t:fits}. The red plus symbols represent upper limit 
cases in Table \ref{t:upper}. The dashed vertical line denotes the region 
where $\rm{N}_{\rm{H,simulated}}>1.6\times 10^{24}$ cm$^{-2}$. These objects 
are designated as Compton-thick AGNs in this work. \label{f:lxlonh}}
\end{figure}

\begin{figure}
\centering
\epsfig{file=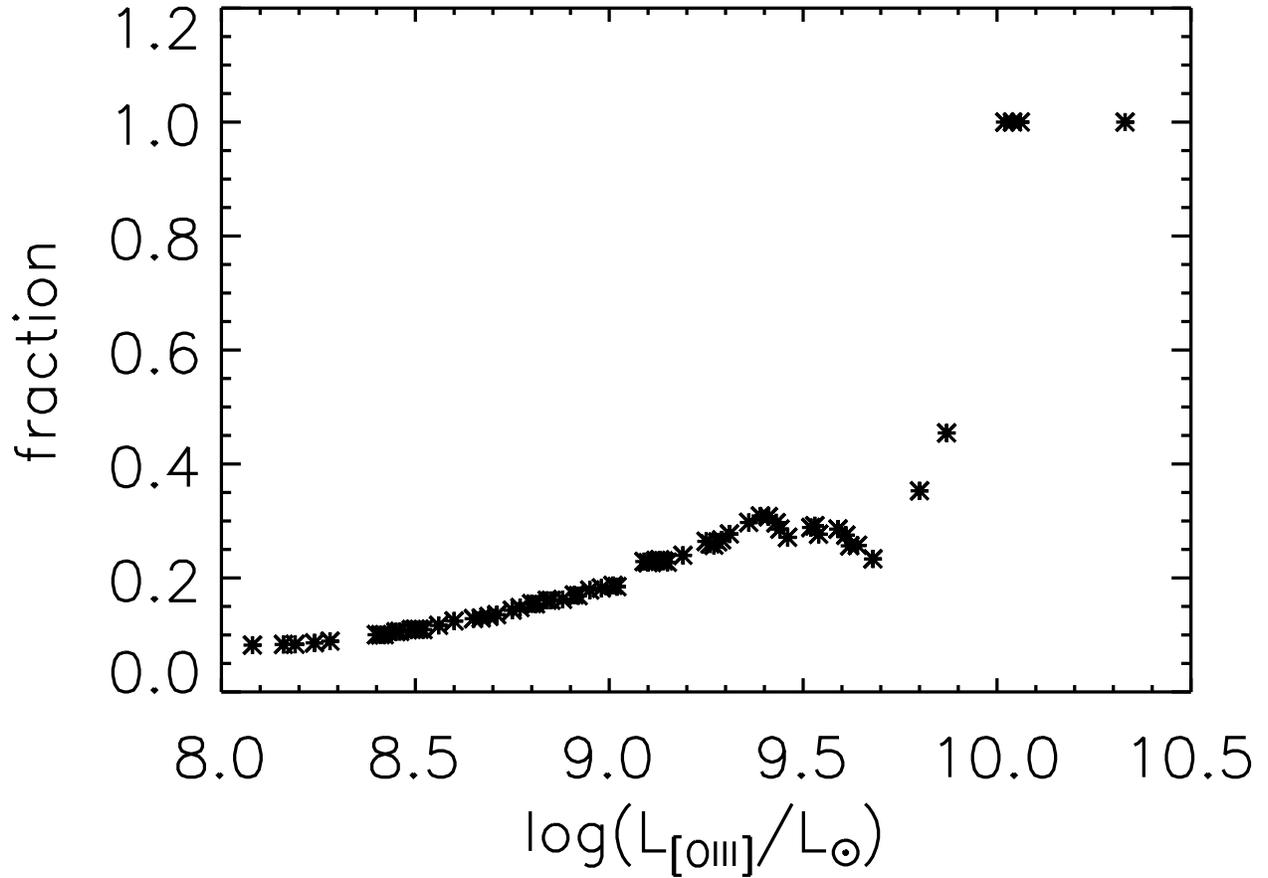,width=1.0\linewidth}
\caption{The completeness of our sample in the catalog of \cite{reyes08} as a 
function of \oiii\ luminosity. The fraction is calculated as the number of 
AGNs in our sample above a given \oiii\ luminosity (X-axis) divided by the 
number of all the AGNs in Reyes' sample above the same \oiii\ 
luminosity.\label{f:completeness}}
\end{figure}

\clearpage
\begin{figure}
\centering
\begin{tabular}{ccc}
\epsfig{file=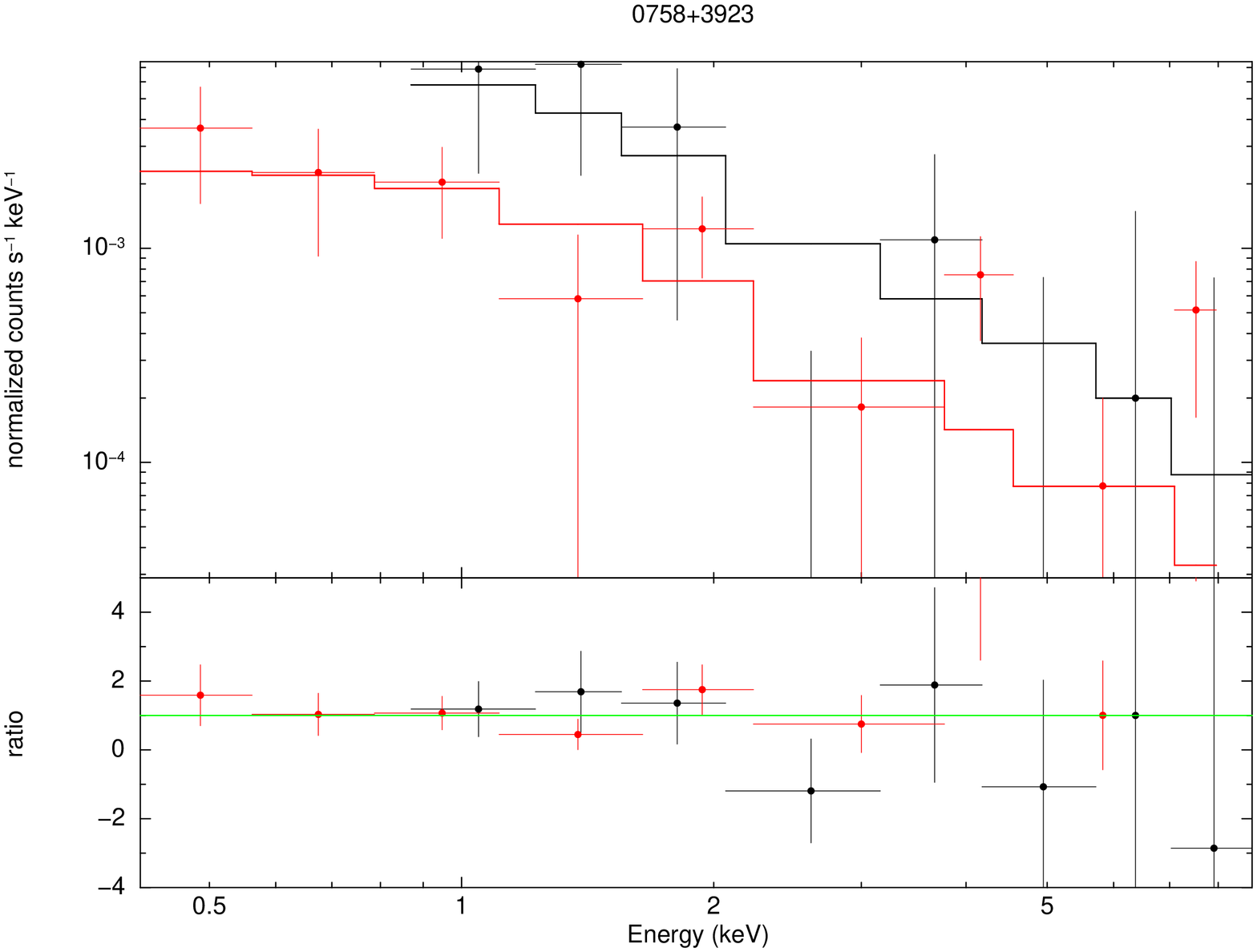,width=0.2\linewidth,angle=-90,clip=} &
\epsfig{file=f13b.eps,width=0.2\linewidth,angle=-90,clip=} &
\epsfig{file=f13c.eps,width=0.2\linewidth,angle=-90,clip=} \\
\epsfig{file=f13d.eps,width=0.2\linewidth,angle=-90,clip=} &
\epsfig{file=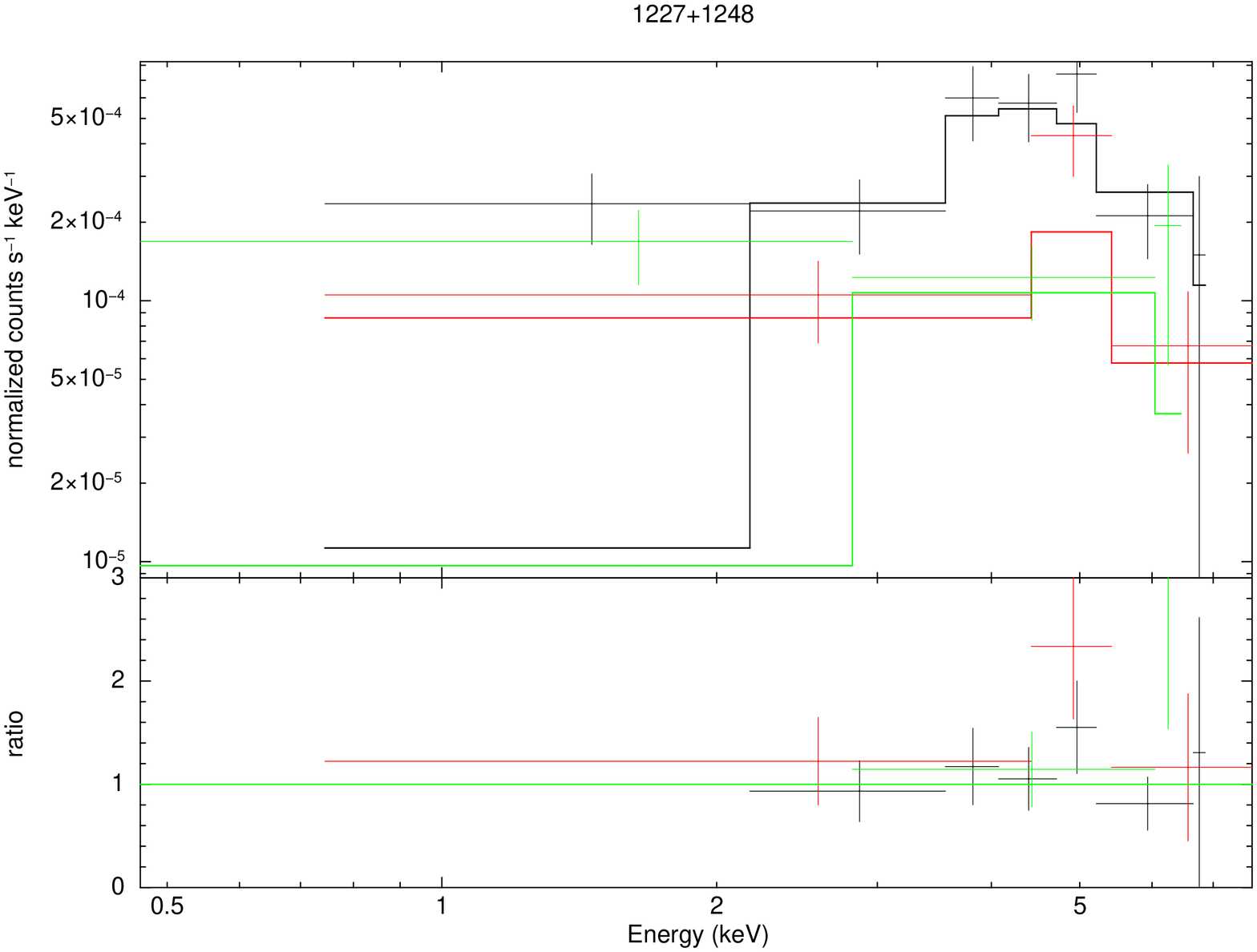,width=0.2\linewidth,angle=-90,clip=} &
\epsfig{file=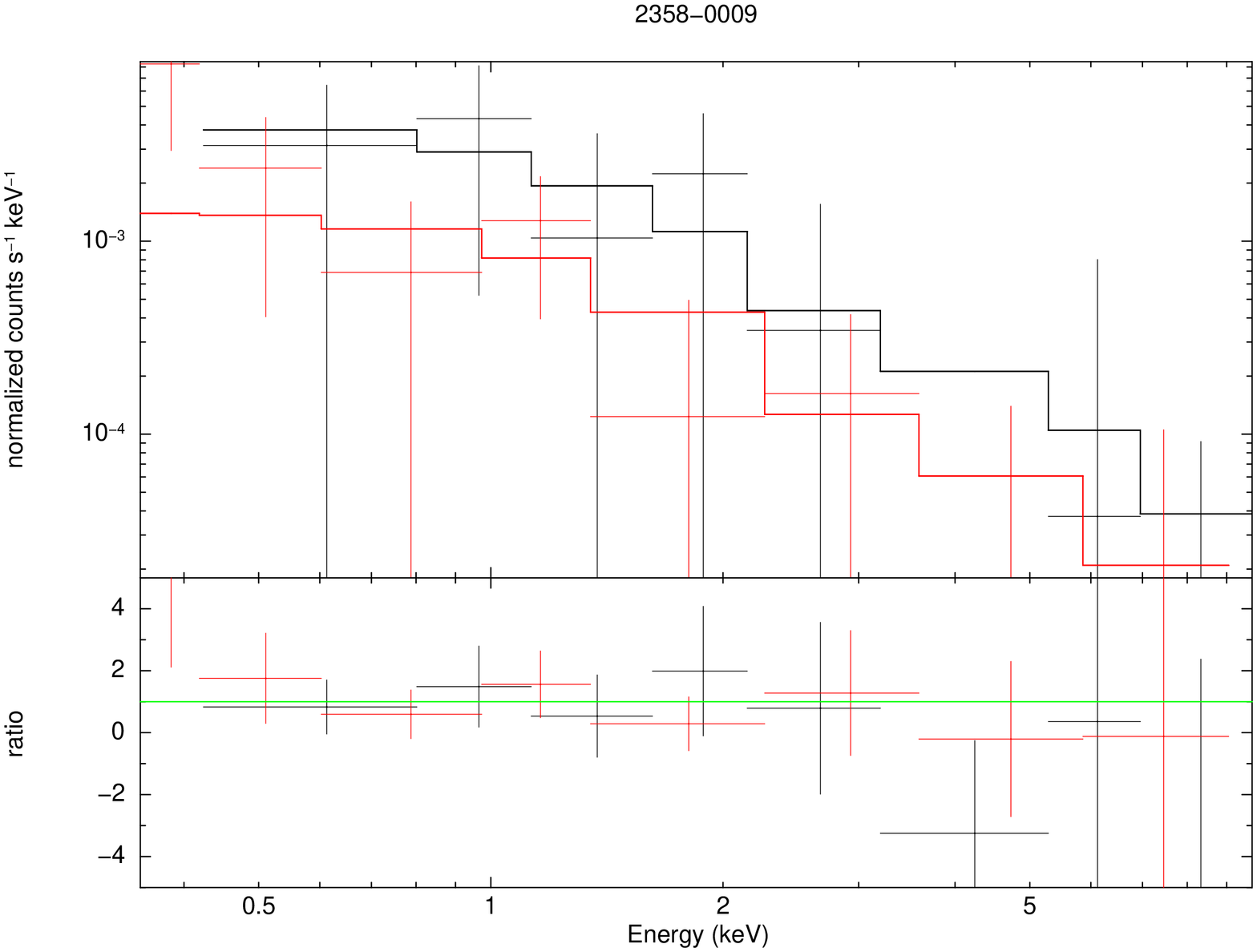,width=0.2\linewidth,angle=-90,clip=} \\
\end{tabular}
\caption{{\it SDSS~J0758+3923}: The symbols in black indicate the data obtained by 
\xmm-0305990101, and the red symbols are from \xmm-0406740101. Only PN 
detections are shown in this plot; {\it SDSS~J0834+5534}: The symbols in black
indicate the data obtained by \ch-4940, the red symbols are from \ch-
1645, and the green ones are PN data of \xmm-0143653901; 
{\it SDSS~J0900+2053}: The symbols in black indicate the data obtained by 
\ch-10463, the red symbols are from \ch-7897, and PN data of \xmm-0402250701
are in green color; {\it SDSS~J0913+4056}: The 
symbols in black and red indicate the data obtained by \ch-10445 and \ch-509, and 
symbols in green indicate the PN data from \xmm-0147671001; 
{\it SDSS~J1227+1248}: The symbols in black, red and green 
indicate the data obtained by \ch-5912, 9509 and 9510, respectively; {\it 
SDSS~J2358-0009}: The symbols in black, red indicate the data obtained by \xmm-
0303110301 and 0303110801, respectively. \label{f:multi}}
\end{figure}

\end{document}